\documentclass[pra, aps, onecolumn, groupedaddress, showpacs, superscriptaddress]{revtex4-2}
\usepackage{amssymb,amsmath,amsthm,color,graphicx,times,graphicx}
\usepackage{hyperref}
\usepackage{subfigure}
\usepackage{times}
\usepackage{bbold}
\usepackage{color}
\usepackage{amsmath}
\usepackage{amssymb}

\newcommand{\bra}[1]{\langle{#1}|}
\newcommand{\ket}[1]{|{#1}\rangle}

\providecommand{\openone}{\leavevmode\hbox{\small1\kern-4.3pt\normalsize1}}

\theoremstyle{plain}

\theoremstyle{definition}

\usepackage{orcidlink}
\makeatletter
\newsavebox{\@brx}
\newcommand{\llangle}[1][]{\savebox{\@brx}{\(\m@th{#1\langle}\)}%
	\mathopen{\copy\@brx\mkern2mu\kern-0.9\wd\@brx\usebox{\@brx}}}
\newcommand{\rrangle}[1][]{\savebox{\@brx}{\(\m@th{#1\rangle}\)}%
	\mathclose{\copy\@brx\mkern2mu\kern-0.9\wd\@brx\usebox{\@brx}}}
\makeatother
\begin{document}
	\title{Vacuum fluctuation induced quantum resource harvesting in triple-layer graphene}
	\author{Yassine Dakir \orcidlink{0009-0005-3408-1309}}\affiliation{LPHE-Modeling and Simulation, Faculty of Sciences, Mohammed V University in Rabat, Rabat, Morocco.}
	\author{Abdallah Slaoui \orcidlink{0000-0002-5284-3240}}\affiliation{LPHE-Modeling and Simulation, Faculty of Sciences, Mohammed V University in Rabat, Rabat, Morocco.}\affiliation{Centre of Physics and Mathematics, CPM, Faculty of Sciences, Mohammed V University in Rabat, Rabat, Morocco.}
	\author{Rachid Ahl Laamara \orcidlink{0000-0001-8410-9983}}\affiliation{LPHE-Modeling and Simulation, Faculty of Sciences, Mohammed V University in Rabat, Rabat, Morocco.}\affiliation{Centre of Physics and Mathematics, CPM, Faculty of Sciences, Mohammed V University in Rabat, Rabat, Morocco.}
	
\begin{abstract}
We examine the non-Markovian dynamics and the generation of quantum coherence and entanglement within a triple-layer graphene (TLG) system embedded in a planar microcavity. Using time-dependent perturbation theory, we derive an exact analytic solution for the system and demonstrate how the confined electromagnetic field mediates quantum correlations between the graphene layers. We employ three complementary measures; the relative entropy of coherence (REC) to quantify quantum coherence, the tangle to assess tripartite entanglement, and a non-Markovianity measure derived from the REC to characterize quantum memory effects. Our analysis reveals that these quantum resources exhibit remarkable sensitivity to various control parameters. Specifically, we demonstrate that the number of cutoff modes, the spatial positioning of the layers, the momentum parameter, and the interlayer rotation angles provide effective control over coherence, entanglement, and memory effects. We further show that these measures exhibit an exceptional sensitivity to the rotation angle between the layers. Ultimately, our results establish cavity-confined TLG as a highly tunable platform for exploring vacuum-mediated quantum phenomena, providing a framework for the precise manipulation of quantum correlations in graphene-based photonic and optoelectronic devices.		
\vspace{0.25cm}
\end{abstract}
\date{\today}
	
\maketitle
\section{Introduction}
Since its isolation more than two decades ago \cite{Novoselov2004}, graphene has garnered immense scientific interest. Its unique two-dimensional structure gives rise to exceptional electronic, mechanical, and optical properties—including high electron mobility, remarkable thermal conductivity, and optical transparency \cite{Zhuo2020,Mao2013,Drissi2011,Balandin2008,Kim2016,Huang2020,Salvo2018}. These attributes position graphene as a transformative material for advanced applications in electronics, optics, and mechanics \cite{Esteghamat2022,Loh2010,Khan2017}. The emergence of phenomena such as negative differential resistance in superlattices \cite{Rahighi2021} and the relativistic behavior of its massless charge carriers underscore graphene's position at the intersection of condensed matter and high-energy physics. Consequently, characterizing graphene \cite{Goerbig2011,McCann2013,Rozhkov2016,Abergel2010} has become a cornerstone of nanostructure research. The unique energy band structure of graphene, which exhibits a low-energy linear profile driven by a massless Dirac-like equation \cite{Neto2009,Novoselov2005}, is responsible for its remarkable electronic properties. Until now, most studies of graphene have focused on its unconventional properties when it comes to transport. However, quantum effects resulting through its interaction with a quantized electromagnetic field have been overlooked.\par

The interaction between graphene electrons and the quantum electromagnetic field has recently garnered considerable interest, given its potential to modify the material's electronic spectrum. It has been demonstrated that exposure to circularly polarized light can open a band gap in graphene, a phenomenon arising from the formation of electron-photon hybrid states analogous to polaritons in ionic crystals and quantum microcavities \cite{Liew2011,Riedrich2012,Low2017}. Similarly, from a quantum electrodynamics perspective, excitonic effects can occur even in the absence of real photons; electrons can interact with vacuum fluctuations by emitting and reabsorbing virtual photons \cite{Berestetskii1982}. This interaction with the quantum vacuum could explain the separation of valence and conduction bands without external optical pumping. These effects are particularly pronounced when the interaction volume is restricted—for instance, by confining the electronic system within a planar microcavity to enhance light-matter coupling at the quantum scale. This fundamental phenomenon has profound implications in various areas of quantum field theory and quantum information physics. It has been demonstrated that two spatially separate detectors can interact locally with the quantum vacuum and become entangled, with the correlations thus collected being accessible and exploitable by external observers. The earliest and most widely used theoretical description of detector–field interactions is provided by the UDW model \cite{DeWitt1979}, which formulates the simplest linear interaction between a localized system of two quantum levels and a scalar quantum field. The structure of the quantum vacuum is defined by its essential and intrinsically non-trivial entanglement characteristics, as reflected in this model. Over time, several extensions and improvements to the UDW model have been proposed.\par

These contributions range from experimental applications in atomic and superconducting circuit systems \cite{Olson2011,Olson2012,Sabin2012} to generalizations involving vector fields. In particular, the electromagnetic field, which inherently carries angular momentum, gives rise to anisotropy and directional dependencies in the mechanisms responsible for entanglement generation. It has also been demonstrated that the UDW model provides a valid approximation of light-matter interaction in regimes where no momentum transfer occurs. More recent work has focused on spatially delocalized coherent matter systems, for which the propagation of the center of mass wave function plays a decisive role in the ability of detectors to generate entanglement between them.\par

In this more general context, the exploration of light-matter interactions within two-dimensional materials, such as graphene, has become increasingly important. The integration of graphene in a microcavity can induce a band gap by circularly polarized illumination or via structural modifications produced by vacuum fluctuations in an optically active medium. These effects become even more pronounced in the ultra-strong coupling regime, where the photons of the cavity and the quasi-particles of the graphene recombine into nonperturbative hybrid states. Numerous studies have subsequently extended this framework to multilayer systems. In the decoupled double layers, exotic phases such as excitonic condensation or the fractional quantum Hall effect can emerge. Casimir-type interactions between layers or between graphene and metamaterials have revealed a strong dependence on optical conductivity, sometimes allowing the appearance of a universal value of conductivity ($e^{2}/4h$) \cite{Drosdoff2010}. The stacked multilayer graphene ABC, for its part, presents a superfluid phase due to a strong electron-hole pairing \cite{Zarenia2014}. Other works have applied the Schrieffer-Wolff transformation to two decoupled graphene layers placed in a microcavity, revealing exchange interactions dependent on distance, photon energy or torsion angle \cite{Profumo2010}. The two-layer graphene twisted at the magic angle has also demonstrated the appearance of superconductivity at low temperature \cite{Drosdoff2010,Bostrom2000,Bordag2000,Beenakker2008,Stauber2012,Fialkovsky2011}. Recently, an experimental framework was introduced to detect the impact of vacuum fluctuations in a microcavity directly. It is shown that two layers, although decoupled, can become entangled by interacting exclusively with the ground state of the confined field \cite{Ardenghi2018}.\par

In this context, several studies have applied the UDW model to two-dimensional materials such as double-layer graphene (DLG) and double-layer silicene, providing natural platforms for exploring confined light-matter interactions. It has been shown that two layers of graphene placed in a planar microcavity can become entangled via fluctuations in the electromagnetic vacuum, leading to entanglement harvesting and non-classical correlations even in the absence of actual photons. Similar results have been obtained for double-layer silicene, where the tunable band structure, strong spin-orbit interaction, and topological properties of the material allow control of entanglement dynamics \cite{Ardenghi2018,Bittencourt2017,Arreyes2024}. \par

Following on from these studies, this work aims to analyze QC, entanglement, and non-Markovian behavior in a TLG system confined within a planar microcavity. The addition of a third layer introduces a new dimension of inter-layer interactions, allowing for a more detailed examination of the role of coherent coupling, electromagnetic confinement, and vacuum fluctuations in the dynamics of coherence and non-Markovianity. By applying time-dependent perturbation theory, we show that the quantized cavity field can induce inter-layer coherence and reciprocal information flow between electrons in the three layers, revealing quantum memory regimes that depend on inter-layer distances and the number of cutoff modes. QC is quantified using the relative entropy of coherence, entanglement is measured through the tangle, and non-Markovianity is evaluated via temporal variations of REC. Our results establish that the TLG system provides a promising platform for investigating information backflow and vacuum memory effects in confined two-dimensional materials, offering new perspectives for manipulating quantum coherence and entanglement in graphene-based photonic devices. This paper is organized to provide a comprehensive analysis of quantum resources in the TLG-cavity system. Section (\ref{sec2}) presents the theoretical framework, while section (\ref{sec3}) applies time-dependent perturbation theory to derive the system dynamics. In this section, we analyze the evolution of coherence, entanglement, and non-Markovianity for various electronic states, examining their dependence on key parameters such as the number of cutoff modes, interlayer positions, momentum, and rotation angles. Finally, section (\ref{clc}) summarizes our main findings and discusses their implications for controlling quantum resources in cavity-confined TLG system.
\section{Theoretical model} \label{sec2}
\begin{figure}[hbtp]
		{{\begin{minipage}[b]{.4\linewidth}
					\centering
					\includegraphics[scale=0.7]{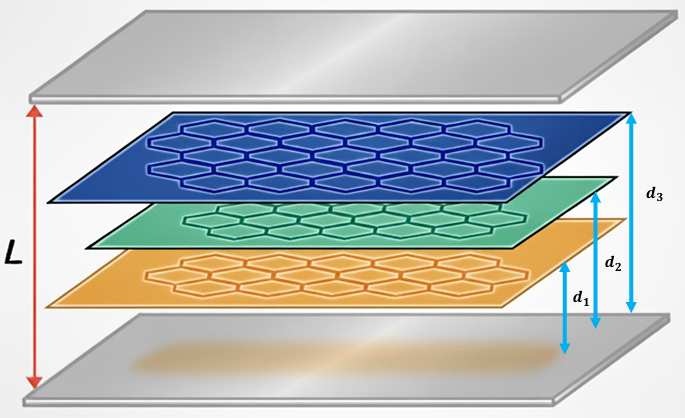} \vfill\vfill\vfill\vfill $\left(a\right)$
				\end{minipage}\hfill
				\begin{minipage}[b]{.45\linewidth}
					\centering
					\includegraphics[scale=0.7]{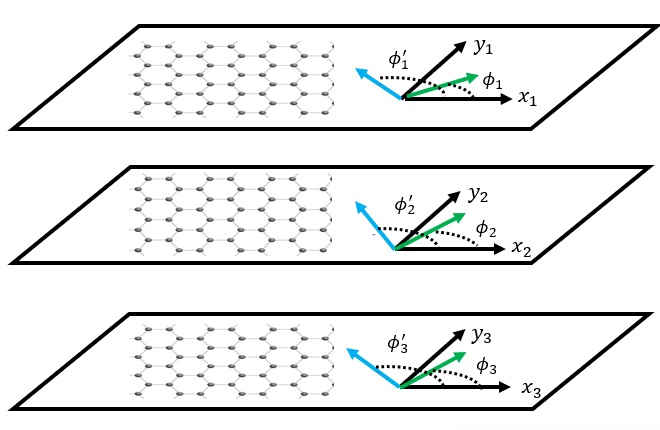} \vfill\vfill \vfill  $\left(b\right)$
		\end{minipage}}}
		\caption{ ($a$) Physical configuration of the triple-layer graphene cavity system and ($b$) the initial/final momentum angles of conduction electrons.} \label{graphene}
\end{figure}
Fig.(\ref{graphene}) depicts the physical configuration of the triple-layer graphene system embedded within a planar microcavity of width $L$. Each graphene layer interacts locally with the quantized electromagnetic field, while the exchange of virtual photons generates effective correlations between layers separated by distances $d_{i}$. The cavity’s confined field modes, determined by the mirror boundary conditions, discrete the spectrum and control the strength of these vacuum-induced interactions. The electronic properties of the honeycomb lattice layers are described within the low-energy Dirac approximation, where the electrons exhibit relativistic-like behavior. Accordingly, the Hamiltonian of the TLG, coupled to the cavity’s electromagnetic field, is expressed as
\begin{equation}
    H=H_{0}+H_{I}+H_{F},
\end{equation}
where,
\begin{equation}
H_{0}=v_{F}\sum_{i=1}^{3}\sigma_{i}\textbf{p}_{i}, \quad\quad H_{I}=-e v_{F} \sum_{i=1}^{3}\sigma_{i} \mathcal{A}_{i}, \quad\quad H_{F}=\sum_{n,q,\nu}\Omega a_{n q \nu}^{\dagger}a_{n q \nu}.
\end{equation}
Here, $H_{0}$ designates the Hamiltonian of the free electrons in each layer, the index $i$ traversing the three electrons, each located in a separate layer of graphene, which can occupy either the valence band or the conduction band. The vectors $\sigma_{i} = (\sigma_{i}^{x}, \sigma_{i}^{y})$ represent the Pauli matrices acting on the corresponding sublattice of each electron. The Fermi velocity $v_{F}$, characteristic of the two-dimensional material ($v_{F} = 10^{6} \mathrm{m/s}$). The potential vector $\mathcal{A}_{i}$ acts on each electron. The second term of the Hamiltonian corresponds to the electromagnetic field of the caity, with 
\begin{equation}
\Omega = c \sqrt{q^2 + \left( \frac{n \pi}{L} \right)^2},
\end{equation}
representing the frequency of the mode $n$ associated with the wave vector $q$, $c$ being the speed of light inside the cavity, $L$ the length of the latter, and $n$ the index of the mode. The operators $a_{n q \nu}$ and $a_{n q \nu}^{\dagger}$ are respectively the annihilation and creation operators of the cavity field, satisfying the canonical commutation relation
\begin{equation}
[a_{n q \nu}, a_{n' q' \nu'}^{\dagger}] = \delta_{n,n'} \, \delta_{q,q'} \, \delta_{\nu,\nu'}.
\end{equation}
Within the minimal coupling formalism $\text{p}\rightarrow \text{p}-e\mathcal{A}$, the interaction Hamiltonian $H_I$ characterizes the coupling between the electromagnetic field confined within the cavity and the electrons residing in the deformed honeycomb lattices. The quantized vector potential inside the cavity can then be expressed in terms of the creation and annihilation operators corresponding to each mode \cite{Ardenghi2018,Arreyes2024,Kibis2013}, according to
\begin{align}
    \mathcal{A}_{i}(\mathbf{r},z,t)=\sum_{\nu=\pm;n,q}\frac{\gamma}{\sqrt{\Omega}}&
    \sin{\left(\frac{n\pi d_{i}}{L}\right)}\left(\mathbf{e}_{n q\nu}a_{n q\nu}e^{i(\mathbf{q}\cdot\mathbf{r}-\Omega t)} + \mathbf{e}_{n q\nu}^{*}a_{n q\nu}^{\dagger}e^{-i(\mathbf{q}\cdot\mathbf{r}-\Omega t)}\right), \label{Ai}
\end{align}
with $\mathbf{e}_{\nu}=(\mathbf{e}_{x}+i\nu \mathbf{e}_{y})/\sqrt{2}$ represent the circular polarization directions ($\nu=\pm$), $d_{i}$ denotes the position of layer $i$ relative to the cavity's lower mirror, $S$ is the system's surface area, and $\gamma=\sqrt{\hbar/(\epsilon L S)}$ with $\varepsilon=\varepsilon_{0}\varepsilon_{r}$. For the unperturbed electronic Hamiltonian $H_{0}$, the eigenstates can be constructed in the sublattice basis as
\begin{equation}
    \ket{\mathbf{\mathbf{k}}_{i},s_{i}}=\frac{e^{i \mathbf{\mathbf{k}}_{i}\cdot\mathbf{r}_{i}}}{\sqrt{2S}}\left(\ket{A_{i}}+s e^{i\phi_{i}}\ket{B_{i}}\right), \label{kk}
\end{equation}
where $\phi_{i}=\arctan(\mathbf{k}_{y_{i}}/\mathbf{k}_{x_{i}})$ specifies the wave vector angle relative to the $x$-axis, and $s=\pm$ indexes the conduction and valence bands. To simplify the electromagnetic field, it is advantageous to use the polarization vectors $\hat{\mathbf{e}}_{x}$ and $\hat{\mathbf{e}}_{y}$ by defining $\mathbf{e}_{\pm}=\frac{1}{\sqrt{2}}(\mathbf{e}_{x}\pm i\mathbf{e}_{y})$, allowing the interaction term to be written as
\begin{equation}
    \boldsymbol{\sigma}_{i} \mathcal{A}_{i}=\sqrt{2}\sum_{\nu=\pm}\sigma_{-\nu}^{(i)}\mathcal{A}_{\nu}^{(i)},
\end{equation}
where $\nu=\pm 1$ corresponds to different helicities, $\sigma_{\nu}^{(i)}=\frac{1}{2}(\sigma_{x}^{(i)}+\nu i \sigma_{y}^{(i)})$, and where
\begin{equation}
\mathcal{A}_{\nu}^{(i)}=\sum_{n,q}\frac{\gamma\sin\left(\frac{n\pi d_{i}}{L}\right)}{\sqrt{\Omega}}\left(a_{n q \nu}e^{i(\mathbf{q}\cdot\mathbf{r} -\Omega t)}+a_{n q \nu}^{\dagger}e^{-i(\mathbf{q}\cdot\mathbf{r} -\Omega t)}\right).
\end{equation}
In these DLG systems, photon absorption and emission are associated with electronic transitions between sublattics, with stationary states forming coherent superpositions described by the eigenvectors of the Dirac Hamiltonian \cite{Ardenghi2018}. Here, we extend this formalism to the case of TLG, where quantum complexity is significantly enriched. While DLG exhibits bipartite coupling between the two sheets, Our three system introduces a network of three-body interactions, allowing for multipartite quantum entanglement and richer coherent interference. Each layer interacts with the confined electromagnetic field via $\sigma_{i}\mathcal{A}_{i}$, creating and a hybrid quantum system where correlations between the three layers are mediated by cavity vacuum fluctuations. Fig. (\ref{graphene}a) illustrates our experimental setup, showing the TLG encapsulated in a planar microacavity.\par

We focus on the most salient results, the time evolution operator $\mathcal{U}$ for the complete system, derived from time-dependent perturbation theory, is given by
\begin{equation}
\mathcal{U}=\mathcal{U}^{(0)}+\mathcal{U}^{(1)}+\mathcal{U}^{(2)}+...,
\end{equation}
where
\begin{align}
	&\mathcal{U}^{(0)}=\mathbf{I}, \quad \mathcal{U}^{(1)}=-i\int_{-\infty}^{t}dt' H_{I}(t'), \notag\\
    &\mathcal{U}^{(2)}=-\int_{-\infty}^{t}dt' H_{I}(t')\int_{-\infty}^{t'}dt'' H_{I}(t''),
\end{align}
with $H_{I}(t)$ denotes the interaction Hamiltonian within the interaction picture
\begin{equation}
	H_{I}(t)=e^{-i(H_{0}+H_{F})t}H_{I}e^{i(H_{0}+H_{F})t}.
\end{equation}
The temporal evolution of the system is entirely determined by the initial density matrix $\varrho_{0}$. The final density matrix $\varrho_{T}$ is then expressed as
\begin{equation}
\varrho_{T}=\mathcal{U}\varrho_{0}\mathcal{U}^{\dagger}= [\mathbf{I}+\mathcal{U}^{(1)}+\mathcal{U}^{(2)}+...]\varrho_{0}[\mathbf{I}+\mathcal{U}^{(1)}+\mathcal{U}^{(2)}+...]^{\dagger}.
\end{equation}
Assuming the interaction is a small disturbance, we expand the final density matrix perturbatively as $\varrho_{T} = \varrho_{0} + \varrho_{T}^{(1)} + \varrho_{T}^{(2)} + \cdots$. Substituting the expansion for $\mathcal{U}$ into $\varrho_{T} = \mathcal{U} \varrho_{0} \mathcal{U}^{\dagger}$ and collecting terms order by order yields
\begin{align}
&\varrho_{T}^{(1)}=\mathcal{U}^{(1)}\varrho_{0}+\varrho_{0}\mathcal{U}^{(1)\dagger}, \notag\\
&\varrho_{T}^{(2)}=\mathcal{U}^{(1)}\varrho_{0}\mathcal{U}^{(1)\dagger}+\mathcal{U}^{(2)}\varrho_{0}+\varrho_{0}\mathcal{U}^{(2)\dagger}.
\label{re}
\end{align}
In the context of studying the vacuum fluctuations of the cavity field, it is reasonable to consider that the initial quantum state of the entire system can be taken as factorizable. In other words, the global state at the initial instant can be written in the form
\begin{equation}
\varrho_{0}=\ket{\chi_{0}}\bra{\chi_{0}}\otimes \varrho_{e},
\end{equation}
where, $\ket{\chi_{0}}=\ket{\chi_{0}^{(+)},\chi_{0}^{(-)}}$ denotes the vacuum state of the cavity field with the circular polarization components $\pm$, and where 
$\varrho_{e}$ represents the initial density matrix of the electronic system, given by
\begin{equation}
\varrho_{e}=\ket{\mathbf{k}_{1},s_{1}}\bra{\mathbf{k}_{1},s_{1}}\otimes \ket{\mathbf{k}_{2},s_{2}}\bra{\mathbf{k}_{2},s_{2}}\otimes \ket{\mathbf{k}_{3},s_{3}}\bra{\mathbf{k}_{3},s_{3}}.
\end{equation}
We now turn our attention to the reduced state of the electrons in the monolayers after their interaction with the quantum field, which is derived through the implementation of the partial trace,
\begin{equation}
	\varrho(t)=\text{Tr}_{a}(\mathcal{U}\varrho_{0}\mathcal{U}^{\dagger}).
\end{equation}
In other words, the non-diagonal terms of an evolving cavity field have no impact on our analysis. In particular, any contribution for which the parities of $i$ and $j$ differ gives a vanishing contribution to the final states of graphene electrons, provided that the initial state of the field is diagonal in the Fock basis. This condition is satisfied, for instance, by the vacuum state or, more generally, by any incoherent mixture of Fock states, such as a thermal state. Therefore, the only remaining contribution is given by the trace over the cavity field basis, i.e., the expression of $\varrho_{T}^{(2)}$. We can then show that, up to the second order, the evolution of the density matrix $\varrho(t)$ is determined by
\begin{equation}
	\varrho(t)=\varrho_{e}+\text{Tr}_{a}(\mathcal{U}^{(1)}\varrho_{0}\mathcal{U}^{(1)\dagger})+\text{Tr}_{a}(\varrho_{0}\mathcal{U}^{(2)\dagger})+\text{Tr}_{a}(\mathcal{U}^{(2)}\varrho_{0}).\label{rt}
\end{equation}
To compute the expression in Eq. (\ref{rt}), we perform an analytical calculation to determine the following traces $\text{Tr}_{a}(\mathcal{U}^{(1)}\varrho_{0}\mathcal{U}^{(1)\dagger})$, $\text{Tr}_{a}(\varrho_{0}\mathcal{U}^{(2)\dagger})$ and $\text{Tr}_{a}(\mathcal{U}^{(2)}\varrho_{0})$,
\begin{widetext}

\begin{align}
	&\text{Tr}_{a}(\mathcal{U}^{(2)}\varrho_{0})=-(ev_{F})^{2}\sum_{i,j=1,2,3}\sum_{\nu,\nu'} \int_{-\infty}^{t}\int_{-\infty}^{t} dt_{1}dt_{2} 
	\mathcal{G}_{\nu,\nu'}^{(i,j)}(r_{i},t_{1},r_{j},t_{2})\sigma_{\nu}^{(i)}(t_{1})\sigma_{-\nu'}^{(j)}(t_{2})\varrho_{e}, \label{term1}\\
	&\text{Tr}_{a}(\varrho_{0}\mathcal{U}^{(2)\dagger})=-(ev_{F})^{2}\sum_{i,j=1,2,3}\sum_{\nu,\nu'}\int_{-\infty}^{t}\int_{-\infty}^{t} dt_{1}dt_{2}\mathcal{G}_{\nu,\nu'}^{(i,j)}(r_{i},t_{1},r_{j},t_{2}) \varrho_{e}\sigma_{-\nu}^{(i)\dagger}(t_{1})\sigma_{-\nu'}^{(j)\dagger}(t_{2}), \label{term2}\\
	&\text{Tr}_{a}(\mathcal{U}^{(1)}\varrho_{0}\mathcal{U}^{(1)\dagger})=2(ev_{F})^{2}\sum_{i,j=1,2,3}\sum_{\nu,\nu'}\int_{-\infty}^{t}\int_{-\infty}^{t} dt_{1}dt_{2}\mathcal{G}_{\nu,\nu'}^{(i,j)*}(r_{i},t_{1},r_{j},t_{2}) \sigma_{-\nu}^{(i)}(t_{1})\varrho_{e}\sigma_{-\nu'}^{(j)\dagger}(t_{2}), \label{term3}
\end{align}
where,
\begin{align}
	&\sigma_{\nu}^{(i)}(t_{1})=e^{i H_{0}t_{1}}\sigma_{-\nu}^{(i)}e^{-i H_{0}t_{1}}, \sigma_{\nu'}^{(j)}(t_{2})=e^{i H_{0}t_{2}}\sigma_{-\nu'}^{(j)}e^{-i H_{0}t_{2}}, \notag\\
	&\mathcal{A}_{\nu}^{(i)}(t_{1})=e^{i H_{F}t_{1}}\mathcal{A}_{\nu}^{(i)}e^{-i H_{F}t_{1}}, \mathcal{A}_{\nu'}^{(j)}(t_{2})=e^{i H_{F}t_{2}}\mathcal{A}_{\nu'}^{(j)}e^{-i H_{F}t_{2}},
\end{align}
with $\mathcal{G}_{\nu,\nu'}^{(i,j)}(r_{i},t_{1},r_{j},t_{2})=\bra{\chi_{0}}\mathcal{A}_{\nu}^{(i)}(t_{1})\mathcal{A}_{\nu'}^{(j)}(t_{2})\ket{\chi_{0}}$. By combining all the expressions (\ref{term1})-(\ref{term3}), the reduced state is written in the form
\begin{align}
	\varrho&=\text{Tr}_{a}(\varrho(t)) \notag\\
    &=-(ev_{F})^{2}\sum_{i,j=1,2,3;\nu}\int_{-\infty}^{t}\int_{-\infty}^{t}dt_{1}dt_{2}\mathcal{G}_{\nu,\nu'}^{(i,j)}(r_{i},t_{1},r_{j},t_{2})\Big(\sigma_{-\nu}^{(i)}(t_{1})\sigma_{-\nu}^{(j)}(t_{2})\varrho_{e}-2\sigma_{-\nu}^{(i)}(t_{1})\varrho_{e}\sigma_{-\nu}^{(j)\dagger}+\varrho_{e}\sigma_{-\nu}^{(i)\dagger}(t_{1})\sigma_{-\nu}^{(j)\dagger}(t_{2})\Big). \label{rh}
\end{align}
\end{widetext}	
Since the operator acts on the quantum field, the only zero term is the one proportional to  $a_{n q \nu} a_{n q \nu'}$. Thus, after calculation, we obtain the following
\begin{equation}
	\mathcal{G}_{\nu,\nu'}^{(i,j)}(r_{i},t_{1},r_{j},t_{2})=\delta_{\nu \nu'}\sum_{n,q}\frac{\gamma^{2}}{\Omega}\sin{(\frac{n\pi d_{i}}{L})}\sin{(\frac{n\pi d_{j}}{L})}
	e^{i(q.r_{i}-\Omega t_{1})}e^{-i(q.r_{j}-\Omega t_{2})}.\label{prop}
\end{equation}
We can compute the photon propagator exactly and the following is the result
$\mathcal{G}_{\nu,\nu'}^{(i,j)}(r_{i},t_{1},r_{j},t_{2}) \equiv\mathcal{G}_{\nu,\nu'}^{(i,j)}(\Delta t,|\Delta r_{ij}|)=\delta_{\nu \nu'} \mathcal{R}_{ij}(|x_{ij}|)$ \cite{Ardenghi2018}, where
\begin{align}
    \mathcal{R}_{ij}(|x_{ij}|)=\frac{\gamma^{2}\sin{(\frac{\pi d_{i}}{L})}\sin{(\frac{\pi d_{j}}{L})}\sinh{(\frac{\pi |x_{ij}|}{L})}}{16 \pi|x_{ij}| \sin{(\frac{\pi(d_{i}-d_{j}-i|x_{ij}|)}{2L})}\sin{(\frac{\pi(d_{i}+d_{j}-i|x_{ij}|)}{2L})}\sin{(\frac{\pi(d_{i}-d_{j}+i|x_{ij}|)}{2L})}\sin{(\frac{\pi(d_{i}+d_{j}+i|x_{ij}|)}{2L})}},
\end{align}	
with $|x_{ij}|=\sqrt{c^{2}\Delta t^{2}-|\Delta r_{ij}|^{2}}$, with $\Delta t= t_{1}-t_{2}$ and $\Delta r_{ij}=r_{i}-r_{j}$. In addition, we emphasize that the above expression corresponds to the closed-form photon propagator obtained in the infinite-mode limit, where the summation over all cavity modes is performed exactly \cite{Ardenghi2018}. In contrast, the numerical results presented in this work are computed using a truncated mode expansion, where the summation is explicitly restricted to $n \leq n_{\max}$. Therefore, the dependence on $n_{\max}$ does not arise from an approximation of the closed-form expression, but rather from a distinct finite-mode representation of the propagator. This approach is physically motivated, as realistic cavity systems effectively involve a finite number of modes contributing to the interaction, and is consistent with previous studies. In particular, similar finite-mode treatments have been employed in the context of entanglement harvesting in confined geometries \cite{Arreyes2024}, where the role of a limited number of cavity modes is explicitly analyzed.\\
The elements of the matrix can be found using Eqs. (\ref{q1}), (\ref{q2}) and  (\ref{q3}) calculated in Appendix (\ref{appA})
\begin{align}
\bra{\mathbf{k}_{1}',s_{1}',\mathbf{k}_{2}',s_{2}',\mathbf{k}_{3}',s_{3}'}\varrho(t)&\ket{\mathbf{k}_{1},s_{1},\mathbf{k}_{2},s_{2},\mathbf{k}_{3},s_{3}}=-(ev_{F})^{2} \delta_{\mathbf{k}_{i},\mathbf{k}_{i}'}\delta_{\mathbf{k}_{j},\mathbf{k}_{j}'}\sum_{i,j=1,2,3;\nu}\int_{0}^{t}\int_{0}^{t}dt_{1}dt_{2}\mathcal{R}_{ij}(\mathbf{k}_{j}-\mathbf{k}_{j}',\Delta t) \notag\\
&\bra{s_{1}',s_{2}',s_{3}'}(\sigma_{-\nu}^{(i)}(t_{1})\sigma_{-\nu}^{(j)}(t_{2})\varrho_{e}-2\sigma_{-\nu}^{(i)}(t_{1})\varrho_{e}\sigma_{-\nu}^{(j)\dagger}(t_{2})+\varrho_{e}\sigma_{-\nu}^{(i)\dagger}(t_{1})\sigma_{-\nu}^{(j)\dagger}(t_{2}))\ket{s_{1},s_{2},s_{3}}, \label{q3}
\end{align}
where $\mathcal{R}_{ij}(\mathbf{k}_{j}-\mathbf{k}_{j}')$ is the Fourier transform of $\mathcal{R}_{ij}(\sqrt{\Delta t^{2}-|\Delta r_{ij}|^{2}})$, respectively
\begin{align}
	\mathcal{R}_{ij}(\mathbf{k}_{j}-\mathbf{k}_{j}',\Delta t)=\int d^{2} \Delta r_{ij} e^{-i(\mathbf{k}_{j}-\mathbf{k}_{j}).\Delta r_{ij}} \mathcal{R}_{ij}(\sqrt{\Delta t^{2}-|\Delta r_{ij}|^{2}}),
\end{align}
and $\ket{s_{1},s_{2},s_{3}}=\ket{s_{1}}\otimes \ket{s_{2}}\otimes \ket{s_{3}}$ the valence conduction band basis (in which cases $s_{i}=\pm$). The Dirac delta $\delta_{\mathbf{k}_{i},\mathbf{k}_{i}'}\delta_{\mathbf{k}_{j},\mathbf{k}_{j}'}$ implies momentum conservation and is the initial (final) momentum of both electrons.
\section{Results and discussions} \label{sec3}
By developing Eq. (\ref{rh}) for small values of $t$, we can determine the critical parameters for which the reduced quantum state is entangled by expanding  $\text{Tr}_{a}(\varrho(t))$ in series. In this analysis, we assume that $\mathbf{k}_i = \mathbf{k}_i'$ for $i=1,2,3$. In the following, we impose $\mathbf{k}i = \mathbf{k}i'$ for $i=1,2,3$, which simplifies the analytical treatment of the diagonal contributions $\varrho{ii}(t)$ of the reduced density matrix. However, for the off-diagonal terms $\varrho{ij}(t)$ with $i \neq j$, the momenta are not constrained to be equal, and a finite momentum difference $\mathcal{K} = |\mathbf{k}_j - \mathbf{k}_j'|$ naturally appears. This momentum transfer is responsible for the dependence on $\mathcal{K}$ explored in the numerical results. Under these conditions, the reduced density operator can be decomposed as follows
\begin{equation}
    \varrho(t)=\varrho_{ii}(t)+\varrho_{ij}(t), \label{rr}
\end{equation}
where, the expressions of $\varrho_{ij}(t)$, evaluated under the condition that the initial and final momenta coincide for the corresponding subsystems, i.e., $\mathbf{k}_{i} = \mathbf{k}_{i}'$ (and similarly for the relevant indices), for the cases $i=j$ and $i\neq j$, are given by
\begin{align} 
\varrho_{ii}(t)=\bra{\mathbf{k}_{1}',s_{1}',\mathbf{k}_{2}',s_{2}',\mathbf{k}_{3}',s_{3}'}\varrho(t)\ket{\mathbf{k}_{1},s_{1},\mathbf{k}_{2},s_{2},\mathbf{k}_{3},s_{3}}&=-(ev_{F})^{2}\delta_{\mathbf{k}_{i},\mathbf{k}_{i}'}\delta_{\mathbf{k}_{j},\mathbf{k}_{j}'}\sum_{\substack{i=1,2,3;\nu;}} \mathcal{R}_{ii}(\mathbf{k}_{i}-\mathbf{k}_{i}',0) \notag\\
&\times \bra{s_{1}',s_{2}',s_{3}'}(\sigma_{-\nu}^{(i)}\sigma_{-\nu}^{(i)}\varrho_{e}-2\sigma_{-\nu}^{(i)}\varrho_{e}\sigma_{-\nu}^{(i)\dagger}+\varrho_{e}\sigma_{-\nu}^{(i)\dagger}\sigma_{-\nu}^{(i)\dagger})\ket{s_{1},s_{2},s_{3}},
\end{align}
and
\begin{align} 
\varrho_{ij}(t)=\bra{\mathbf{k}_{1}',s_{1}',\mathbf{k}_{2}',s_{2}',\mathbf{k}_{3}',s_{3}'}\varrho(t)\ket{\mathbf{k}_{1},s_{1},\mathbf{k}_{2},s_{2},\mathbf{k}_{3},s_{3}}&=-(ev_{F})^{2}\delta_{\mathbf{k}_{i},\mathbf{k}_{i}'}\delta_{\mathbf{k}_{j},\mathbf{k}_{j}'} \sum_{\substack{i,j=1,2,3;\nu;\\i\neq j}}\mathcal{R}_{ij}(\mathbf{k}_{j}-\mathbf{k}_{j}',0) \notag\\
&\times \bra{s_{1}',s_{2}',s_{3}'}(\sigma_{-\nu}^{(i)}\sigma_{-\nu}^{(j)}\varrho_{e}-2\sigma_{-\nu}^{(i)}\varrho_{e}\sigma_{-\nu}^{(j)\dagger}+\varrho_{e}\sigma_{-\nu}^{(i)\dagger}\sigma_{-\nu}^{(j)\dagger})\ket{s_{1},s_{2},s_{3}}.
\end{align}
In order to investigate how the cavity field influences the electronic properties and harvesting the quantum resources in a TLG system, we consider the initial electronic state
\(\varrho_{e} = \ket{A,A,A}\bra{A,A,A}\), 
where each electron in the graphene layers has a nonzero amplitude in the sublattice \(A\). The normalized reduced quantum state can then be expressed in the basis $\{\ket{A,A,A}, \ket{A,A,B}, \ket{A,B,A}, \ket{A,B,B},\ket{B,A,A}, \ket{B,B,A}, \ket{B,A,B},\ket{B,B,B}\}$,
\begin{equation}
	\varrho(t)=\begin{pmatrix}
	1-2 \alpha^{2}(t)( \mathcal{R}_{11}+\mathcal{R}_{22}+\mathcal{R}_{33}) & 0 & 0 & -2\alpha^{2}(t) \mathcal{R}_{23} & 0  & -2 \alpha^{2}(t) \mathcal{R}_{13} & -2 \alpha^{2}(t) \mathcal{R}_{12} & 0
	\\0 & 2 \alpha^{2}(t) \mathcal{R}_{33} & 2 \alpha^{2}(t) \mathcal{R}_{23} & 0 & 2 \alpha^{2}(t) \mathcal{R}_{13} & 0 & 0 & 0
	\\0 & 2 \alpha^{2}(t) \mathcal{R}_{23} & 2 \alpha^{2}(t) \mathcal{R}_{22} & 0 & 2 \alpha^{2}(t) \mathcal{R}_{12} & 0 & 0 & 0
	\\-2\alpha^{2}(t) \mathcal{R}_{23} & 0 & 0 & 0 & 0 & 0 & 0 & 0
	\\0 & 2 \alpha^{2}(t) \mathcal{R}_{13} & 2 \alpha^{2}(t) \mathcal{R}_{12} & 0 & 2 \alpha^{2}(t) \mathcal{R}_{11} & 0 & 0 & 0
	\\-2\alpha^{2}(t) \mathcal{R}_{13} & 0 & 0 & 0 & 0 & 0 & 0 & 0
	\\-2 \alpha^{2}(t) \mathcal{R}_{12} & 0 & 0 & 0 & 0 & 0 & 0 & 0
	\\0 & 0 & 0 & 0 & 0 & 0 & 0 & 0
	\end{pmatrix}, \label{rhot}
\end{equation}
where $\alpha(t)=e v_{F} t$, $\mathcal{K}_{22}=|\mathbf{k}_{2}-\mathbf{k}_{2}'|$ and $\mathcal{K}_{33}=|\mathbf{k}_{3}-\mathbf{k}_{3}'|$. To derive analytical results for the case where $\mathcal{K}_{11}=\mathcal{K}_{22}=\mathcal{K}_{33}=0$, we can perform the integration over $\Delta r_{ij}$ rather than evaluating the sum over $n$, as was done in \cite{Ardenghi2018}. Consequently, $\mathcal{R}_{ij}$ (with $i\neq j$) can be rewritten as
\begin{equation}
	\mathcal{R}_{ij}(\mathcal{K})=\frac{\gamma^{2}}{16 \pi^{2}}\sum_{n=1}^{\infty}\frac{\sin{(\frac{n\pi d_{i}}{L})}\sin{(\frac{n\pi d_{i}}{L})}}{\sqrt{\mathcal{K}^{2}+ (\frac{n\pi}{L})^{2}}}. \label{rij}
\end{equation}
To evaluate the term $\mathcal{R}_{ii}(\mathcal{K}=0)$, a direct substitution of $\mathcal{K}=0$ into Eq. (\ref{rij}) leads to ill-defined results due to the improper handling of the discrete mode summation and the associated normalization factor. In particular, the summation over confined modes requires a careful treatment of the density of states, as commonly encountered in the calculation of propagators in bounded geometries. To address this issue, we follow a standard regularization procedure by reformulating the summation with properly normalized modes and performing an appropriate change of variables. This approach allows us to evaluate the series consistently and obtain a finite, well-defined expression, and using this method, and following the procedure outlined in \cite{Arreyes2022}, we arrive at the result given
\begin{equation}
    \mathcal{R}_{ii}(\mathcal{K}=0)=\sum_{n,q} \frac{1}{\Omega}\sin^{2}(\frac{n\pi d_{i}}{L})=\frac{S}{48cL}\Bigg[\frac{3}{\sin(\frac{\pi d_{i}}{L})}-1\Bigg].
\end{equation}
Similarly, we consider the TLG system to be initially prepared in the pure state $\varrho_{e} = \ket{+,+,+}\bra{+,+,+}$, where $\ket{+}$ is a state vector in the sublattice basis (as shown in Fig.(\ref{graphene}b)). The interaction with the cavity field is mediated by a detector that couples in this same basis, a choice which mixes the energy eigenstates of $H_{0}$ and drives the quantum dynamics. This interaction is characterized by the matrix element
\begin{equation}
\bra{s'}\sigma_{\nu}^{'}\ket{s} =
\begin{cases}
e^{i\phi}, & \text{if } s = s' = +1 \text{ and } \nu = +1, \\
e^{-i\phi'}, & \text{if } s = s' = -1 \text{ and } \nu = -1, \\
0, & \text{otherwise}.
\end{cases}
\end{equation}
Refer to the detailed derivation provided in Appendix (\ref{EM1}), the reduced density matrix $\varrho$ for the TLG system, expressed in the computational basis $\{\ket{+++}, \ket{++-}, \ket{+-+}, \ket{+--}, \ket{-++}, \ket{-+-}, \ket{--+}, \ket{---}\}$, is found to have the explicit form
\begin{equation}
	\varrho(t) =\begin{pmatrix}
1-\alpha^{2}(t)\mathcal{W}_{11} & -\alpha^{2}(t)\mathcal{M}_{+} & -\alpha^{2}(t)\mathcal{N}_{+} & -\alpha^{2}(t)\mathcal{P} & -\alpha^{2}(t)\mathcal{J}_{+}  & -\alpha^{2}(t)\mathcal{T}_{-} & -\alpha^{2}(t)\mathcal{Z} & 0
	\\-\alpha^{2}(t)\mathcal{M}_{-} & -\alpha^{2}(t)\mathcal{W}_{22} & -\alpha^{2}(t)\mathcal{L} & 0 & -\alpha^{2}(t)\mathcal{B}_{-} & 0 & 0 & 0
	\\-\alpha^{2}(t)\mathcal{N}_{-} & -\alpha^{2}(t)\mathcal{L} & -\alpha^{2}(t)\mathcal{W}_{33} & 0 & -\alpha^{2}(t)\mathcal{Y}_{+} & 0 & 0 & 0
	\\-\alpha^{2}(t)\mathcal{P} & 0 & 0 & 0 & 0 & 0 & 0 & 0
	\\-\alpha^{2}(t)\mathcal{J}_{-} & -\alpha^{2}(t)\mathcal{B}_{+} & -\alpha^{2}(t)\mathcal{Y}_{-} & 0 & -\alpha^{2}(t)\mathcal{W}_{55} & 0 & 0 & 0
	\\-\alpha^{2}(t)\mathcal{T}_{+} & 0 & 0 & 0 & 0 & 0 & 0 & 0
	\\-\alpha^{2}(t)\mathcal{Z} & 0 & 0 & 0 & 0 & 0 & 0 & 0
	\\0 & 0 & 0 & 0 & 0 & 0 & 0 & 0
	\end{pmatrix}, \label{rhot2}
\end{equation}
where
\begin{align}
&\mathcal{W}_{11}=2\sum_{i=1}^{3}\mathcal{R}_{ii}e^{i\Delta \phi_{ii}}, \quad\quad
\mathcal{W}_{22}=-2\mathcal{R}_{33}e^{i\Delta \phi_{33}}, \quad \quad\mathcal{W}_{33}=-2\mathcal{R}_{22}e^{i\Delta \phi_{22}}, \quad\quad \mathcal{W}_{55}=-2\mathcal{R}_{11} e^{i\Delta \phi_{11}}, \notag\\
&\mathcal{M}_{\pm}=-2\mathcal{R}_{13}\Big[ \pm e^{i\Delta \phi_{31}}\mp e^{i\Delta \phi_{13}} \mp \frac{1}{2}e^{i\phi_{13}} \pm \frac{1}{2}e^{-i \phi_{13}^{'}}\Big]-2\mathcal{R}_{23}\Big[ \pm e^{i\Delta \phi_{32}} \mp e^{i\Delta \phi_{23}}\mp\frac{1}{2} e^{i \phi_{23}}\pm \frac{1}{2}e^{-i \phi_{23}^{'}}\Big], \notag\\
&\mathcal{N}_{\pm}=-2\mathcal{R}_{12}\Big[ \pm e^{i\Delta \phi_{21}} \mp e^{i\Delta \phi_{12}}\mp\frac{1}{2} e^{i\phi_{12}} \pm \frac{1}{2} e^{-i \phi_{12}^{'}}\Big]-2\mathcal{R}_{23}\Big[\mp e^{i\Delta \phi_{32}} \pm e^{i\Delta \phi_{23}} \mp \frac{1}{2} e^{i \phi_{23}} \pm \frac{1}{2} e^{-i \phi_{23}^{'}}\Big], \notag\\
&\mathcal{J}_{\pm}=-2\mathcal{R}_{12}\Big[ \mp e^{i\Delta \phi_{21}} \pm e^{i\Delta \phi_{12}}\mp \frac{1}{2} e^{i \phi_{12}} \pm \frac{1}{2} e^{-i \phi_{12}^{'}}\Big]-2\mathcal{R}_{13}\Big[ \mp e^{i\Delta \phi_{31}}\pm e^{i\Delta \phi_{13}}\mp \frac{1}{2} e^{i \phi_{13}}\pm \frac{1}{2} e^{-i \phi_{13}^{'}}\Big], \notag \\
&\mathcal{P}=\mathcal{R}_{23}\Big[e^{i \phi_{23}}+e^{-i \phi_{23}^{'}}\Big], \quad\quad \mathcal{T}_{\pm}=\mathcal{R}_{12}\Big[e^{i\phi_{12}}\pm e^{-i\phi_{12}^{'}}\Big], \quad\quad \mathcal{Z}=\mathcal{R}_{13}\Big[e^{i\phi_{13}}+e^{-i\phi_{13}^{'}}\Big], \notag\\
&\mathcal{L}=-2\mathcal{R}_{23}\Big[e^{i\Delta \phi_{32}}+e^{i\Delta \phi_{23}}\Big], \quad\quad \mathcal{B}_{\pm}=-2\mathcal{R}_{13}\Big[ \pm e^{i\Delta \phi_{31}}\mp e^{i\Delta \phi_{13}}\Big], \quad\quad \mathcal{Y}_{\pm}=-2\mathcal{R}_{12}\Big[e^{i\Delta \phi_{21}}+e^{i\Delta \phi_{12}}\Big].
\end{align}
This study analyzes fundamental quantum properties, quantum coherence, entanglement, and non-Markovianity in a TLG system subjected to vacuum fluctuations in a microcavity. We focus specifically on three quantitative metrics, REC, tangle to measure entanglement, and a measure of non-Markovianity based on REC. Our main objective is to characterize the evolution of these physical quantities during the systematic variation of key parameters, the number of cutoff modes $n_{\max}$, the relative position of the layers $d_{i}/L$, time $t$ and the momentum $\mathcal{K}$.\par

{\bf Non-Markovianity via quantum coherence:} A key feature of open dynamics is the potential emergence of non-Markovianity. This phenomenon is characterized by the system's evolution depending not only on its present state, but also on its history under the influence of the memory carried by the environment. This contrasts with Markovian processes, whose dynamics are entirely determined by the instantaneous state and have no memory. Quantifying non-Markovianity enables the nature and intensity of system-environment interactions to be specified, particularly by tracing the reversible information flows that cross them. Several measures have been developed for this purpose, such as the trace distance \cite{Breuer2009}, local quantum Fisher information \cite{Dakir2025a}, quantum coherence via Kirkwood-Dirac quasiprobability \cite{Dakir2025b} and $l_{1}$-norm of coherence \cite{Chanda2016}. These tools can detect signatures of non-Markovian dynamics, such as temporary increases in distinguishability of states, reflux of information from environment to system or oscillations of entanglement — all manifestations of active quantum memory. In this work, we use the measure based on relative entropy of coherence, $\mathcal{N}_{\text{REC}}$, which is defined as follows \cite{He2017}
\begin{equation}
\mathcal{N}_{\text{REC}}=\sup_{\varrho(0)}\int_{\frac{d}{dt}C_{r}(\varrho (t))>0}\frac{d}{dt}C_{r}(\varrho(t))dt, \label{MNMN}
\end{equation}
where, $d\mathcal{C}_{r}(\varrho (t))/dt$ denotes the rate of change of the REC. It is defined for the density matrix $\varrho$, and is given by \cite{Baumgratz2014}
\begin{equation}
\mathcal{C}_{r}(\varrho)=\min_{\delta \in I} \mathcal{S}(\varrho||\delta)=\mathcal{S}(\varrho_{d})-\mathcal{S}(\varrho), \label{REC}
\end{equation}
here, $\varrho_d$ is the dephased state in the computational basis $\{\ket{i}\}$, i.e., it is the diagonal part of $\varrho$ with all coherence erased. Finally, $\mathcal{S}(\varrho)=-\text{Tr} \varrho \log_{2} \varrho$ corresponds to the von Neumann entropy. Note that the original definition of the non-Markovianity measure $\mathcal{N}_{\mathrm{REC}}$ was formulated in an extended Hilbert space that included an ancillary system \cite{He2017}. In this work, however, we evaluate this quantity directly from the reduced density matrix of the system. This is justified by the fact that the detection of non-Markovianity relies on the non-monotonic behavior of the relative entropy of coherence, i.e., temporary increases of REC, which signal information backflow from the environment to the system. Consequently, we employ $\mathcal{N}_{\mathrm{REC}}$ as an operational witness of non-Markovianity, capturing memory effects without the need for an explicit ancillary system.

{\bf Quantum entanglement quantified by tangle:} The tangle is a way of measuring entanglement that is specific to tripartite systems, which is neither a simple generalization nor a direct combination of bipartite entanglement measures \cite{Coman2000}. Quantifies the degree of global entanglement between the three subsystems, which cannot be adequately described by bipartite correlations alone. In particular, when one of the subsystems (say A) is entangled with the combination of the other two (BC), this entanglement can be interpreted as a bipartite entanglement between the two parts A and BC, and evaluated using a well-known measure of bipartite entanglement. In this case, the tangle for a tripartite system is written as
\begin{equation}
    \mathcal{T}_{g}= \max_{\left\lbrace i,j,k\right\rbrace=\left\lbrace \text{A,B,C}\right\rbrace } \left\lbrace \mathcal{C}_{i|jk}^{2}-\mathcal{C}_{i|j}^{2}-\mathcal{C}_{i|k}^{2}\right\rbrace,
\end{equation}
where $\mathcal{C}_{i|jk}$ denotes the bipartite competition between subsystem $i$ and combined subset $jk$, defined by
\begin{equation}
    \mathcal{C}_{i|jk}=\sqrt{2(1-Tr(\varrho_{i}^{2})},
\end{equation}
and $\mathcal{C}_{i|j}$ is the competition between pairs of qubits $i$ and $j$, which is defined by
\begin{equation}
    \mathcal{C}(\varrho)=\max \left\lbrace 0, \sqrt{\xi_{1}}-\sqrt{\xi_{2}}-\sqrt{\xi_{3}}-\sqrt{\xi_{4}}\right\rbrace,
\end{equation}
where, the $\xi_{i}$ are the eigenvalues in order of decreasing magnitude of the following matrix
\begin{equation}
    \tilde{\varrho}= (\sigma_{y} \otimes \sigma_{y}) \varrho^{*} (\sigma_{y} \otimes \sigma_{y}).
\end{equation}

\begin{figure}[h]
\centering

\begin{minipage}[b]{.23\linewidth}
    \centering
    \includegraphics[scale=0.24]{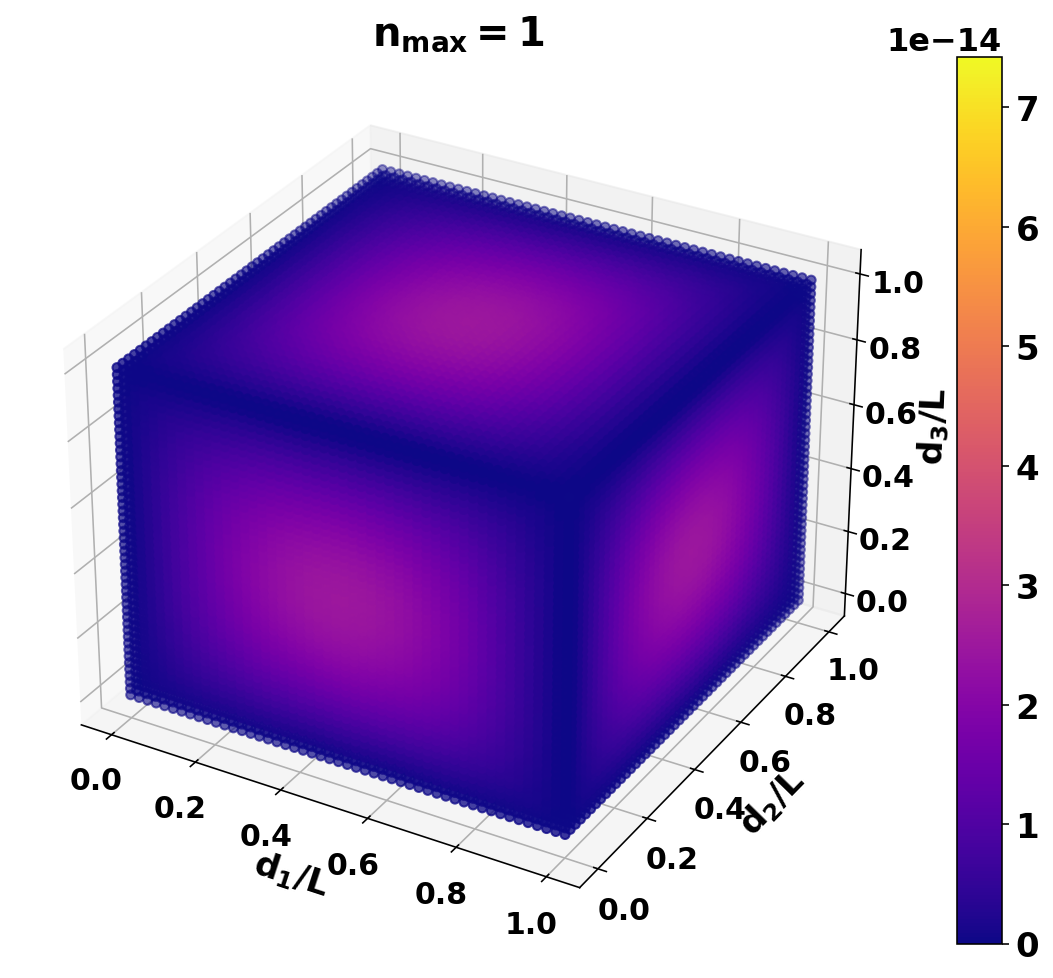}
     \vfill \vfill  $\left(a\right)$
\end{minipage}
\hfill
\begin{minipage}[b]{.23\linewidth}
    \centering
    \includegraphics[scale=0.24]{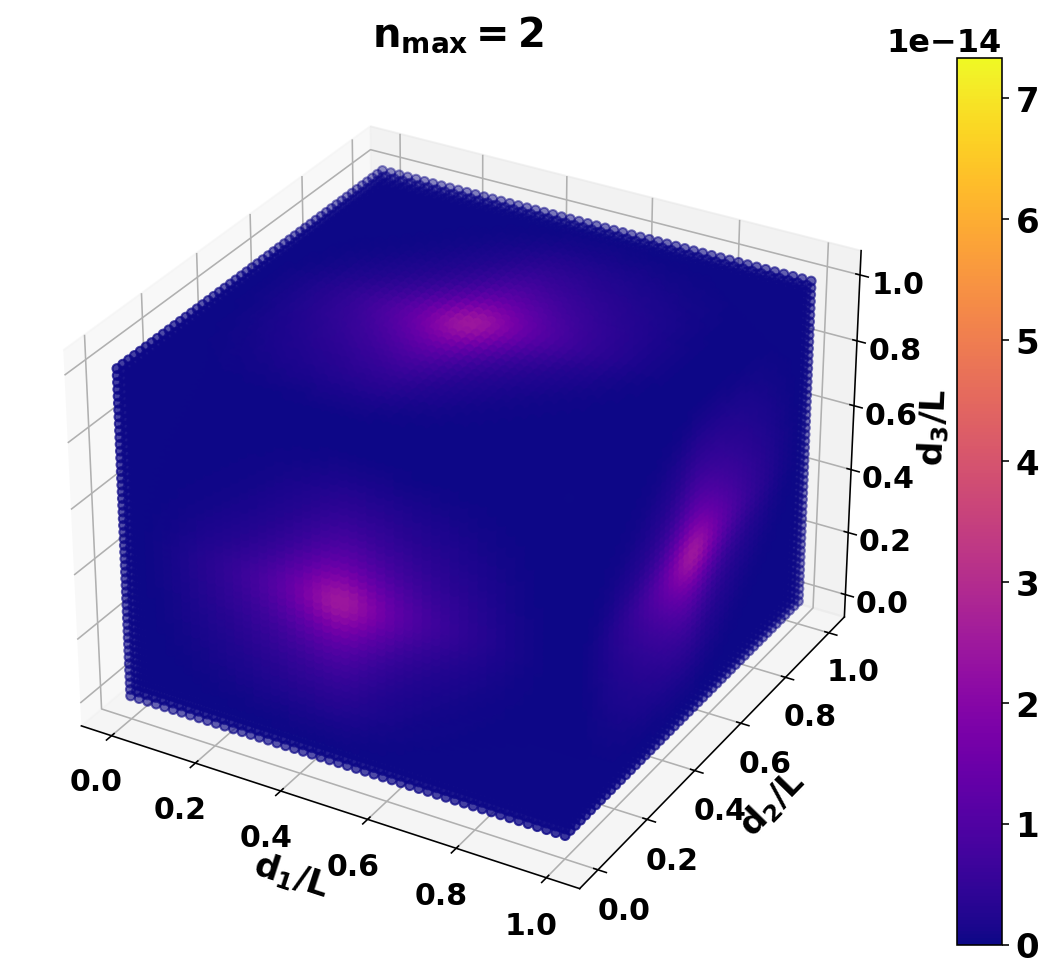}
     \vfill \vfill  $\left(b\right)$
\end{minipage}
\hfill
\begin{minipage}[b]{.23\linewidth}
    \centering
    \includegraphics[scale=0.24]{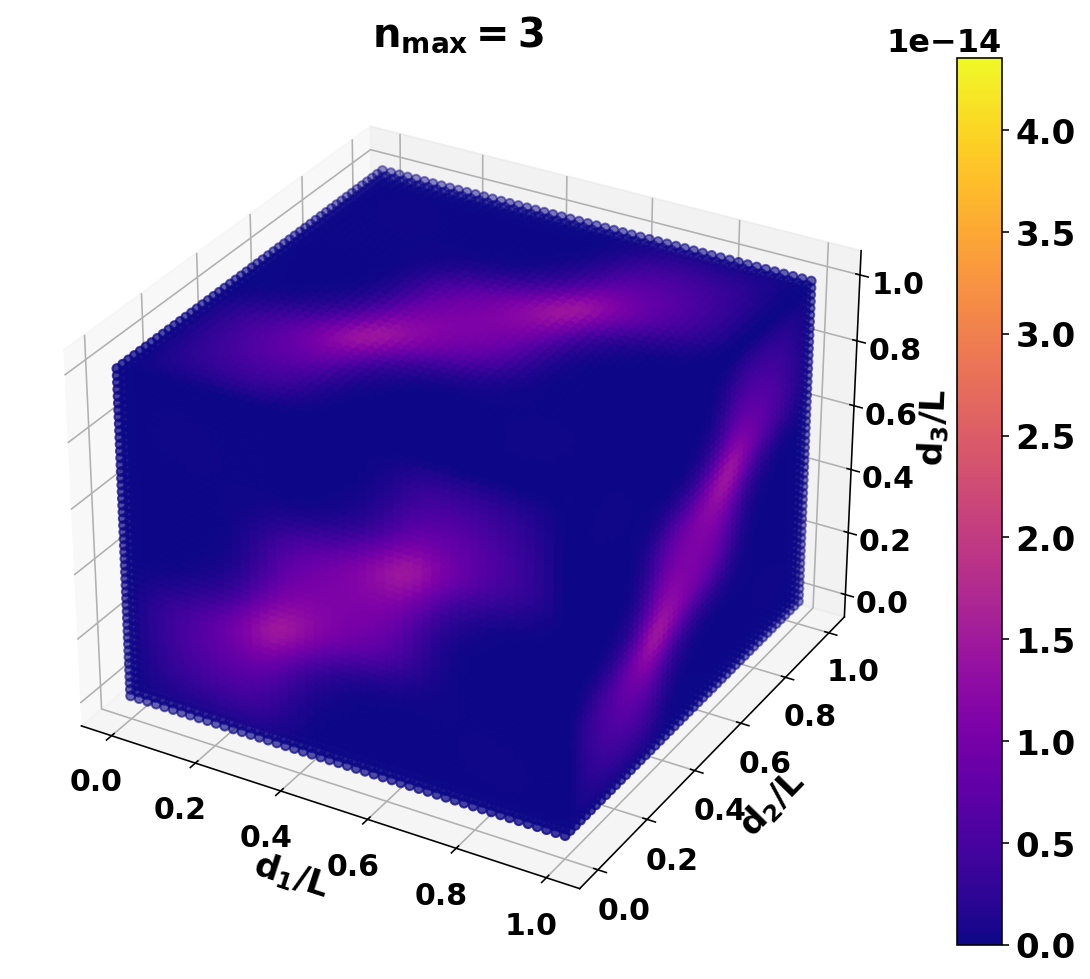}
     \vfill \vfill  $\left(c\right)$
\end{minipage}
\hfill
\begin{minipage}[b]{.23\linewidth}
    \centering
    \includegraphics[scale=0.24]{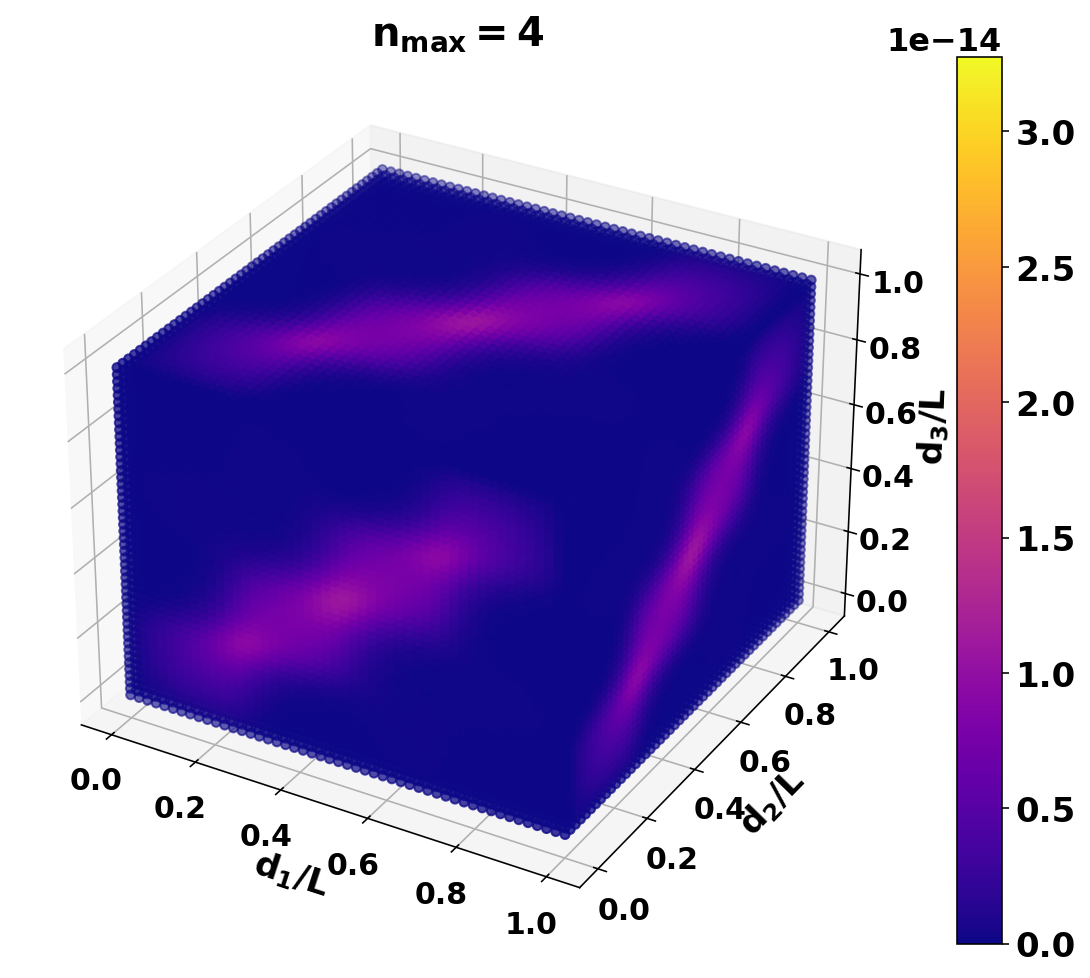}
     \vfill \vfill  $\left(d\right)$
\end{minipage}

\vspace{1cm} 

\begin{minipage}[b]{.23\linewidth}
    \centering
    \includegraphics[scale=0.24]{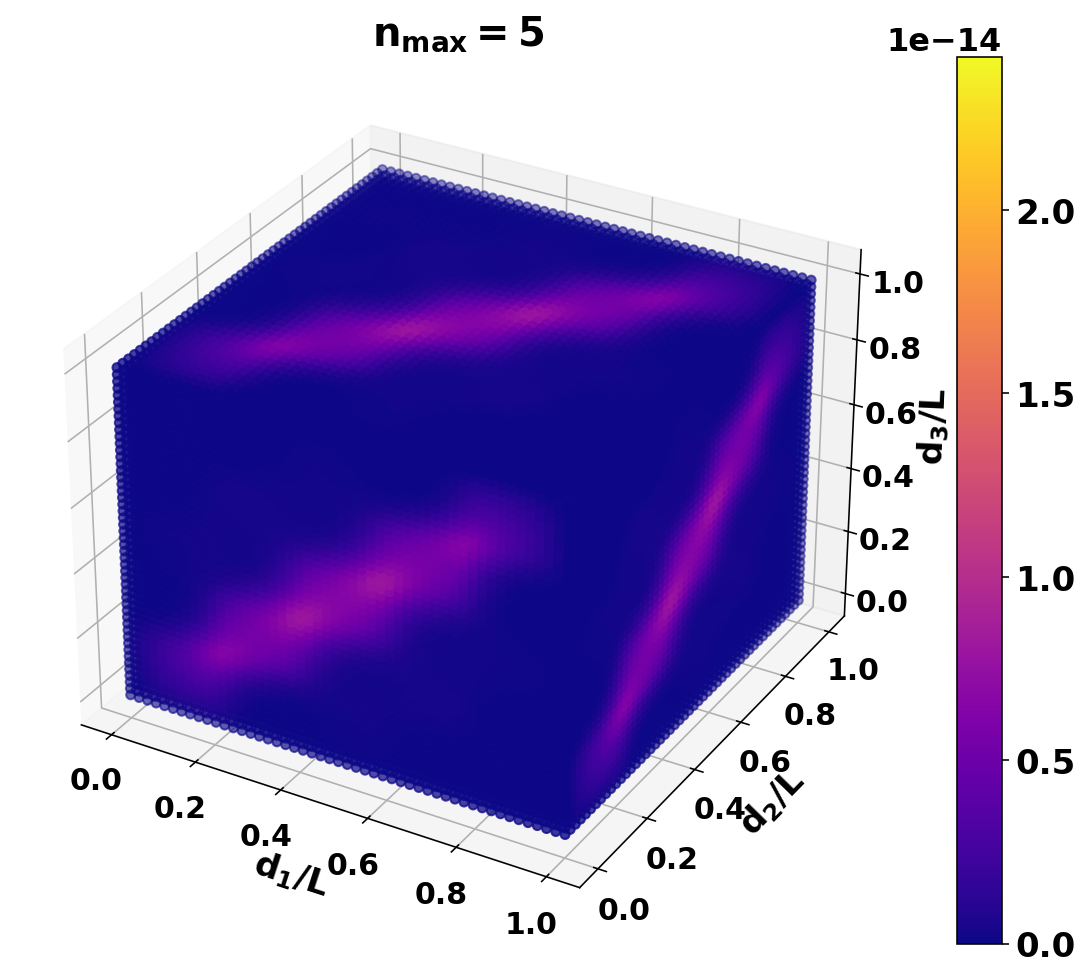}
     \vfill \vfill  $\left(e\right)$
\end{minipage}
\hfill
\begin{minipage}[b]{.23\linewidth}
    \centering
    \includegraphics[scale=0.24]{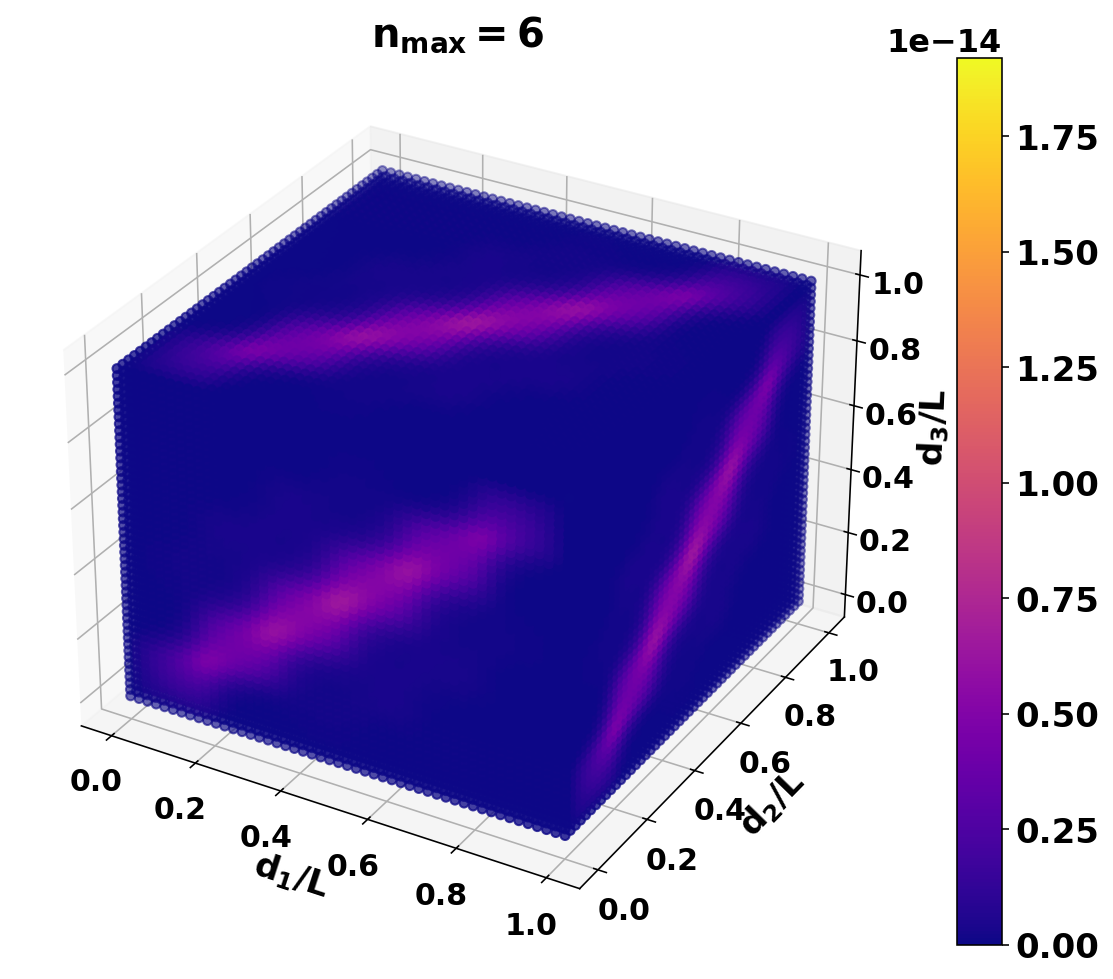}
     \vfill \vfill  $\left(f\right)$
\end{minipage}
\hfill
\begin{minipage}[b]{.23\linewidth}
    \centering
    \includegraphics[scale=0.24]{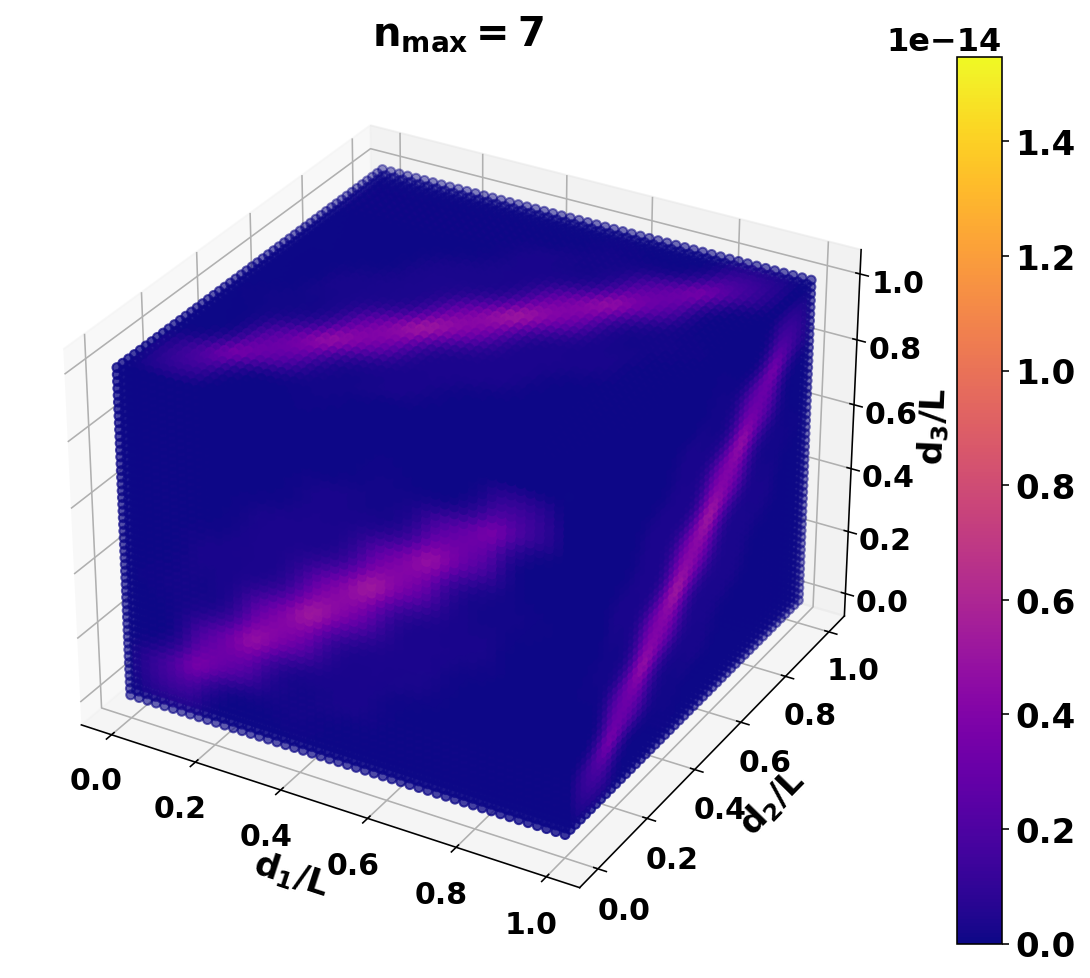}
     \vfill \vfill  $\left(g\right)$
\end{minipage}
\hfill
\begin{minipage}[b]{.23\linewidth}
    \centering
    \includegraphics[scale=0.24]{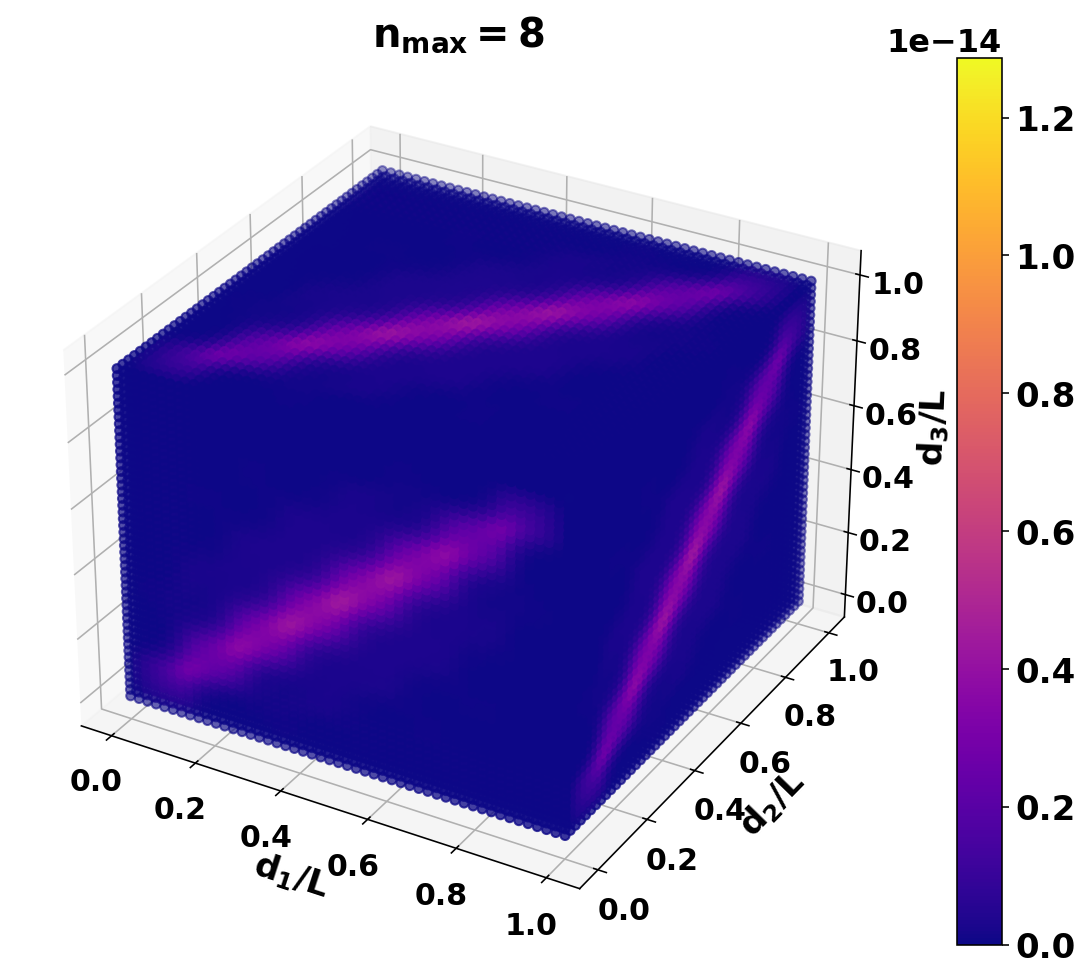}
     \vfill \vfill  $\left(h\right)$
\end{minipage}

\caption{Relative entropy of coherence $\mathcal{C}_{r}$ over $\alpha^{2}(t)$ as a function of $\frac{d_{i}}{L}$, for different number of cutoff modes $n_{\max}$, with $\mathcal{KL}=0$}
\label{figREC}
\end{figure}

\begin{figure}[hbtp]
		{{\begin{minipage}[b]{.33\linewidth}
					\centering
					\includegraphics[scale=0.28]{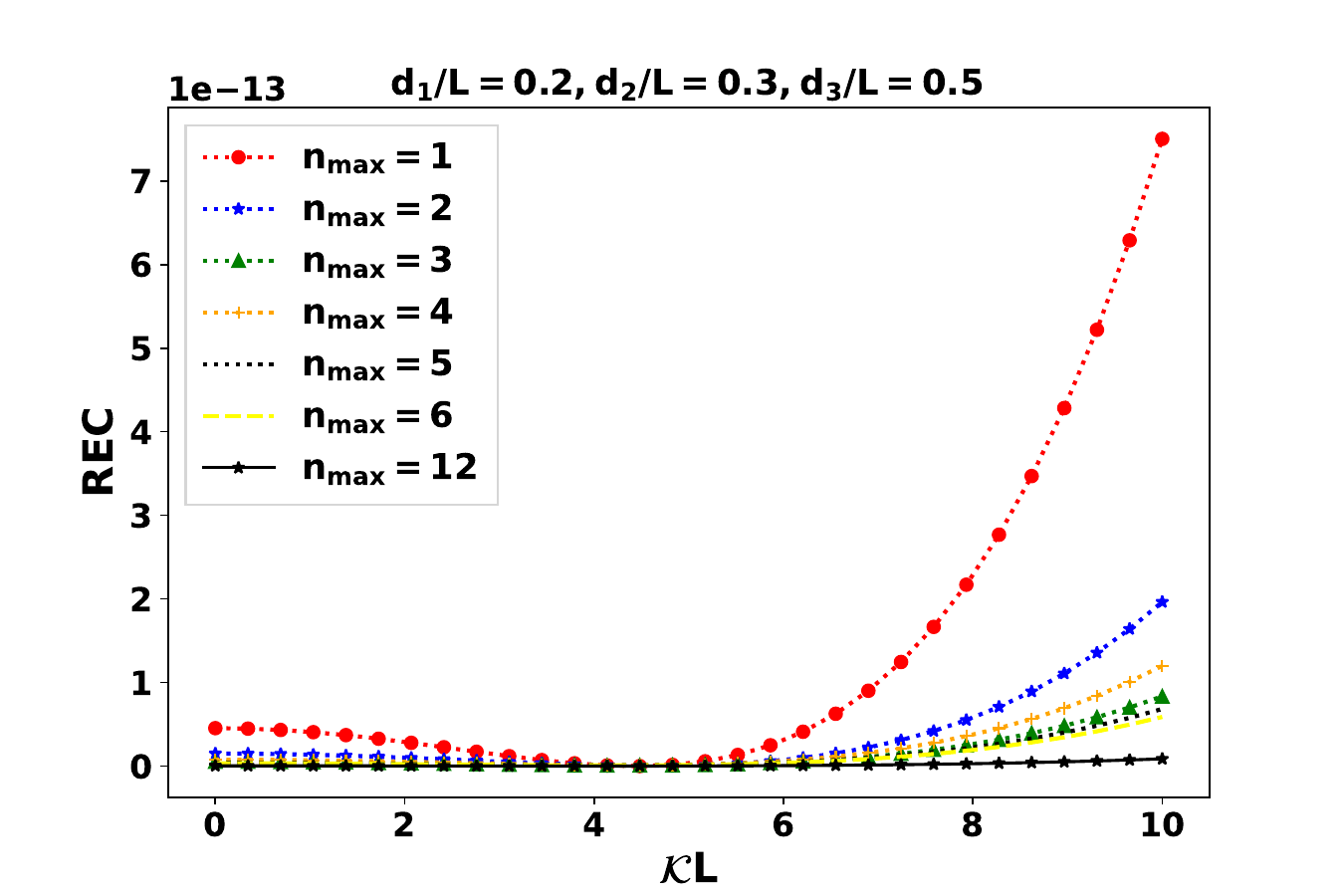} \vfill $\left(a\right)$
				\end{minipage}\hfill
				\begin{minipage}[b]{.33\linewidth}
					\centering
					\includegraphics[scale=0.28]{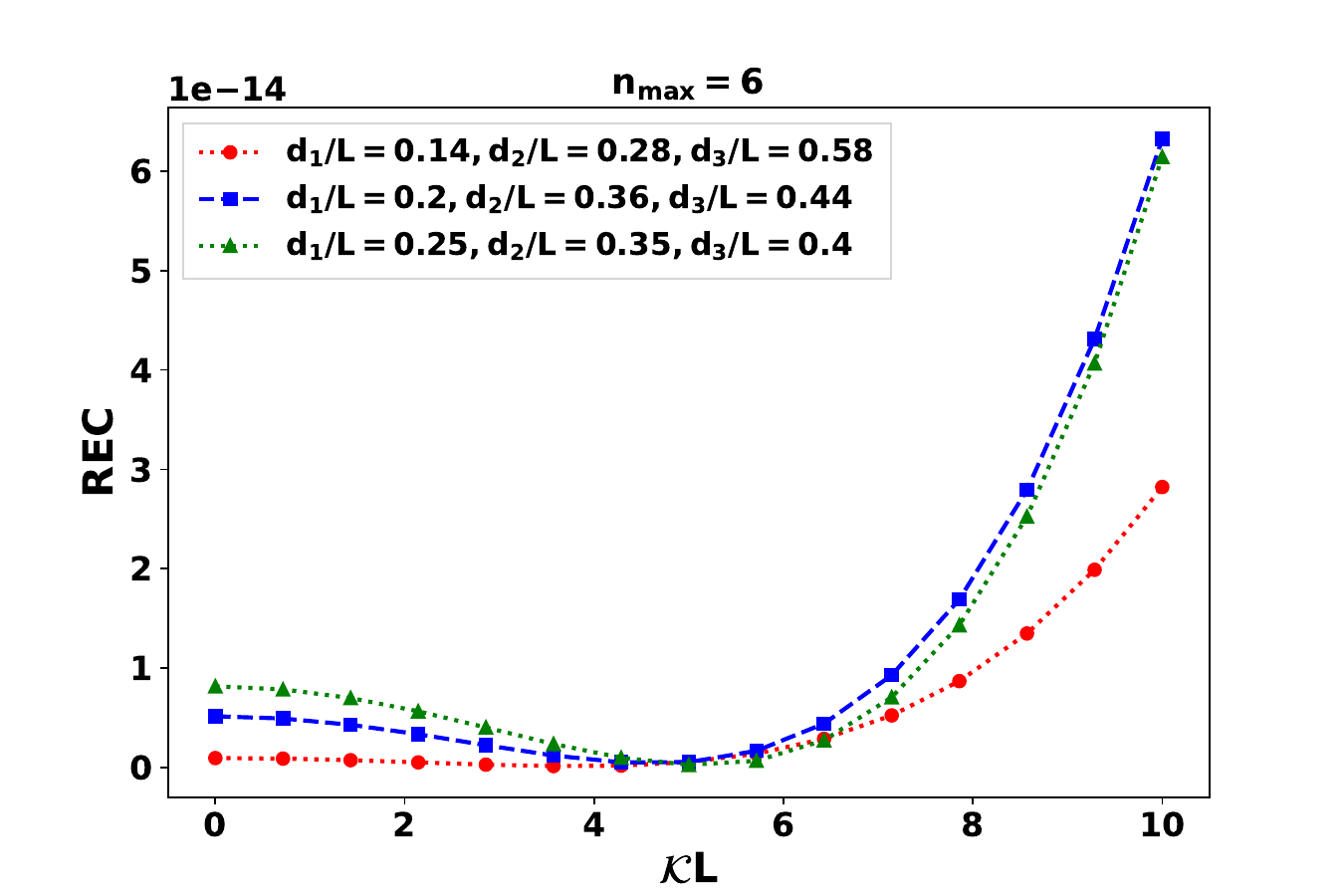} \vfill \vfill  $\left(b\right)$
		\end{minipage}}
	\begin{minipage}[b]{.33\linewidth}
		\centering
		\includegraphics[scale=0.28]{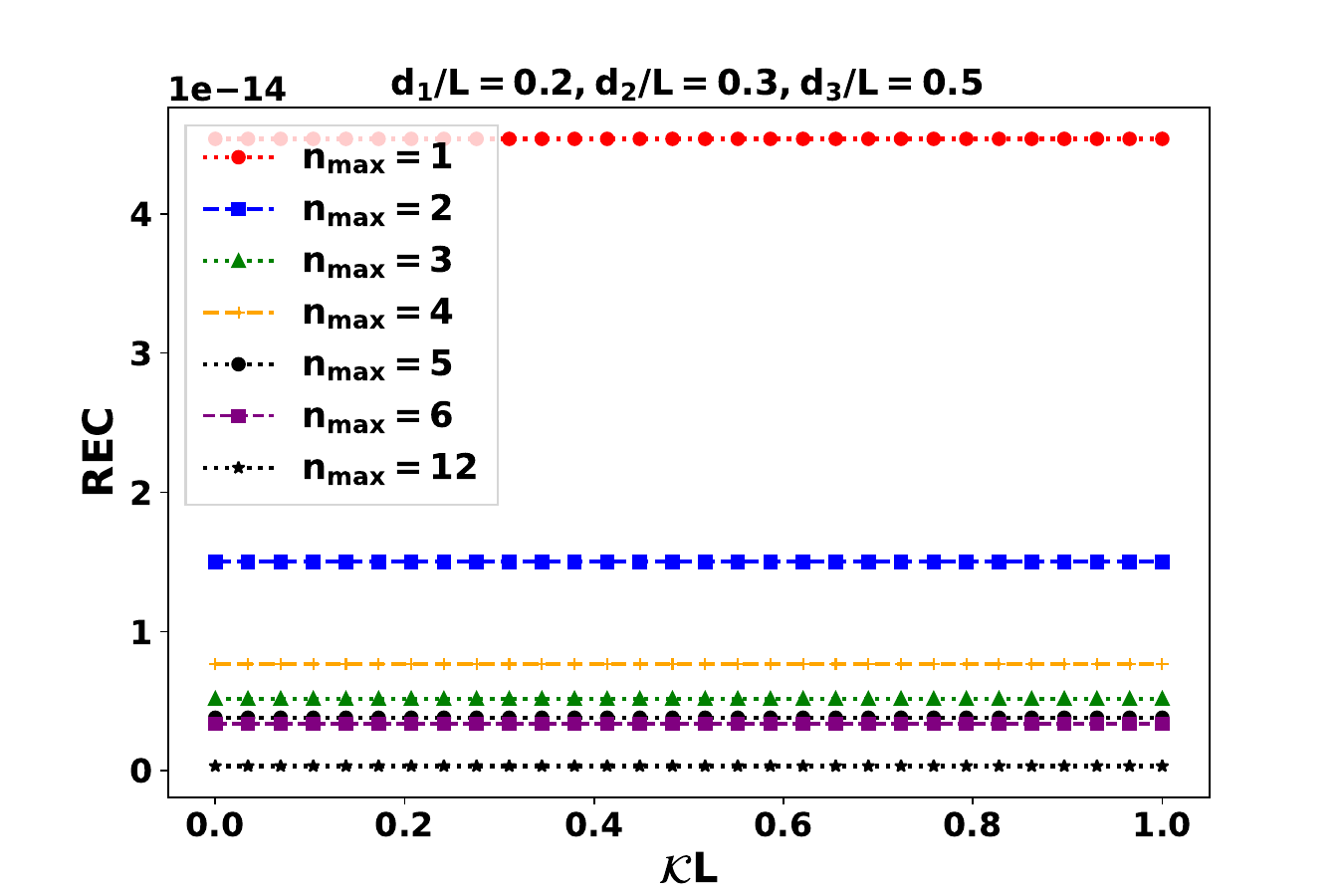} \vfill \vfill  $\left(c\right)$
\end{minipage}}
		\caption{REC $\mathcal{C}_{r}$ over $\alpha^{2}(t)$  as a function of the momentum $\mathcal{K}$ (a) for different values of $n_{\max}$ (b) for different positions of the layers $\frac{d_{i}}{L}$ and for different values of the momentum $\mathcal{K}$ and (c) as a function time $t$ for different values of $n_{\max}$} \label{REC_2D}
\end{figure}
Fig. (\ref{figREC}) shows how the REC $\mathcal{C}_{r}/\alpha^{2}(t)$, evolves in the TLG system within a planar microcavity, parameterized by the position of the layers $d_{i}/L$ and the number of cutoff modes $n_{\max}$. REC quantifies the distillable QC present in the system. We observe that increasing $n_{\max}$ significantly improves QC. This effect is most pronounced when the layers are in close proximity $(\frac{d_{1}}{L} \sim \frac{d_{2}}{L} \sim \frac{d_{3}}{L})$, a configuration that subjects them to a nearly identical confined electromagnetic field. In this regime, the vacuum fluctuations of the cavity, structured by its discrete modes, induce a collective and coherent coupling between the layers, thereby generating and protecting a high degree of quantum superposition within the TLG system. Conversely, a large spatial separation introduces a geometric phase difference in each layer's interaction with the vacuum field, leading to distinct local environments (cavity) that suppress the global quantum coherence.\par

A comprehensive analysis of the REC is presented in Fig.(\ref{REC_2D}). For Fig.(\ref{REC_2D}a), with fixed layer positions, the REC grows with the momentum $\mathcal{K}$ after an initial plateau at low values. The amplitude and slope of this growth are enhanced with increasing $n_{\max}$. This is a direct consequence of the enriched vacuum fluctuation spectrum: a higher cutoff mode number increases the number of structured environmental modes that can interact coherently with the electronic degrees of freedom in the graphene layers. This provides more pathways for the system to maintain quantum superpositions, thus increasing the QC. The dimensionless parameter $\mathcal{K}L$ represents the product of the effective momentum $\mathcal{K}$ and the cavity length $L$. For a fixed cavity size, varying $\mathcal{K}L$ corresponds to exploring different low-energy momenta around the Dirac point. In Fig.(\ref{REC_2D}b), we show the critical influence of the position of the layers for a fixed number of modes ($n_{\max}=6$). The REC increases sharply as the interlayer distances are reduced, indicating an increase in QC behavior. This is due to the intensification of dipole–dipole interactions, which are mediated by the cavity's vacuum modes. At short distances, the layers experience strong mutual coupling to the same field modes, enhancing the global coherence resource. In contrast, increased layer separation isolates the layers, decreasing QC. Finally, Fig.(\ref{REC_2D}c) shows the temporal evolution of the REC. We find that the steady-state value of the REC is a function of $n_{\max}$, yet it remains constant in time for each cutoff mode. This temporal stability indicates that the system reaches a steady state. The dependence of the steady-state value on $n_{\max}$ confirms that the spectral structure of the cavity environment is the primary factor determining the long-time QC of the TLG system. In summary, our results demonstrate that QC in a TLG system, as measured by the REC, can be precisely engineered. The cutoff mode controls the maximum achievable coherence, while the position of the layers determines how much QC is improved.

\begin{figure}[hbtp]
		{{\begin{minipage}[b]{.33\linewidth}
					\centering
					\includegraphics[scale=0.26]{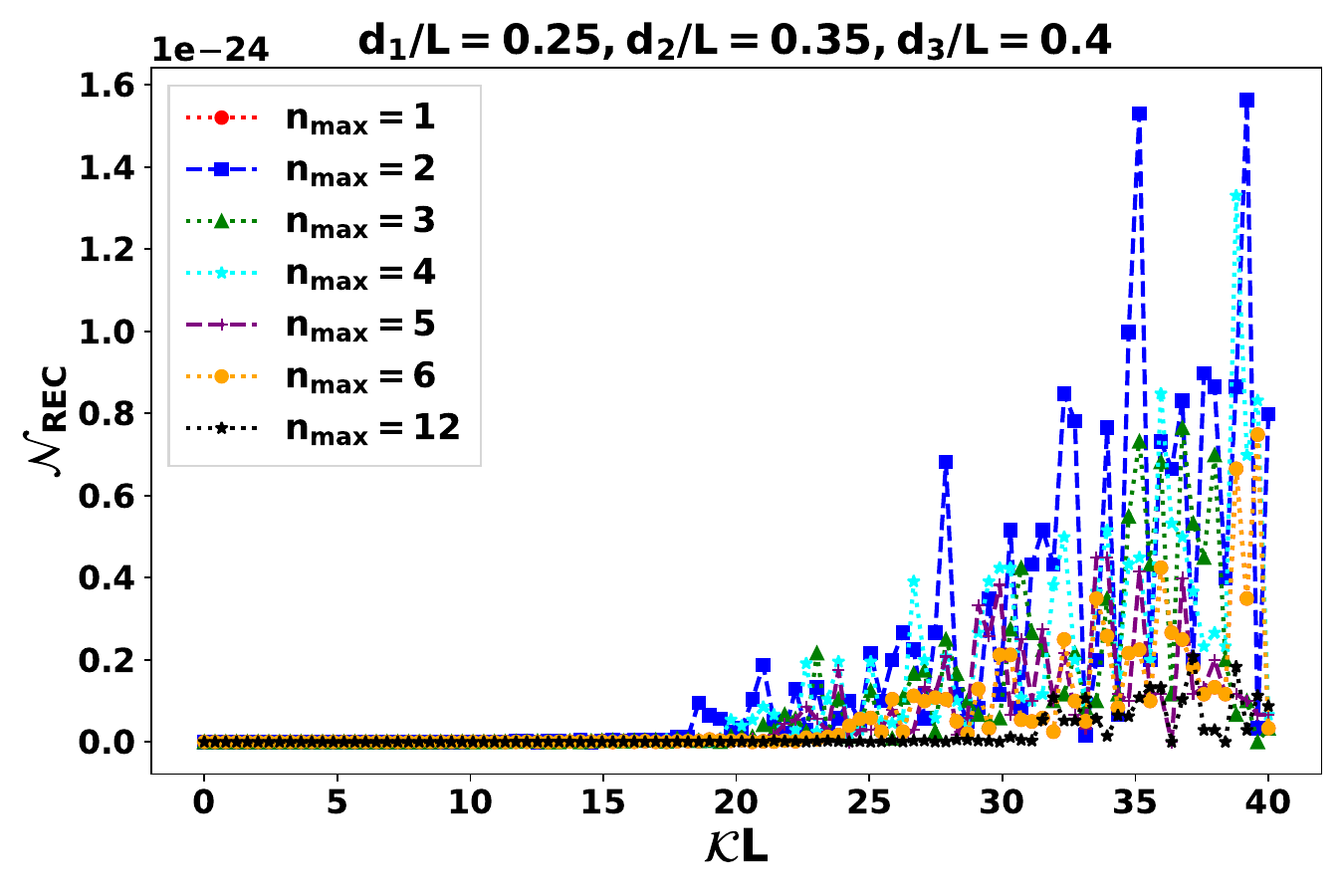} \vfill $\left(a\right)$
				\end{minipage}\hfill
				\begin{minipage}[b]{.33\linewidth}
					\centering
					\includegraphics[scale=0.26]{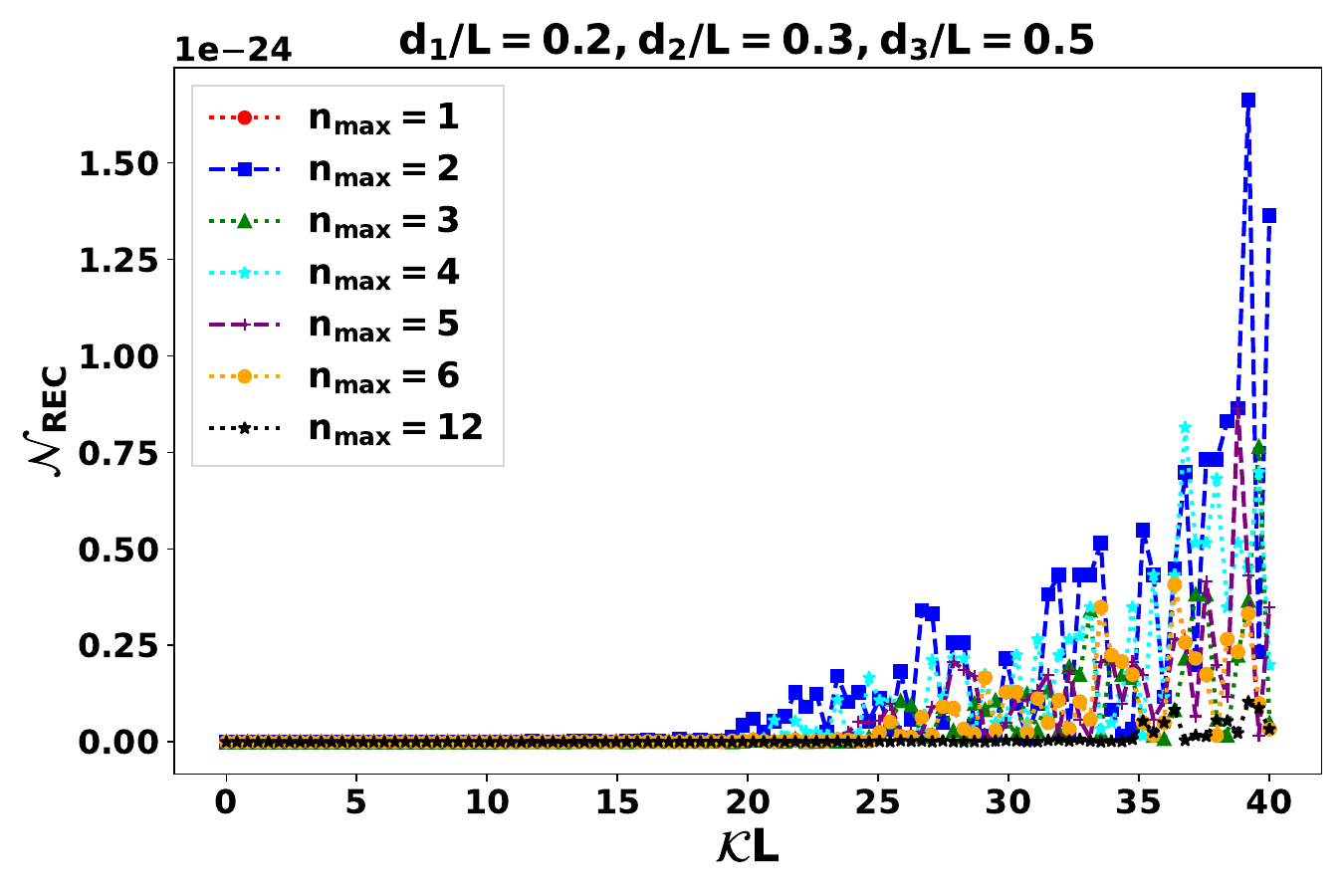} \vfill \vfill  $\left(b\right)$
		\end{minipage}}
	\begin{minipage}[b]{.33\linewidth}
		\centering
		\includegraphics[scale=0.26]{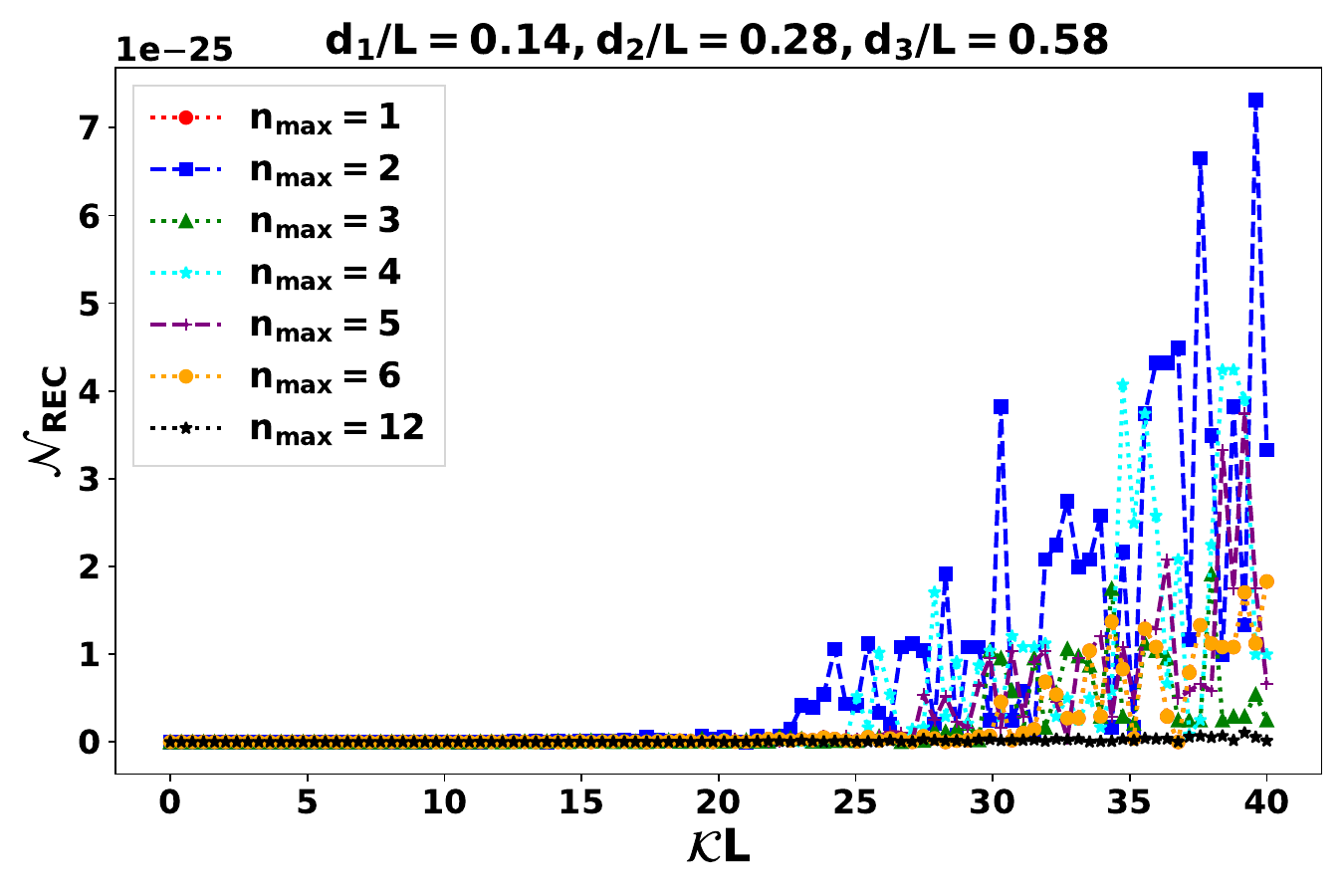} \vfill \vfill  $\left(c\right)$
\end{minipage}}
		\caption{Non-Markovianity dynamics measured by $\mathcal{N}_{\text{REC}}$ as a function of the momentum $\mathcal{K}$ for different values of $n_{\max}$, with varying values of (a) $(\frac{d_{1}}{L}=0.25, \frac{d_{2}}{L}=0.35, \frac{d_{3}}{L}=0.4)$, (b) $(\frac{d_{1}}{L}=0.2, \frac{d_{2}}{L}=0.36, \frac{d_{3}}{L}=0.44)$ and (c) $(\frac{d_{1}}{L}=0.14, \frac{d_{2}}{L}=0.28, \frac{d_{3}}{L}=0.58)$.} \label{NM}
\end{figure}
Fig. (\ref{NM}) presents the non-Markovianity dynamics measured by $\mathcal{N}_{\text{REC}}$ (\ref{MNMN}) as a function of the momentum $\mathcal{K}$ for different cutoff modes and three scenarios of positions of the layers for TLG system. Our results reveal that the non-Markovian character of the quantum dynamics can be precisely controlled by cutoff modes and the position of the layers. The systematic suppression of $\mathcal{N}_{\text{REC}}$ with increasing $n_{\max}$ across all layer configurations indicates that a richer electromagnetic environment drives the system toward Markovian behavior. This occurs because additional high-frequency vacuum modes create a continuum of decay channels, effectively dispersing quantum information and decoherence revivals necessary for information backflow.\\
The spatial arrangement of the graphene layers provides another powerful control mechanism. Comparing Figs. (\ref{NM}a), (\ref{NM}b), and (\ref{NM}c), we observe significant enhancement of non-Markovianity with increasing interlayer asymmetry. The configuration in (c) ($\frac{d_{1}}{L} = 0.14, \frac{d_{2}}{L} = 0.28, \frac{d_{3}}{L} = 0.58$), where layers are most spatially distinct, exhibits the strongest non-Markovian character. This enhancement stems from the formation of individual resonant interaction channels between each layer and the cavity field. The spatial separation prevents the formation of collective dark states and instead enables periodic information backflow from the cavity to the system. In contrast, the more symmetric configuration in (a) ($\frac{d_{1}}{L} = 0.25, \frac{d_{2}}{L} = 0.35, \frac{d_{3}}{L} = 0.4$) promotes collective coupling that leads to efficient irreversible energy dissipation, thereby damped $\mathcal{N}_{\text{REC}}$. The universal emergence of significant non-Markovianity for $\mathcal{K}L > 20$ signals a dynamical cross-over governed by value of the momentum $\mathcal{K}$. Below this threshold, the system exhibits Markovian relaxation, while above it, the interplay between electronic excitations and the confined field becomes dominated by memory effects and information backflow.

Furthermore, increasing the mode cutoff attenuates $\mathcal{N}_{\text{REC}}$. A denser spectrum of high-frequency modes acts as a more Markovian bath, dispersing information and not decoherence revivals. The momentum $\mathcal{K}L$ controls a dynamical transition, the system is Markovian ($\mathcal{N}_{\text{REC}} \approx 0$) for $\mathcal{K}L < 20$, and becomes distinctly non-Markovian for $\mathcal{K}L > 20$. In summary,  maximal non-Markovianity is achieved by combining large interlayer separation, a low mode cutoff, and high $\mathcal{K}L$, thereby optimizing the cavity's memory capacity.

\begin{widetext}

\begin{figure}[h]
\centering

\begin{minipage}[b]{.23\linewidth}
    \centering
    \includegraphics[scale=0.24]{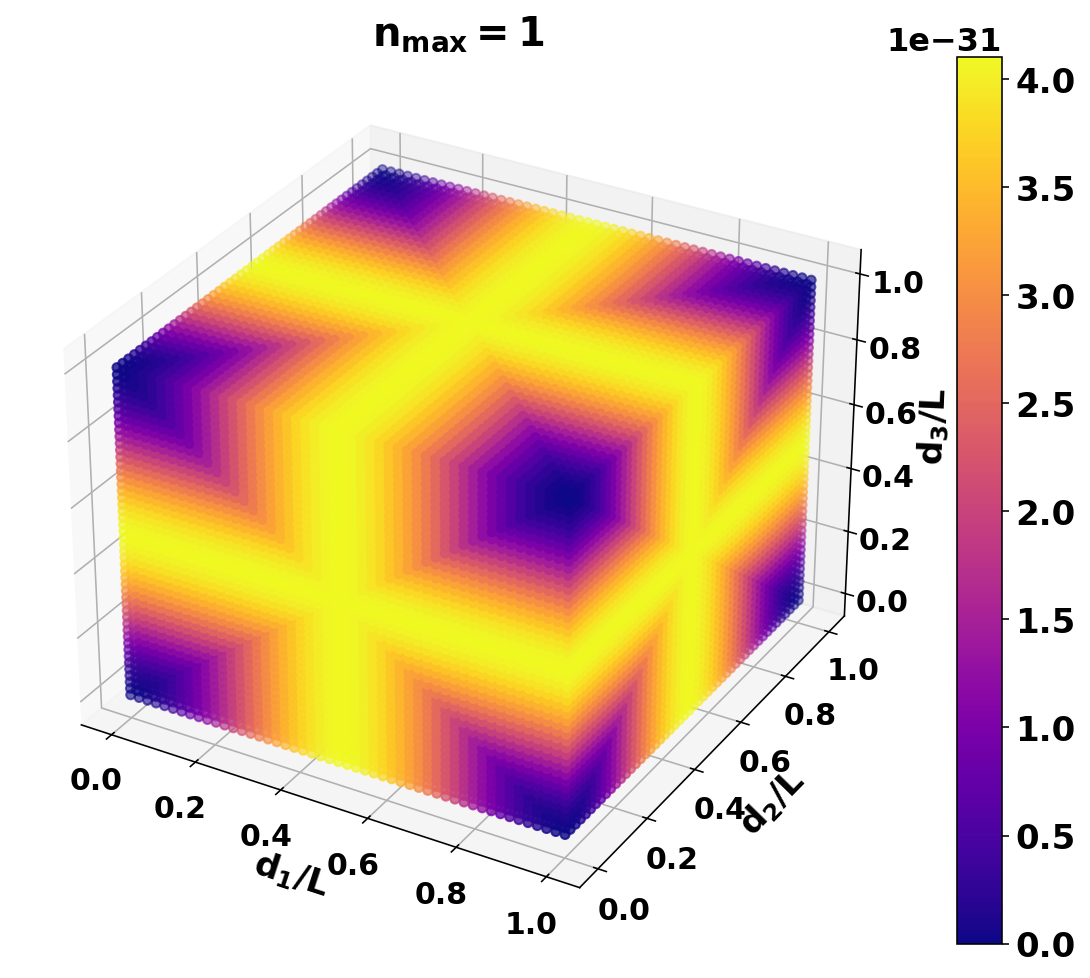}
     \vfill \vfill  $\left(a\right)$
\end{minipage}
\hfill
\begin{minipage}[b]{.23\linewidth}
    \centering
    \includegraphics[scale=0.24]{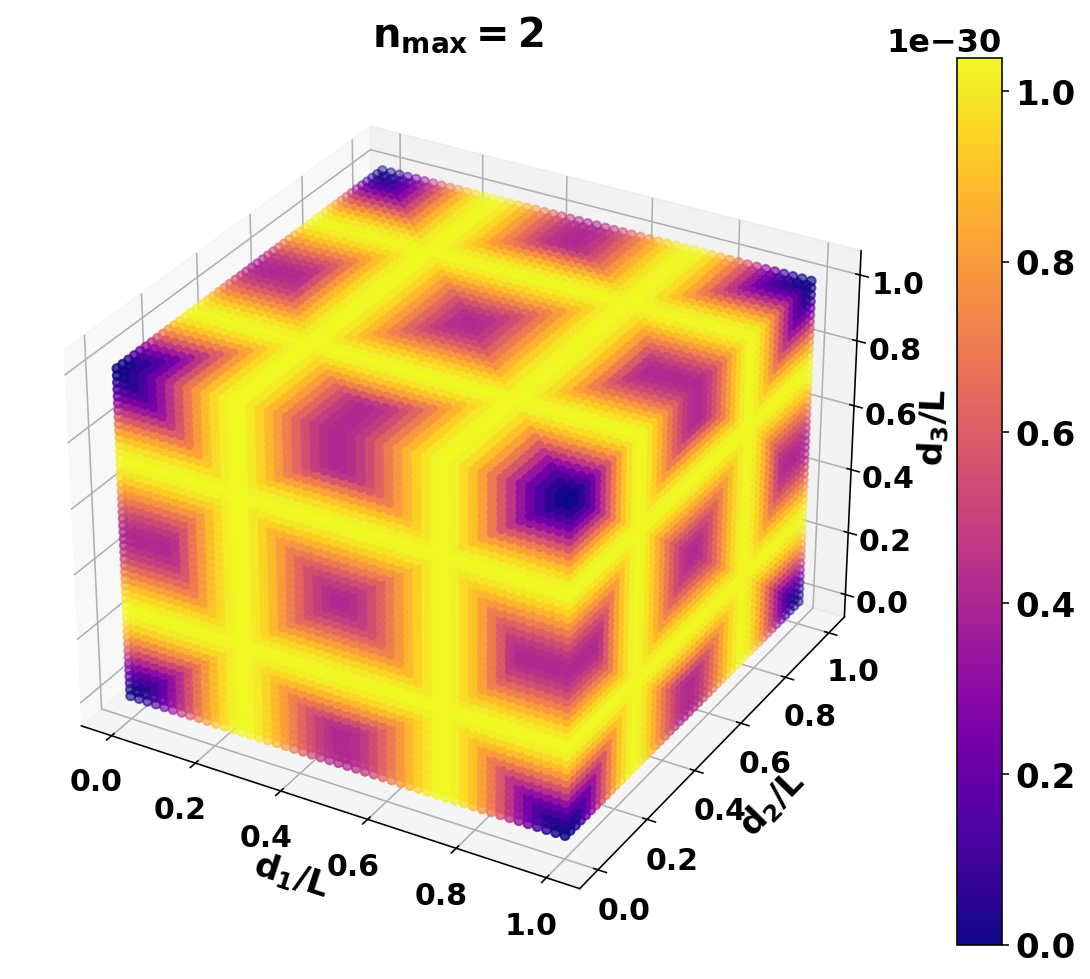}
     \vfill \vfill  $\left(b\right)$
\end{minipage}
\hfill
\begin{minipage}[b]{.23\linewidth}
    \centering
    \includegraphics[scale=0.24]{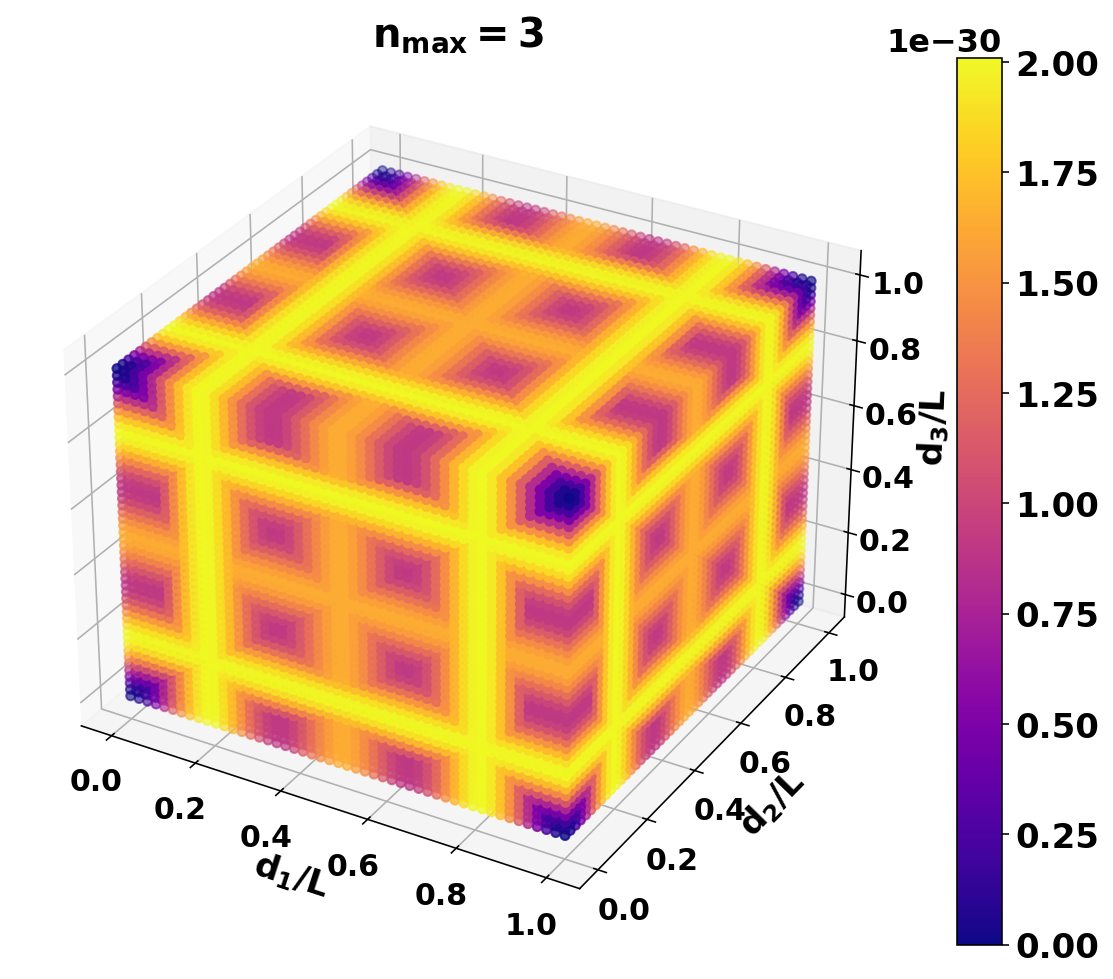}
     \vfill \vfill  $\left(c\right)$
\end{minipage}
\hfill
\begin{minipage}[b]{.23\linewidth}
    \centering
    \includegraphics[scale=0.24]{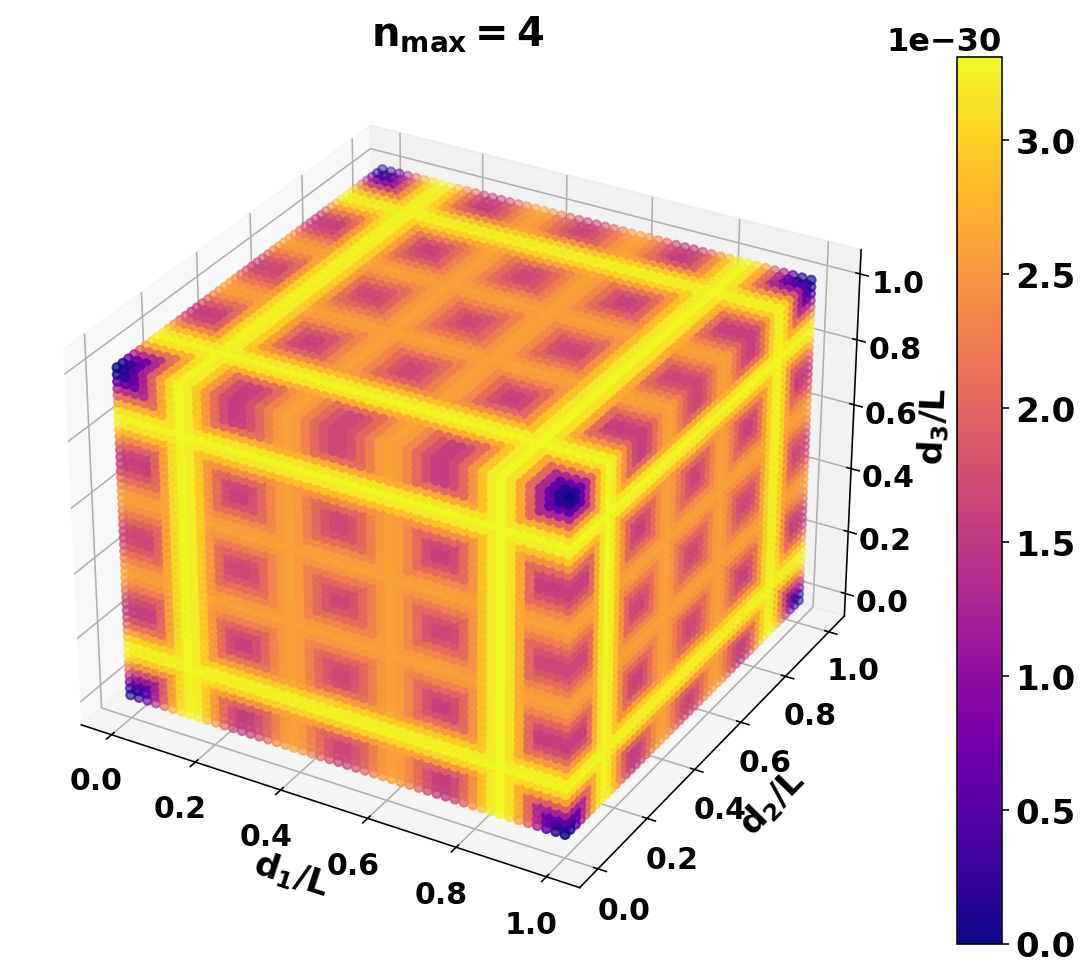}
     \vfill \vfill  $\left(d\right)$
\end{minipage}

\vspace{1cm} 

\begin{minipage}[b]{.23\linewidth}
    \centering
    \includegraphics[scale=0.24]{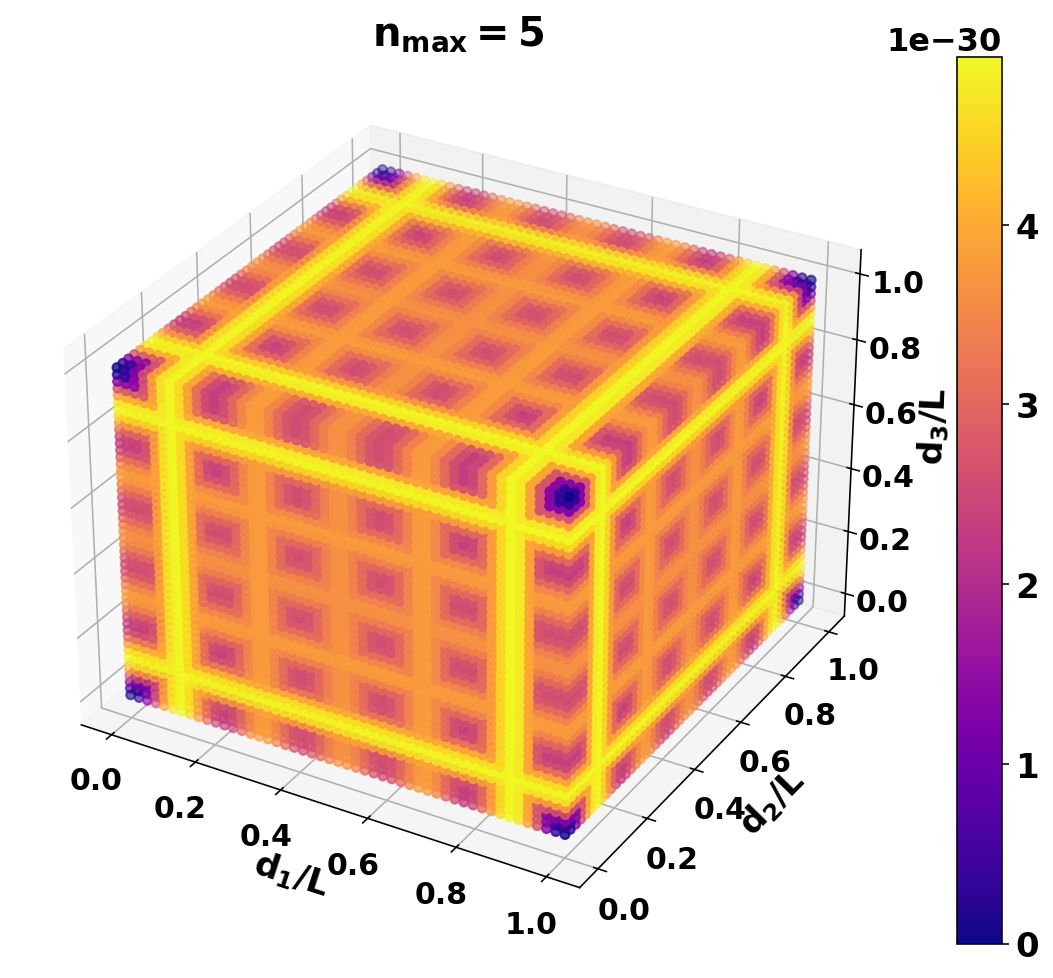}
     \vfill \vfill  $\left(e\right)$
\end{minipage}
\hfill
\begin{minipage}[b]{.23\linewidth}
    \centering
    \includegraphics[scale=0.24]{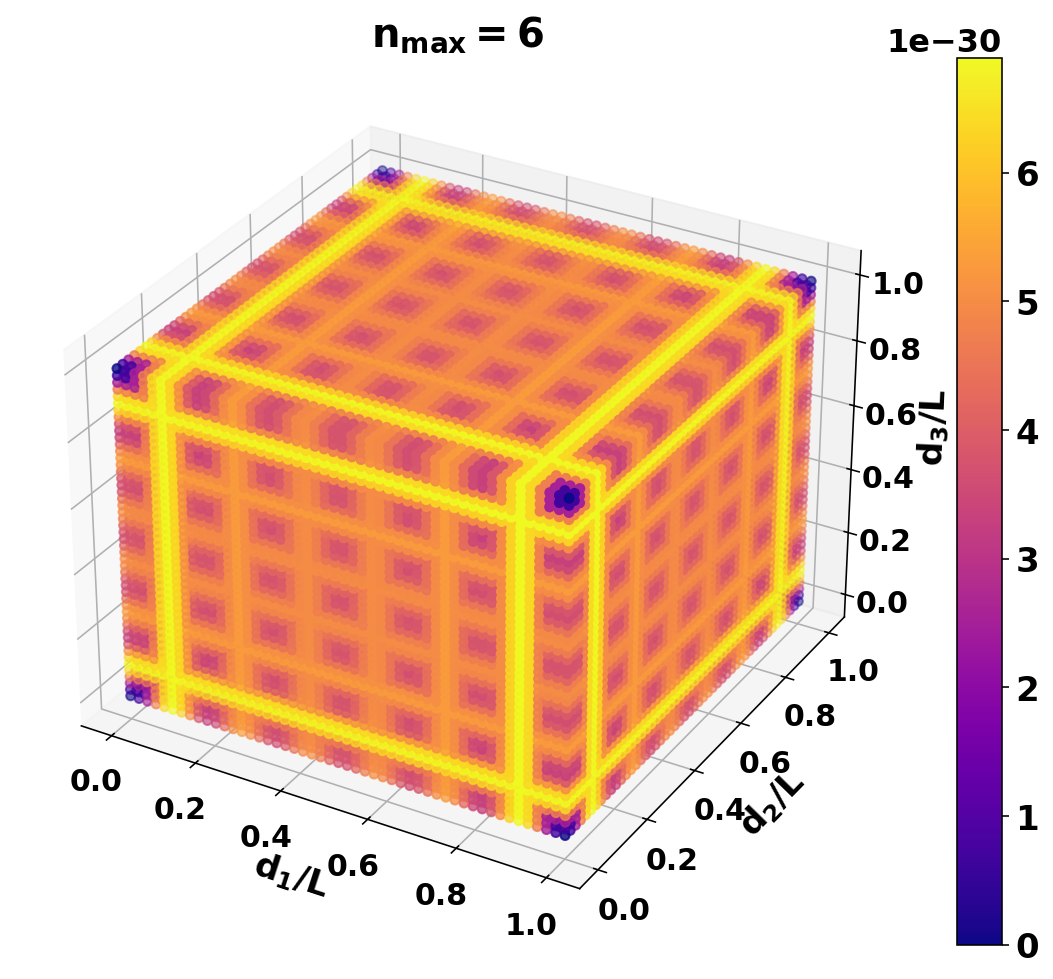}
     \vfill \vfill  $\left(f\right)$
\end{minipage}
\hfill
\begin{minipage}[b]{.23\linewidth}
    \centering
    \includegraphics[scale=0.24]{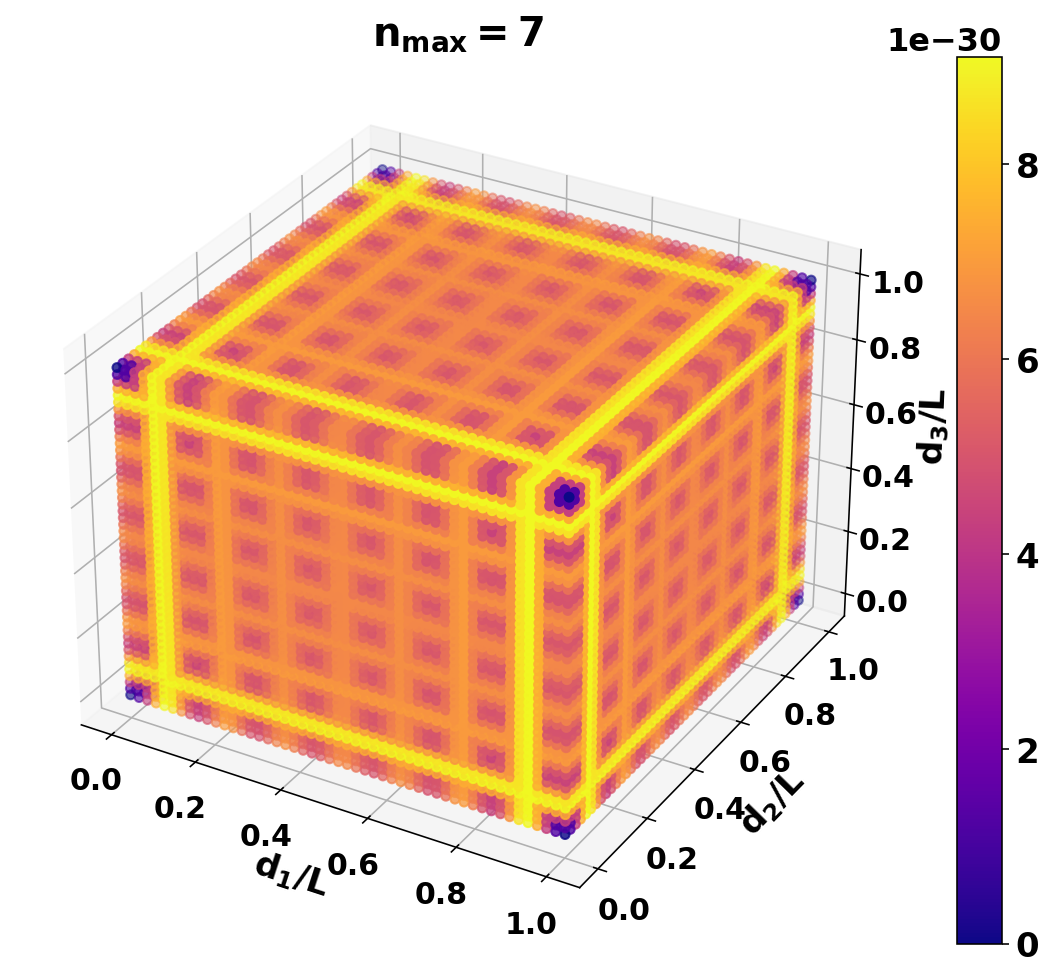}
     \vfill \vfill  $\left(g\right)$
\end{minipage}
\hfill
\begin{minipage}[b]{.23\linewidth}
    \centering
    \includegraphics[scale=0.24]{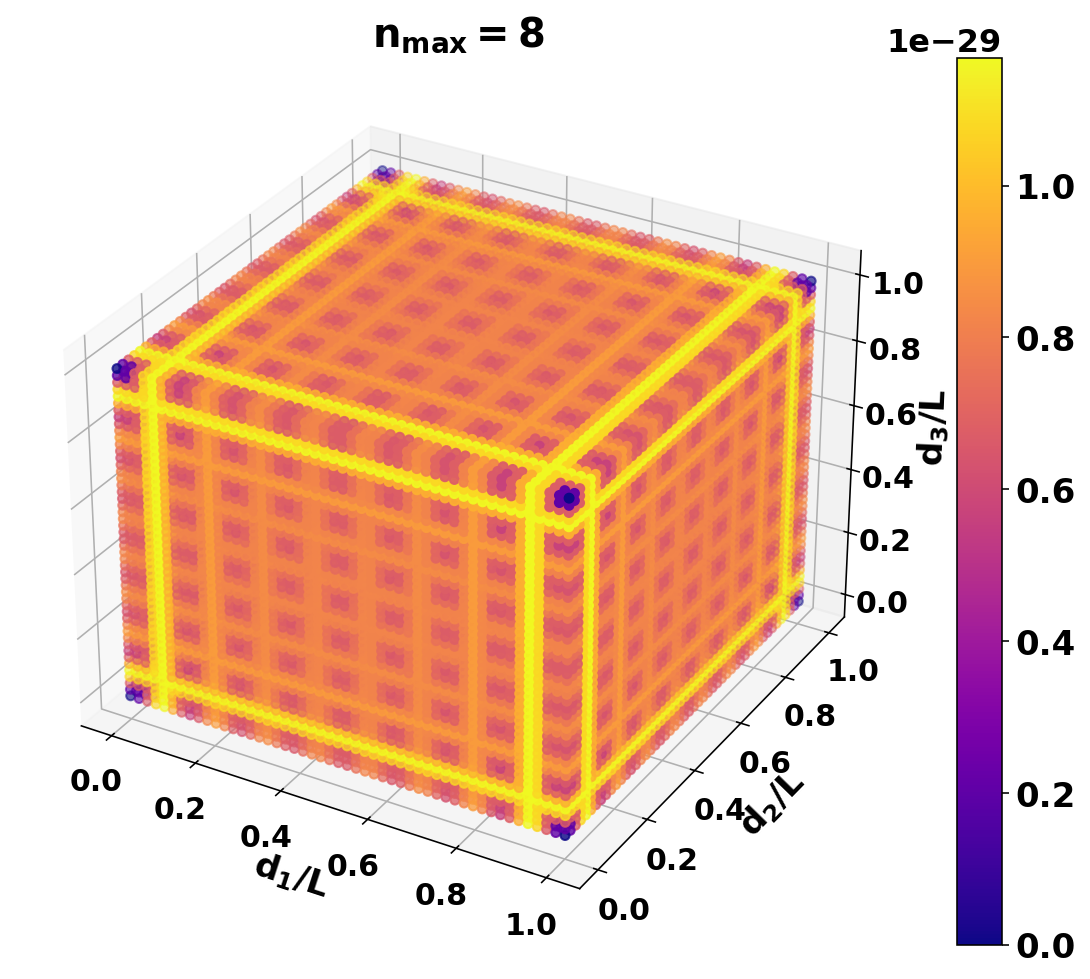}
     \vfill \vfill  $\left(h\right)$
\end{minipage}

\caption{Tangle $\mathcal{T}_{g}$ over $\alpha^{2}(t)$ as a function of $\frac{d_{i}}{L}$, for different cutoff modes $n_{\max}$, with $\mathcal{K} L=0$.}
\label{fig:tangle1}
\end{figure}
\begin{figure}[hbtp]
		{{\begin{minipage}[b]{.33\linewidth}
					\centering
					\includegraphics[scale=0.28]{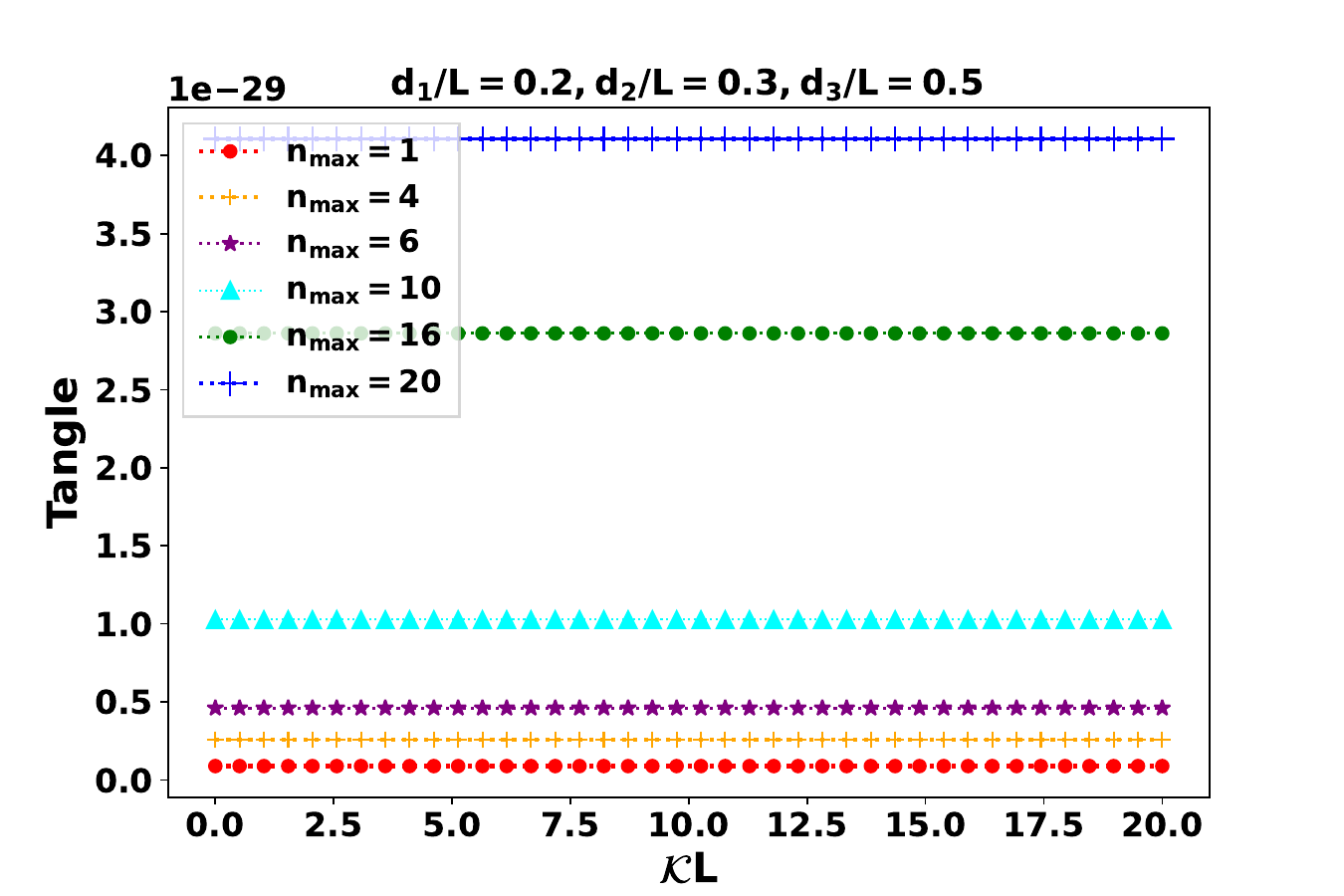} \vfill $\left(a\right)$
				\end{minipage}\hfill
				\begin{minipage}[b]{.33\linewidth}
					\centering
					\includegraphics[scale=0.28]{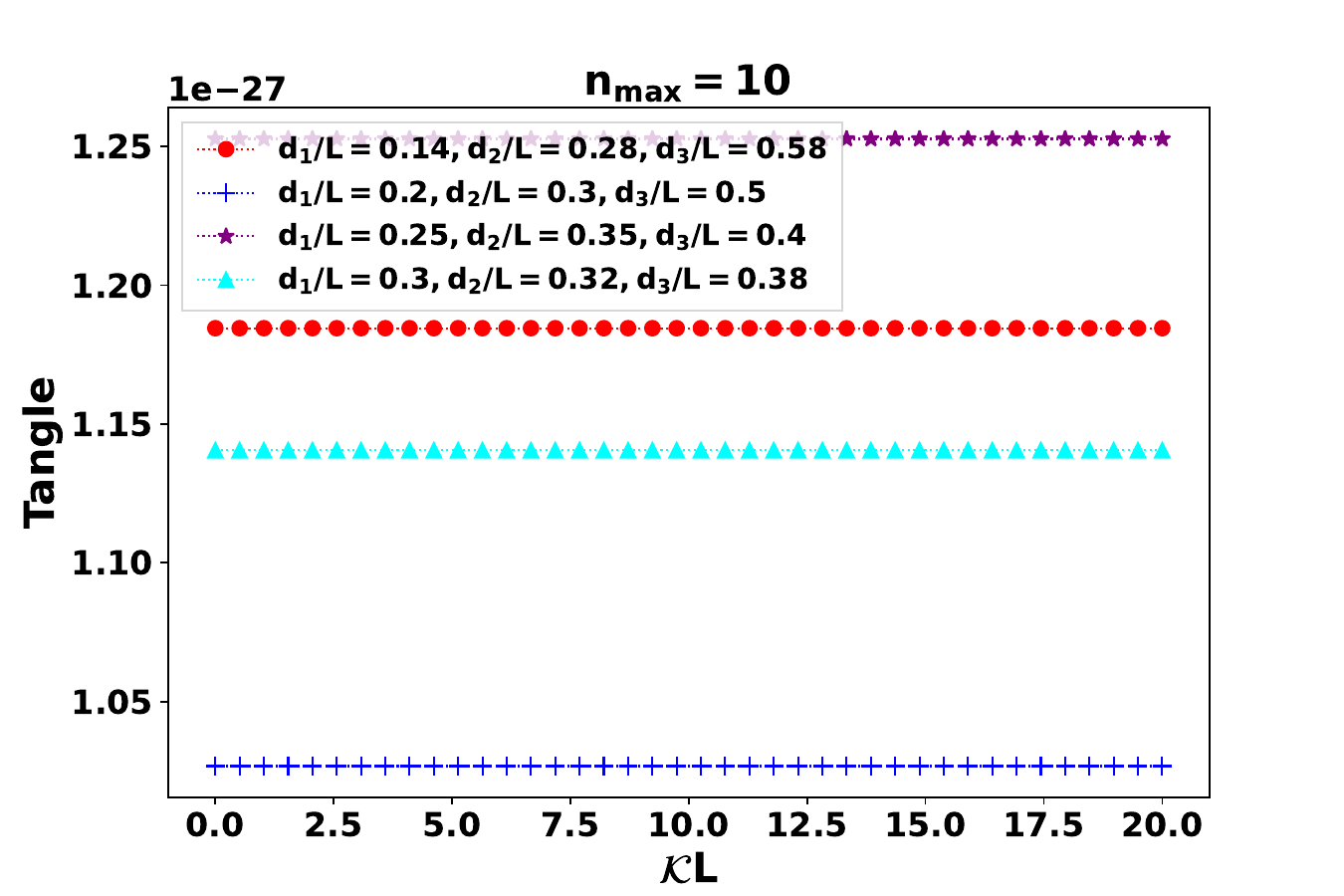} \vfill \vfill  $\left(b\right)$
		\end{minipage}}
	\begin{minipage}[b]{.33\linewidth}
		\centering
		\includegraphics[scale=0.28]{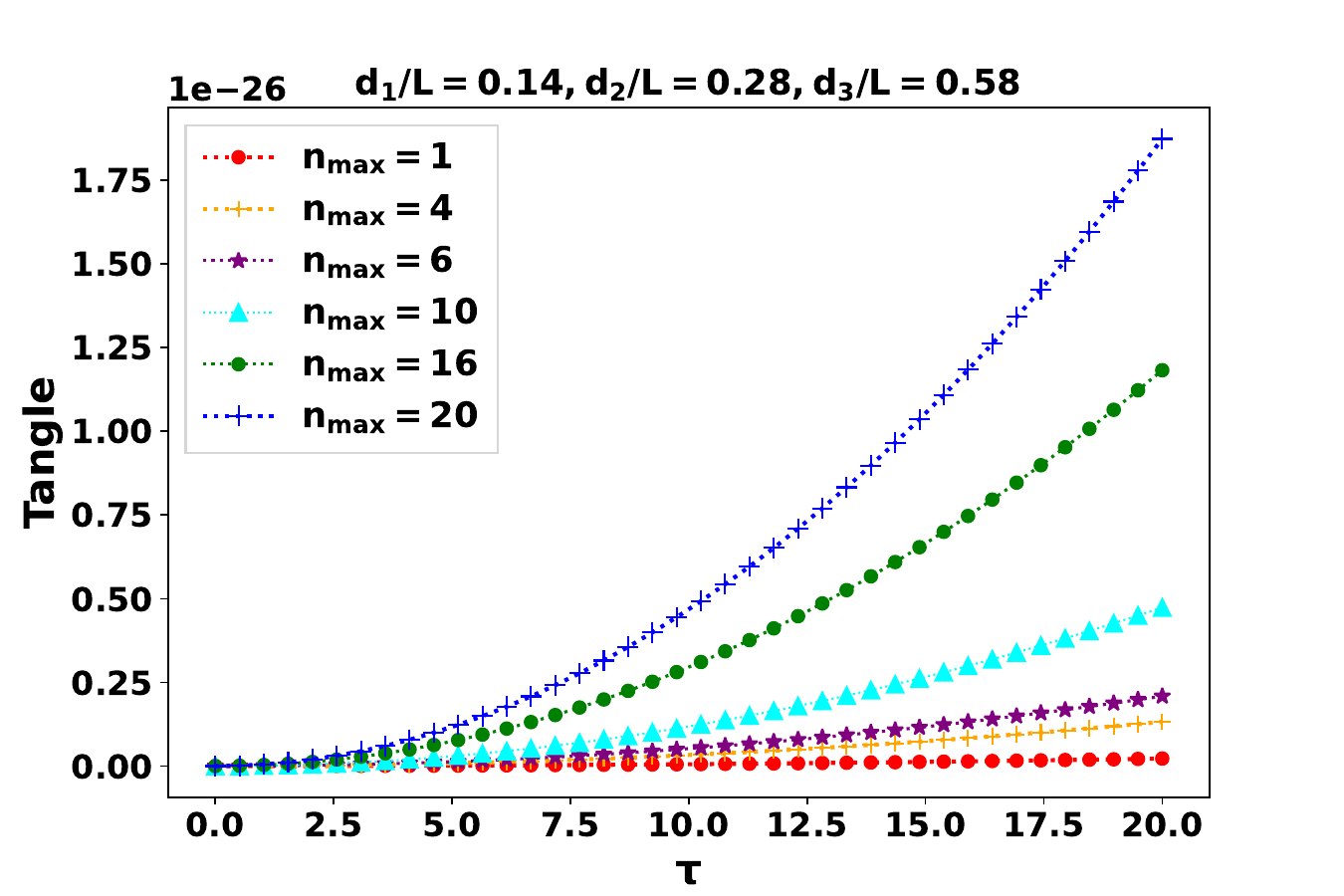} \vfill \vfill  $\left(c\right)$
\end{minipage}}
		\caption{Tangle $\mathcal{T}_{g}$ over $\alpha^{2}(t)$ (a) as a function $\mathcal{K}L$ for different cutoff modes $n_{\max}$, (b) for different layer positions $\frac{d_{i}}{L}$, and (c) as function time $t$ for different cutoff modes $n_{\max}$.} \label{figtang2}
\end{figure}
Fig. (\ref{fig:tangle1}) illustrates the evolution of the tangle $\mathcal{T}_{g}/\alpha^{2}(t)$ as a function of the normalized layer positions $\frac{d_{i}}{L}$ for different cutoff modes at fixed $\mathcal{K}L=0$. We observe a fundamental contrast with coherence behavior, the tangle amplitude increases with $n_{\max}$, suggesting distinct physical mechanisms govern entanglement generation versus QC preservation. A comprehensive analysis in Fig. (\ref{figtang2}) reveals the entanglement controllability through cavity and geometric parameters. Fig.(\ref{figtang2}a) demonstrates that increasing the number of cutoff mode $n_{\max}$ significantly enhances entanglement generation, particularly in the intermediate the momentum regime. This enhancement originates from the enriched vacuum environment, where additional electromagnetic modes provide more efficient channels for mediating quantum correlations between the graphene layers. The critical role of interlayer geometry is shown in Fig.(\ref{figtang2}b), the position of the layer ($\frac{d_{1}}{L}=0.25, \frac{d_{2}}{L}=0.35, \frac{d_{3}}{L}=0.4$) sustains stronger entanglement compared to more separated configurations. For Fig. (\ref{figtang2}c) reveals the temporal dynamics, showing that high $n_{\max}$ values not only enhance initial entanglement but also slow its decay. In summary, quantum entanglement in the TLG system can be optimized through dual control of the electromagnetic field (via high $n_{\max}$) and the position of the layers. This approach provides a powerful strategy for managing quantum resources in integrated nanophotonic devices, with entanglement and coherence exhibiting complementary responses to environmental structuring.

Both quantities are governed by the propagators $\mathcal{R}{ij}$, which encode the effective interaction mediated by the cavity modes. However, quantum coherence is mainly determined by the off-diagonal elements of the density matrix and is therefore highly sensitive to dephasing effects. As $n{\max}$ increases, the contribution of higher-order modes leads to enhanced destructive interference, resulting in a suppression of coherence. In contrast, entanglement is associated with correlations between different subsystems and depends on the cross terms $\mathcal{R}_{ij}$ with $i \neq j$. The inclusion of additional modes effectively strengthens these correlations, leading to an increase in entanglement. This explains the distinct trends observed in Figs. (\ref{REC_2D}) and (\ref{RECoh}).

\end{widetext}

\begin{figure}
\begin{minipage}[b]{.23\linewidth}
    \centering
    \includegraphics[scale=0.23]{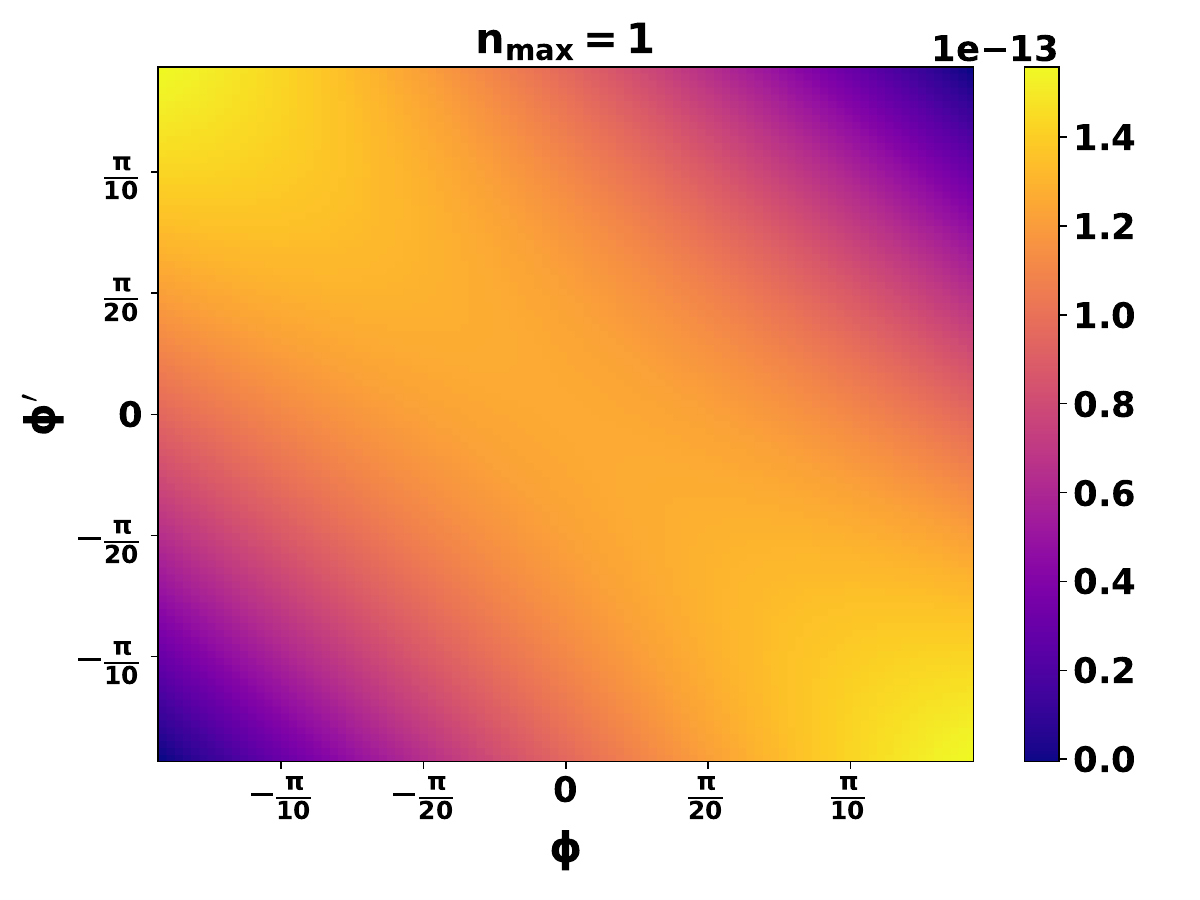}
    \vfill \vfill  $\left(a\right)$
\end{minipage}
\hfill
\begin{minipage}[b]{.23\linewidth}
    \centering
    \includegraphics[scale=0.23]{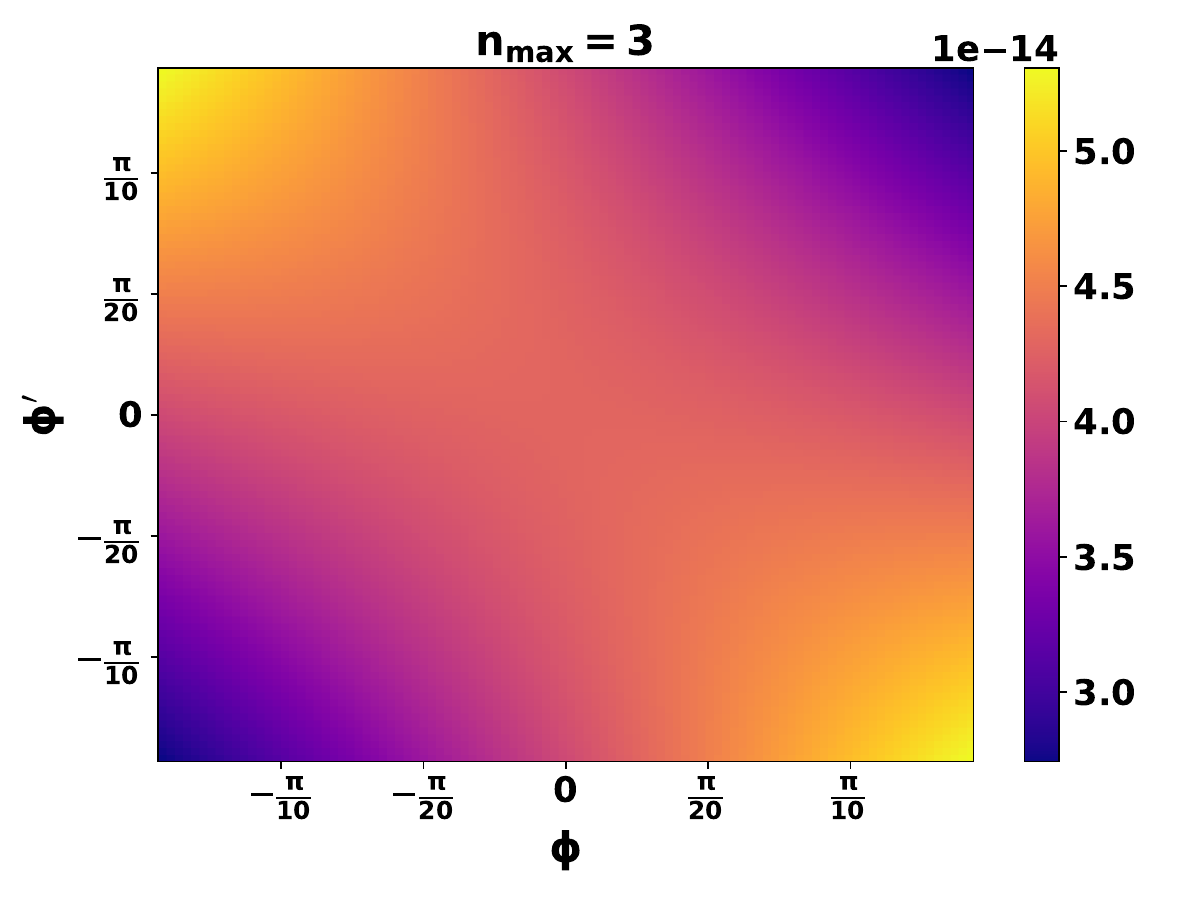}
    \vfill \vfill  $\left(b\right)$
\end{minipage}
\hfill
\begin{minipage}[b]{.23\linewidth}
    \centering
    \includegraphics[scale=0.23]{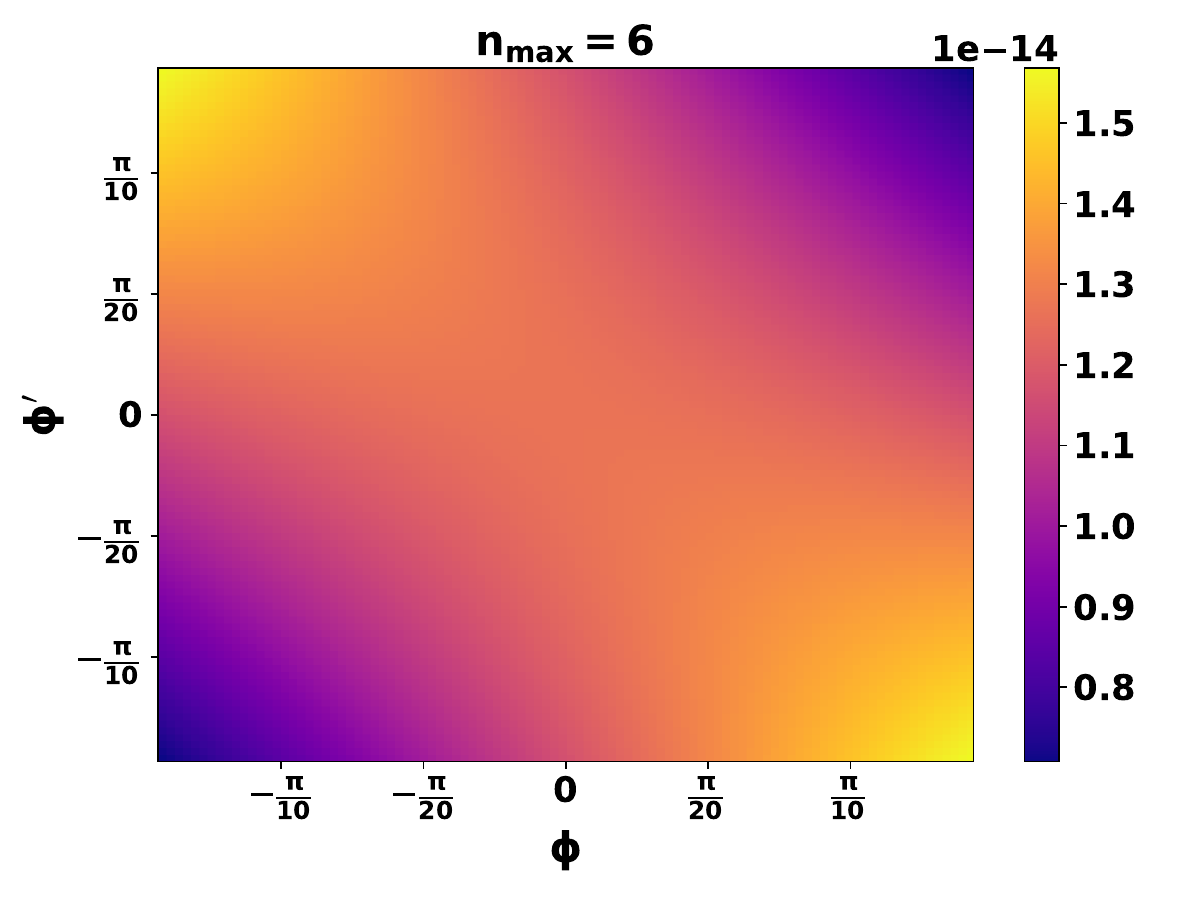}
    \vfill \vfill  $\left(c\right)$
\end{minipage}
\hfill
\begin{minipage}[b]{.23\linewidth}
    \centering
    \includegraphics[scale=0.23]{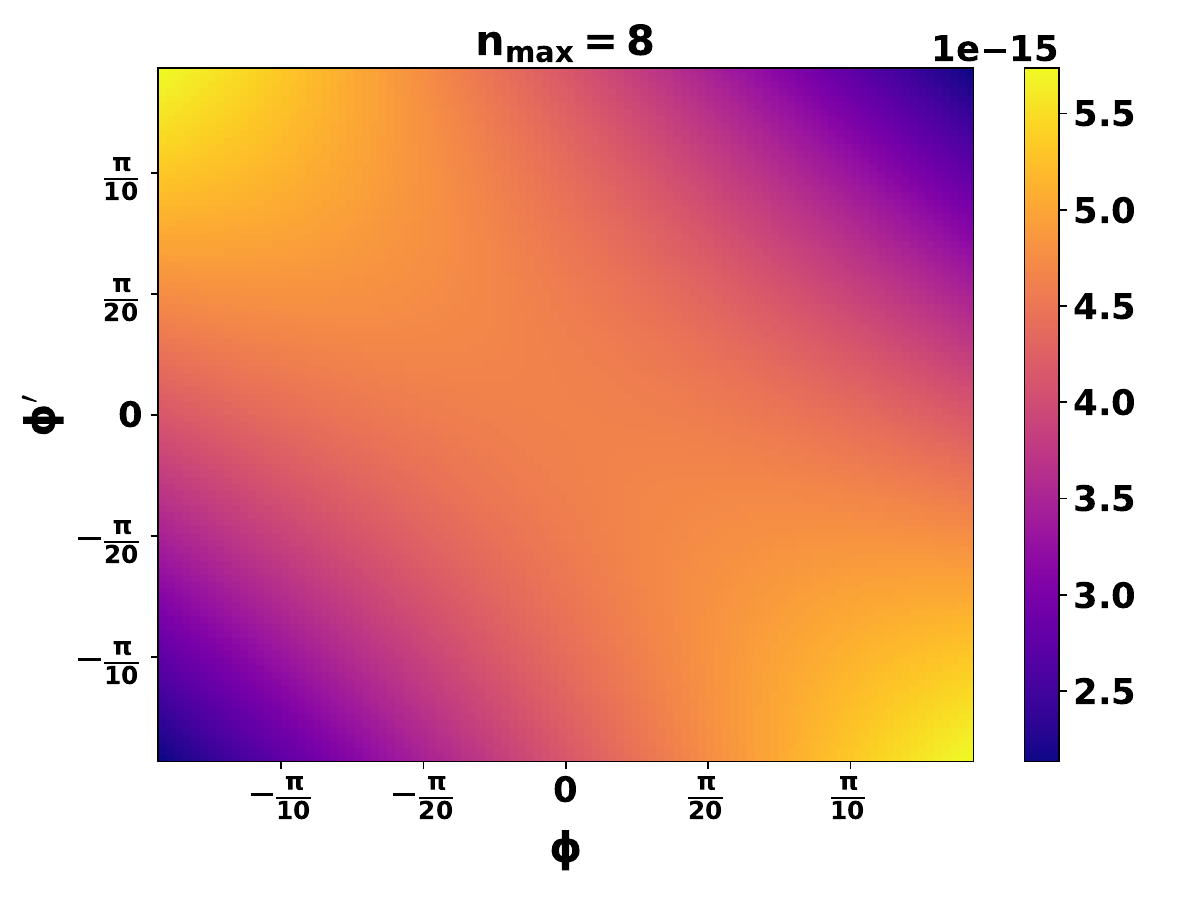}
    \vfill \vfill  $\left(d\right)$
\end{minipage}
\caption{REC $\mathcal{C}_{r}$ over $\alpha^{2}(t)$ as a function of $\phi$ (where $\phi=\phi_{1}=\phi_{2}=\phi_{3})$ and $\phi^{'}$ (where $\phi^{'}=\phi_{1}^{'}=\phi_{2}^{'}=\phi_{3}^{'}$) for different cutoff mode $n_{\max}$, with $\mathcal{K}L=0$ and $\frac{d_{1}}{L}=0.2, \frac{d_{2}}{L}=0.3, \frac{d_{3}}{L}=0.5$.} \label{RECoh}
\end{figure}

\begin{figure}
\begin{minipage}[b]{.23\linewidth}
    \centering
    \includegraphics[scale=0.23]{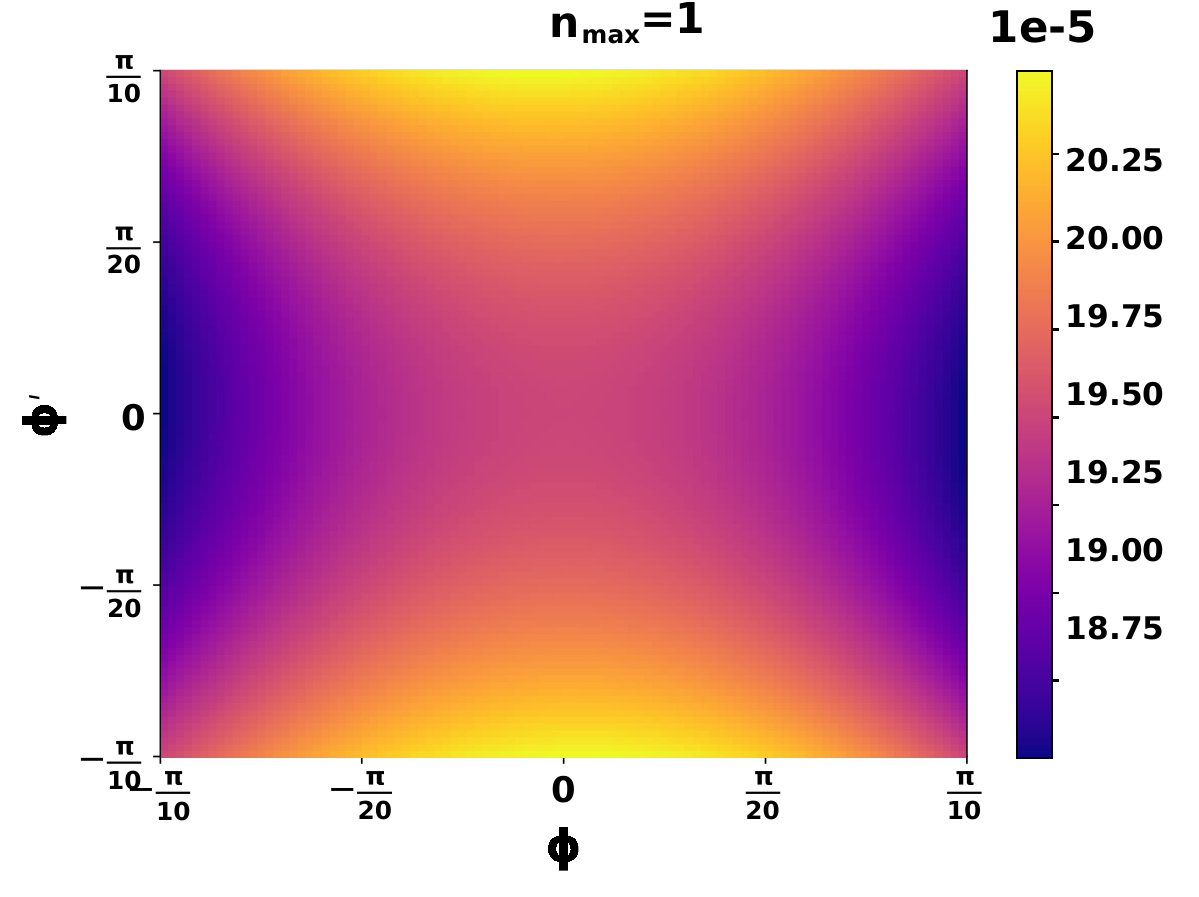}
    \vfill \vfill  $\left(a\right)$
\end{minipage}
\hfill
\begin{minipage}[b]{.23\linewidth}
    \centering
    \includegraphics[scale=0.23]{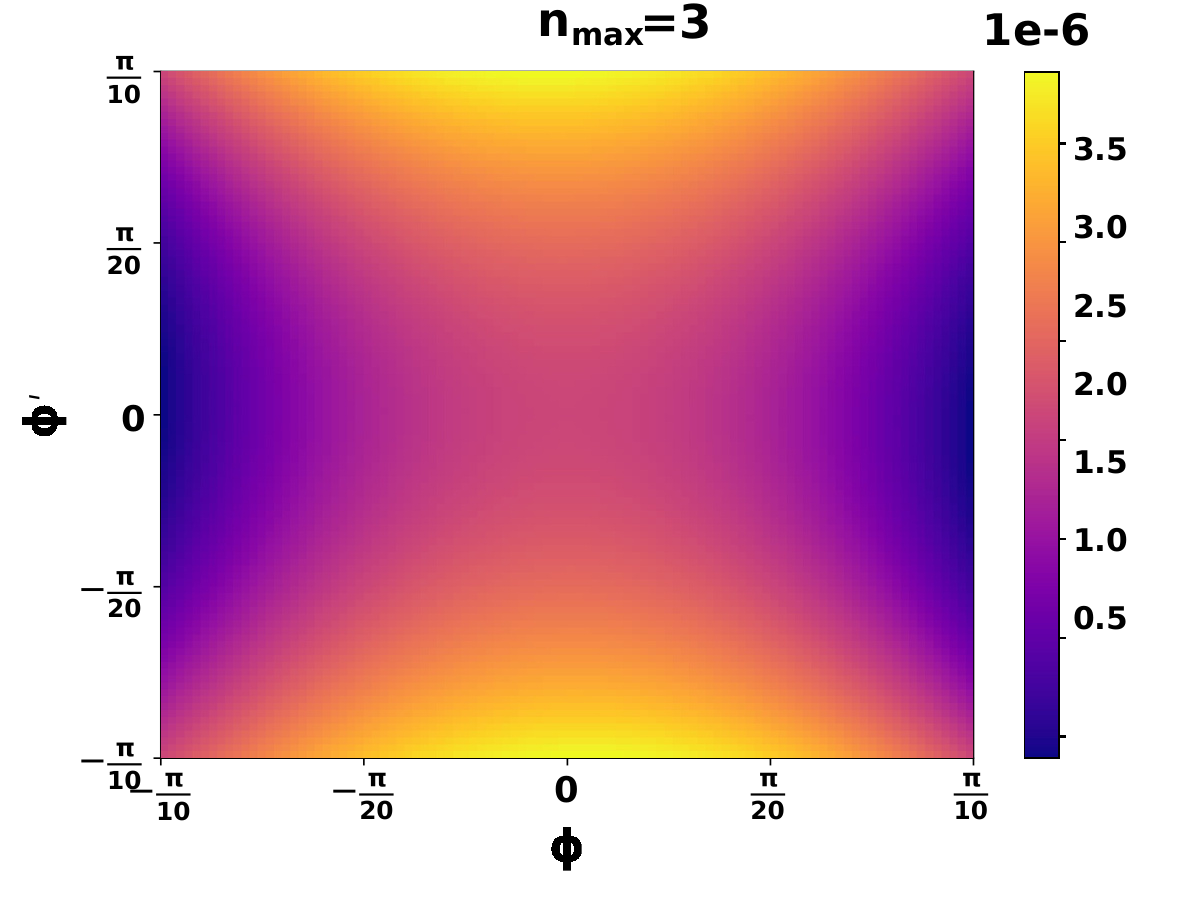}
    \vfill \vfill  $\left(b\right)$
\end{minipage}
\hfill
\begin{minipage}[b]{.23\linewidth}
    \centering
    \includegraphics[scale=0.23]{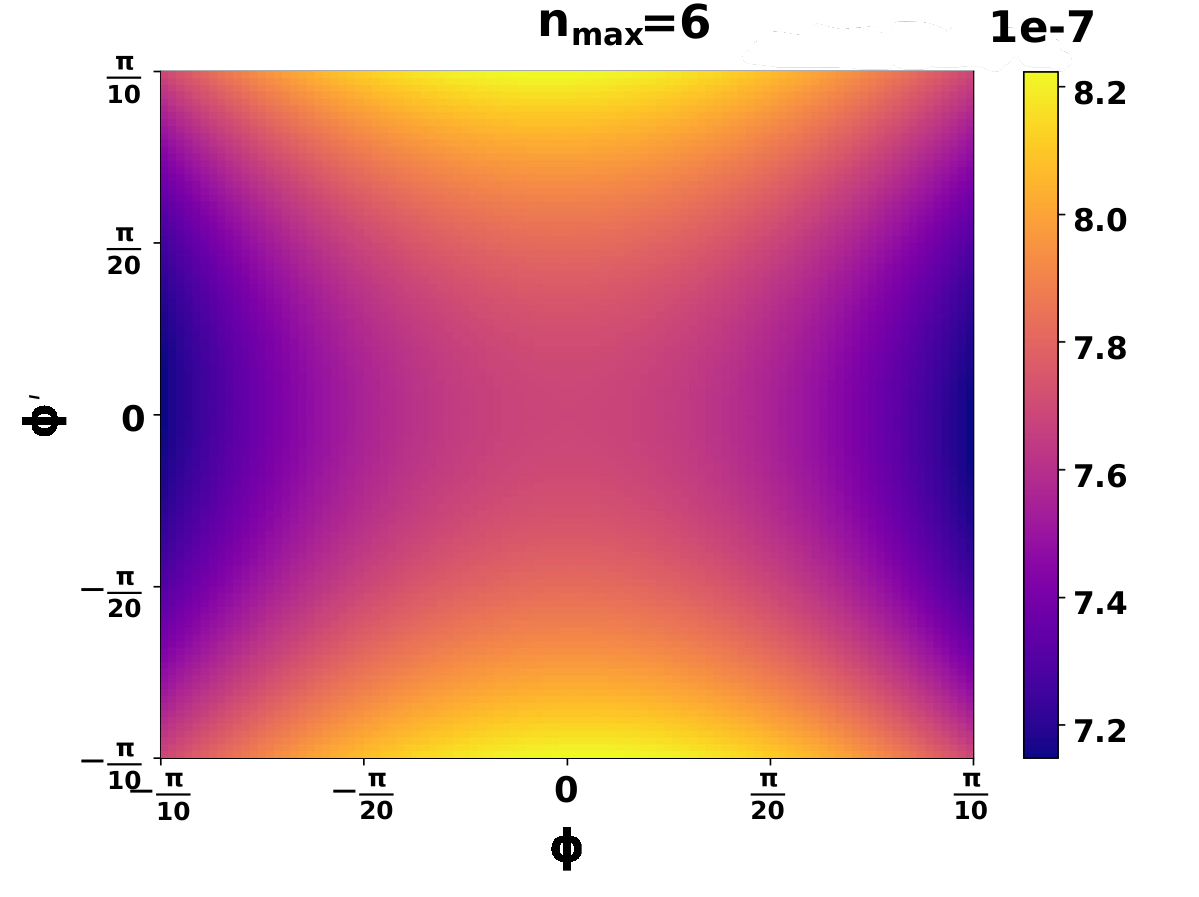}
    \vfill \vfill  $\left(c\right)$
\end{minipage}
\hfill
\begin{minipage}[b]{.23\linewidth}
    \centering
    \includegraphics[scale=0.23]{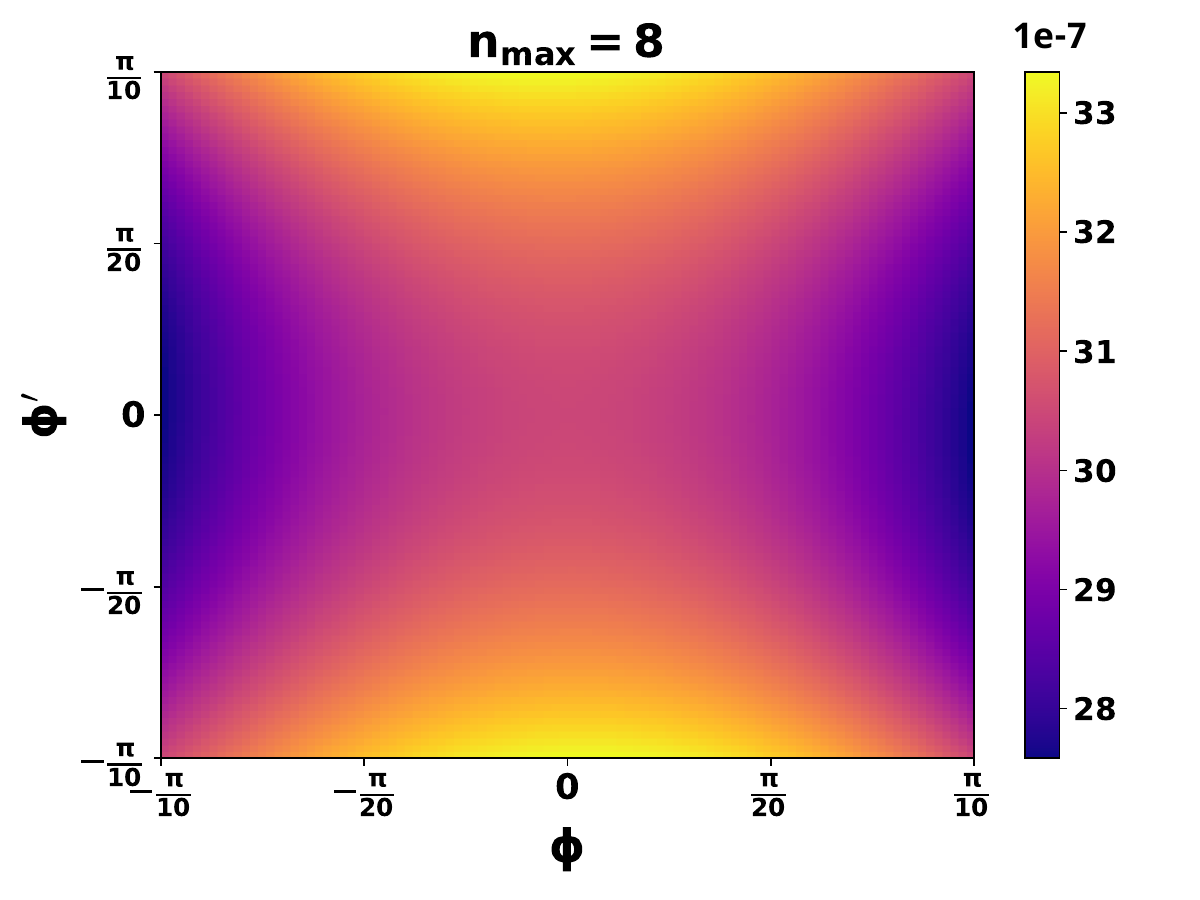}
    \vfill \vfill  $\left(d\right)$
\end{minipage}
\caption{Tangle $\mathcal{T}_{g}$ over $\alpha^{2}(t)$ as a function of $\phi$ (where $\phi=\phi_{1}=\phi_{2}=\phi_{3})$ and $\phi^{'}$ (where $\phi^{'}=\phi_{1}^{'}=\phi_{2}^{'}=\phi_{3}^{'}$) for different cutoff mode $n_{\max}$, with $\mathcal{K}L=0$ and $\frac{d_{1}}{L}=0.2, \frac{d_{2}}{L}=0.3$, $\frac{d_{3}}{L}=0.5$.} \label{Tangeldd}
\end{figure}

\subsection*{Effect of electron angles $\phi$ and $\phi'$ on quantum coherence, entanglement, and non-Markovian dynamics}
This section systematically examines the combined influence of critical physical parameters on the quantum behavior of the system. We study how the cavity parameters,  the number of cutoff modes $n_{\max}$, time $t$, the position of the layers $d_{i}/L$ and the momentum $\mathcal{K}$ encoded in the reduced density matrix $\varrho(t)$ (\ref{rhot2}), as well as the quantum configuration angles ($\phi$ and $\phi^{'}$) that determine the initial state, simultaneously modulate the evolution of three quantifies, REC, tangle, and non-Markovianity. This multidimensional analysis aims to establish quantitative relationships between these controllable degrees of freedom and the emerging quantum properties of the TLG system.\par

Here, we study the evolution of the REC as a function of the rotation angles $\phi$ and $\phi^{'}$ for different values of the cutoff mode in Fig. (\ref{RECoh}). Analysis of the angular dependence reveals two fundamental characteristics of the system. On the one hand, the coherent rotation of the three graphene layers according to angles $\phi$ and $\phi^{'}$ generates a structured quantum landscape in which the REC exhibits a pronounced maximum at $\phi=\phi^{'} = \pi/10$. On the other hand, we observe that this coherence systematically decreases with increasing $n_{\max}$, indicating a particular sensitivity of coherent coupling to the modal density in this specific angular regime. This trend is consistent with the behavior observed for the tangle in Fig. (\ref{Tangeldd}), where quantum entanglement also decreases with increasing cutoff mode number.
\begin{figure}[hbtp]
		{{\begin{minipage}[b]{.33\linewidth}
					\centering
					\includegraphics[scale=0.26]{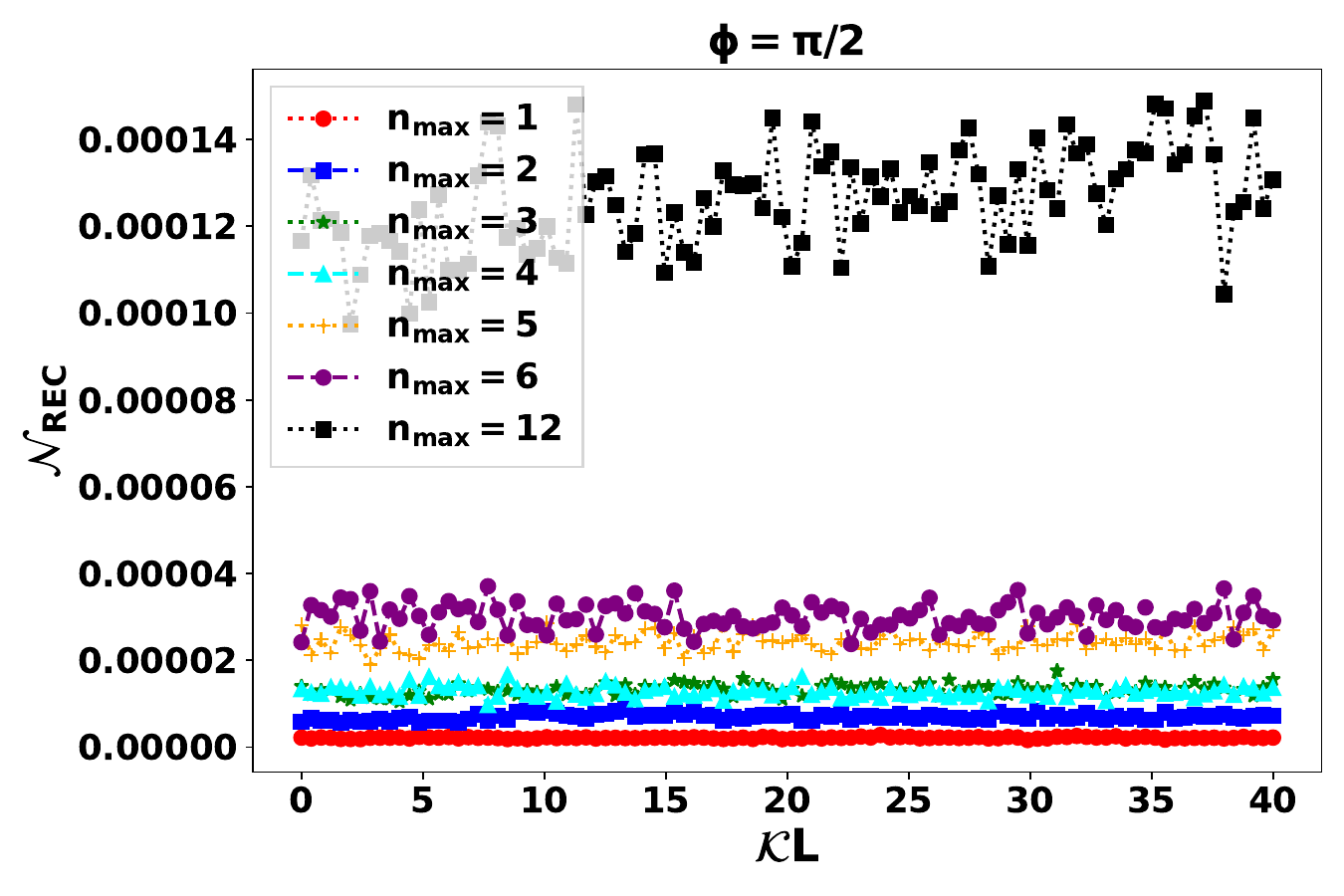} \vfill $\left(a\right)$
				\end{minipage}\hfill
				\begin{minipage}[b]{.33\linewidth}
					\centering
					\includegraphics[scale=0.26]{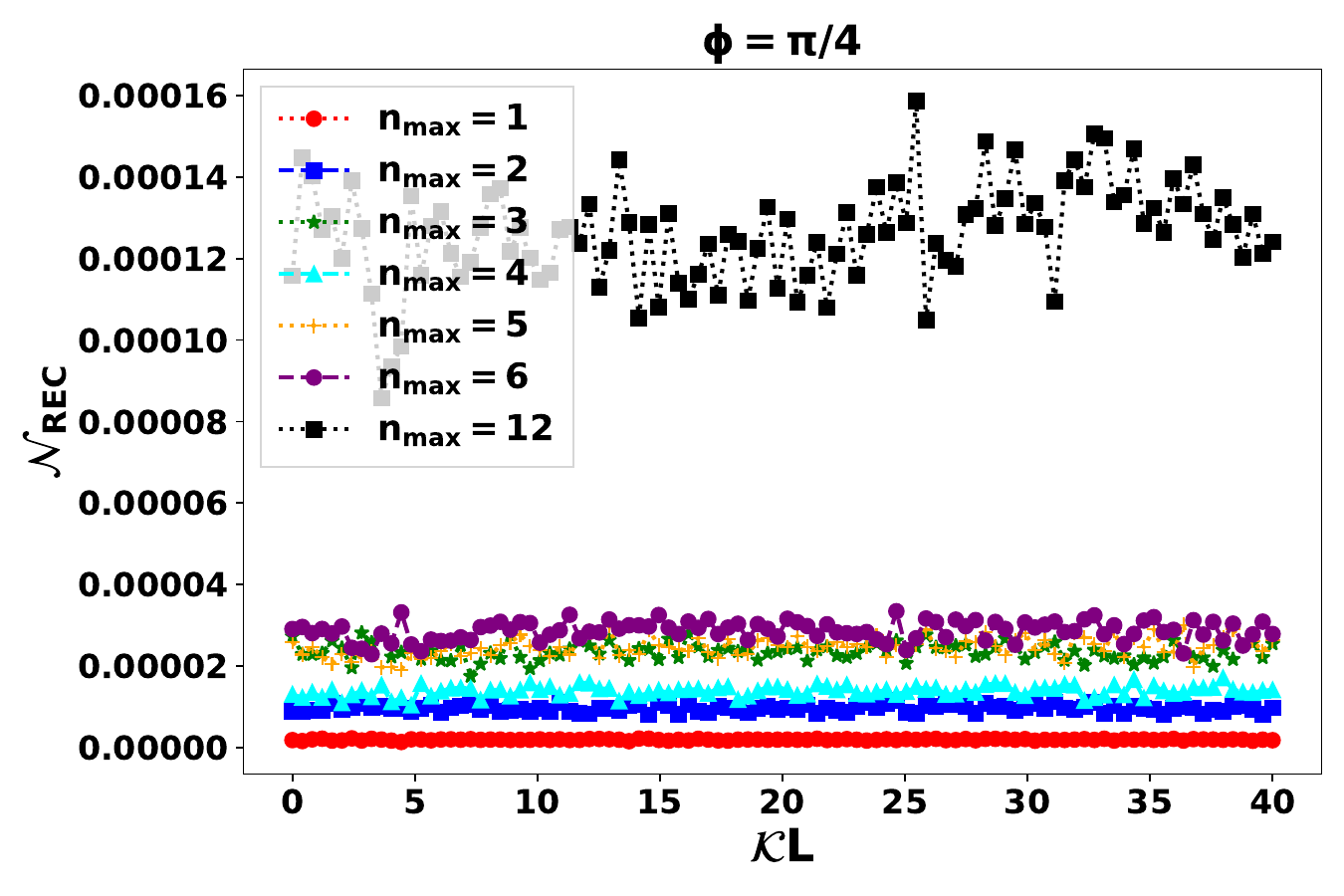} \vfill \vfill  $\left(b\right)$
		\end{minipage}}
	\begin{minipage}[b]{.33\linewidth}
		\centering
		\includegraphics[scale=0.26]{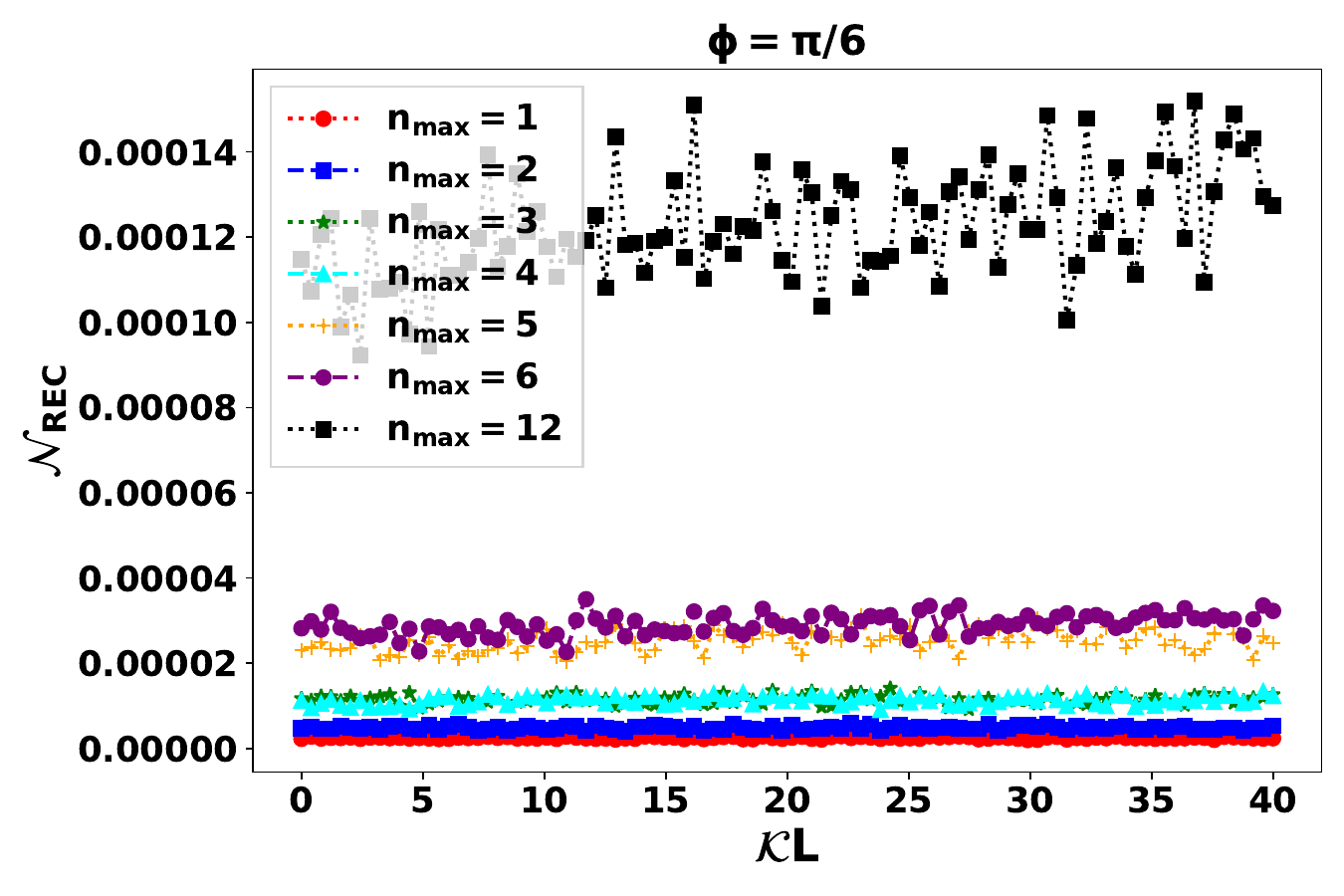} \vfill \vfill  $\left(c\right)$
\end{minipage}}
		\caption{Non-Markovianity $\mathcal{N}_{\text{REC}}$ as a function of momentum $\mathcal{K}$ for different values of $n_{\max}$, with varying values of $\phi$ (a) $\phi=\frac{\pi}{2}$ (b) $\phi=\frac{\pi}{4}$ and (c) $\phi=\frac{\pi}{6}$, with $\phi^{'}=-\frac{\pi}{4}$ and $\frac{d_{1}}{L}=0.2, \frac{d_{2}}{L}=0.3$, $\frac{d_{3}}{L}=0.5$.} \label{NM2}
\end{figure}
Fig.(\ref{NM2}) in a complementary study shows the evolution of non-Markovianity as a function of the momentum $\mathcal{K}$ for different cutoff modes $n_{\max}$, when the angle $\phi$ is varied from $\frac{\pi}{2}$ to $\frac{\pi}{4}$ and then to $\frac{\pi}{6}$, while keeping $\phi^{'}$ fixed at $-\frac{\pi}{4}$. Our observations reveal that the behavior of non-Markovianity fluctuates for each value of $n_{\max}$, but follows two main trends. First, the amplitude of non-Markovianity increases with $n_{\max}$. Secondly, it also increases with a decrease in value from the transition from $\mathcal{N}_{\text{REC}}(\phi=\pi/2)$ to $\mathcal{N}_{\text{REC}}(\phi=\pi/4)$ followed by a rise from $\mathcal{N}_{\text{REC}}(\phi=\pi/4)$ to $\mathcal{N}_{\text{REC}}(\phi=\pi/6)$. This double dependence indicates a complex coupling between the conditional structure of the cavity and the angular geometry of electronic states in non-Markovian system dynamics. Second, it also increases as the value of $\phi$ decreases. This dual dependence suggests complex coupling between the modal structure of the cavity and the angular geometry of the electronic states in the system's non-Markovian dynamics.\par

Our results demonstrate that the TLG system within a planar microcavity allows the robust generation of quantum coherence, entanglement and non-Markovian dynamics via the exchange of virtual photons with the confined vacuum. These effects have a critical dependence on the parameters of the cavity, number of cutoff modes ($n_{\max}$), on the inter-layer geometry ($d_{i}/L$), and on the angular configuration ($\phi, \phi'$). Unlike atomic systems or point detectors where the entanglement harvest is limited by the spatial location, our configuration exploits the natural delocalization of Dirac electrons and the graphene band structure to amplify quantum correlations. Future work could explore the role of an external electric field to modulate the valley-spin separation and thus control the entanglement channels. The richness of the controllable parameters (position of the layers, number of cutoff modes and angular configuration) makes this system a versatile platform for the engineering of quantum states and the study of the fundamental mechanisms of quantum memory.
\section{Conclusion} \label{clc}
In this paper, through a systematic investigation, we have examined the behaviors of quantum coherence, entanglement and non-Markovian dynamics in a triple-layer graphene system, as mediated by interaction with the vacuum state of a quantized electromagnetic field in a microcavity. Through the application of time-dependent perturbation theory and by tracing out the cavity field degrees of freedom, we have calculated and analyzed three key quantities, the relative entropy of coherence, the tangle a measure of entanglement and a non-Markovianity measure developed by REC. Our results demonstrate that these quantum properties exhibit strong and tunable dependence on three fundamental parameters, the number of cutoff modes, the relative positions of the layers, and the momentum. Specifically, we have shown that quantum coherence and entanglement can be enhanced through strategic configuration of the cavity environment and layer geometry, while non-Markovianity emerges under specific conditions of interlayer separation and the momentum. Furthermore, our systematic examination of the combined influence of critical physical parameters reveals additional insights into the quantum behavior of the system. The cavity parameters encoded in the reduced density matrix, along with the quantum configuration angles ($\phi$ and $\phi'$) defining the initial state, collectively modulate the evolution of the three fundamental observables. This multidimensional analysis establishes quantitative relationships between these controllable degrees of freedom and the emerging quantum properties of the TLG system. \par

It is important to emphasize that the magnitudes of the quantum measures obtained in this work remain relatively small, which is a direct consequence of the perturbative regime considered. In this framework, quantum coherence, entanglement, and memory effects arise as weak contributions induced by the interaction and are therefore naturally limited. Nevertheless, their strictly non-zero values demonstrate the emergence of genuine quantum correlations in the system. Their evolution is governed by the interaction time, leading to an increase at short times, although this growth remains constrained by the validity of the perturbative expansion. Consequently, the present results should be interpreted as capturing the leading-order behavior and parameter dependence of quantum correlations, rather than their absolute magnitude. Accessing larger values would require going beyond the perturbative regime, for instance by considering stronger coupling or longer interaction times.\par

Overall, our findings suggest that TLG within a cavity is a flexible platform for manipulating quantum coherence, entanglement and non-Markovianity via vacuum-mediated interactions. Our ability to manipulate these quantum features via experimentally accessible parameters demonstrates the potential of multilayer graphene systems for future graphene-based quantum technologies.
\appendix
\section{Analytical expressions for the matrix elements of the reduced density matrix} \label{appA}
To derive Eq. (\ref{q3}), we evaluate the matrix elements of the reduced density matrix $\varrho(t)$, which characterize the evolution of the subsystems under consideration. Then, the quantum states can be expressed $\bra{\mathbf{k}_{1}',s_{1}',\mathbf{k}_{2}',s_{2}',\mathbf{k}_{3}',s_{3}'}$ and $\ket{\mathbf{k}_{1},s_{1},\mathbf{k}_{2},s_{2},\mathbf{k}_{3},s_{3}}$ in coordinate representation as follows
\begin{align}
\bra{\mathbf{k}_{1}',s_{1}',\mathbf{k}_{2}',s_{2}',\mathbf{k}_{3}',s_{3}'}\text{Tr}_{a}(\mathcal{U}^{(2)}\varrho_{0})&\ket{\mathbf{k}_{1},s_{1},\mathbf{k}_{2},s_{2},\mathbf{k}_{3},s_{3}}= -(ev_{F})^{2}\sum_{i,j=1,2,3;\nu} \int_{0}^{t}\int_{0}^{t}dt_{1}dt_{2} e^{i \mathbf{k}_{i}.r_{i}}e^{i \mathbf{k}_{j}.r_{j}}e^{-i k'_{i}.r_{i}}e^{-i k'_{j}.r_{j}} \notag\\
&\times \bra{\chi_{0}}\mathcal{A}_{\nu}(r_{i},d_{i},t_{1})\mathcal{A}_{\nu'}(r_{j},d_{j},t_{2})\ket{\chi_{0}}\bra{s_{1}',s_{2}',s_{3}'}\sigma_{-\nu}^{(i)}(t_{1})\sigma_{-\nu'}^{(j)}(t_{2})\ket{s_{1},s_{2},s_{3}},
\end{align}
we have $\bra{\chi_{0}}A_{\nu}^{(i)}(t_{1})\mathcal{A}_{\nu'}^{(j)}(t_{2})\ket{\chi_{0}}=\delta_{\nu \nu'}\mathcal{R}_{ij}(|x|)$. Then,
\begin{align}
\bra{\mathbf{k}_{1}',s_{1}',\mathbf{k}_{2}',s_{2}',\mathbf{k}_{3}',s_{3}'}\text{Tr}_{a}(\mathcal{U}^{(2)}\varrho_{0})&\ket{\mathbf{k}_{1},s_{1},\mathbf{k}_{2},s_{2},\mathbf{k}_{3},s_{3}} =-(ev_{F})^{2}\sum_{i,j=1,2,3;\nu} \int_{0}^{t}\int_{0}^{t}dt_{1}dt_{2}\int d^{2}r_{i}\int d^{2}r_{j} \notag\\
& \times e^{i(\mathbf{k}_{i}-\mathbf{k}_{i}')r_{i}}e^{i(\mathbf{k}_{j}-\mathbf{k}_{j}')r_{j}}\mathcal{R}_{ij}(|x_{ij}|)\bra{s_{1}',s_{2}',s_{3}'}\sigma_{-\nu}^{(i)}(t_{1})\sigma_{-\nu'}^{(j)}(t_{2})\varrho_{e}\ket{s_{1},s_{2},s_{3}}.
\end{align}
The change of variables $\Delta r_{ij}=r_{i}-r_{j}$ is performed, resulting in
\begin{align}
\bra{\mathbf{k}_{1}',s_{1}',\mathbf{k}_{2}',s_{2}',\mathbf{k}_{3}',s_{3}'}&Tr_{a}(\mathcal{U}^{(2)}\varrho_{0})\ket{\mathbf{k}_{1},s_{1},\mathbf{k}_{2},s_{2},\mathbf{k}_{3},s_{3}} =-(ev_{F})^{2}\sum_{i,j=1,2,3;\nu} \int_{0}^{t}\int_{0}^{t}dt_{1}dt_{2}\int d^{2}r_{i}\int d^{2}r_{j} \notag\\
& \times e^{i(\mathbf{k}_{i}+\mathbf{k}_{i}'+\mathbf{k}_{j}-\mathbf{k}_{j}')r_{i}}e^{(\mathbf{k}_{j}-\mathbf{k}_{j}').\Delta r_{ij}} \mathcal{R}_{ij}(\sqrt{\Delta t^{2}-|\Delta r_{ij}}|^{2})\bra{s_{1}',s_{2}',s_{3}'}\sigma_{-\nu}^{(i)}(t_{1})\sigma_{-\nu'}^{(j)}(t_{2})\varrho_{e}\ket{s_{1},s_{2},s_{3}}.
\end{align}
For the first term, we have to integrate over $r_{i}$,
\begin{align} \bra{\mathbf{k}_{1}',s_{1}',\mathbf{k}_{2}',s_{2}',\mathbf{k}_{3}',s_{3}'}Tr_{a}(\mathcal{U}^{(2)}\varrho_{0})&\ket{\mathbf{k}_{1},s_{1},\mathbf{k}_{2},s_{2},\mathbf{k}_{3},s_{3}}= -(ev_{F})^{2} \delta_{\mathbf{k}_{i},\mathbf{k}_{i}'}\delta_{\mathbf{k}_{j},\mathbf{k}_{j}'}\sum_{i,j=1,2,3;\nu} \mathcal{R}_{ij}(\mathbf{k}_{2}-\mathbf{k}_{2}') \notag\\
& \times \int_{0}^{t}\int_{0}^{t}dt_{1}dt_{2} \bra{s_{1}',s_{2}',s_{2}'}\sigma_{-\nu}^{(i)}(t_{1})\sigma_{-\nu}^{(j)}(t_{2})\varrho_{e}\ket{s_{1},s_{2},s_{2}}\mathcal{R}_{ij}(\mathbf{k}_{j}-\mathbf{k}_{j}'), \label{q1}
\end{align}
where $\mathcal{R}_{ij}(\mathbf{k}_{j}-\mathbf{k}_{j}')$ denotes, respectively, the Fourier transform of 
$\mathcal{R}_{ij}\big(\sqrt{\Delta t^{2}-|\Delta r_{ij}|^{2}}\big)$,
\begin{align}
	\mathcal{R}_{ij}(\mathbf{k}_{j}-\mathbf{k}_{j}',\Delta t)=\int d^{2} \Delta r_{ij} e^{-i(\mathbf{k}_{j}-\mathbf{k}_{j}).\Delta r_{ij}} \mathcal{R}_{ij}(\sqrt{\Delta t^{2}-|\Delta r_{ij}|^{2}}).
\end{align}
The same technique can be used for $\bra{\mathbf{k}_{1}',s_{1}',\mathbf{k}_{2}',s_{2}',\mathbf{k}_{3}',s_{3}'}\text{Tr}_{a}(\mathcal{U}^{(1)}\varrho_{0}\mathcal{U}^{(1)\dagger})\ket{\mathbf{k}_{1},s_{1},\mathbf{k}_{2},s_{2},\mathbf{k}_{3},s_{3}}$ and \\$\bra{\mathbf{k}_{1}',s_{1}',\mathbf{k}_{2}',s_{2}',\mathbf{k}_{3}',s_{3}'}\text{Tr}_{a}(\varrho_{0}\mathcal{U}^{(2)\dagger})\ket{\mathbf{k}_{1},s_{1},\mathbf{k}_{2},s_{2},\mathbf{k}_{3},s_{3}}$,
\begin{align}
\bra{\mathbf{k}_{1}',s_{1}',\mathbf{k}_{2}',s_{2}',\mathbf{k}_{3}',s_{3}'}\text{Tr}_{a}(\varrho_{0}\mathcal{U}^{(2)\dagger})&\ket{\mathbf{k}_{1},s_{1},\mathbf{k}_{2},s_{2},\mathbf{k}_{3},s_{3}}=-(ev_{F})^{2}\delta_{\mathbf{k}_{i},\mathbf{k}_{i}'}\delta_{\mathbf{k}_{j},\mathbf{k}_{j}'}\sum_{i,j=1,2,3;\nu}\mathcal{R}_{ij}(\mathbf{k}_{2}-\mathbf{k}_{2}') \notag\\
&\quad\quad\quad\quad\quad\quad\times \int_{0}^{t}\int_{0}^{t}dt_{1}dt_{2}\bra{s_{1}',s_{2}',s_{2}'}\varrho_{e}\sigma_{-\nu}^{(i)\dagger}(t_{1})\sigma_{-\nu}^{(j)\dagger}(t_{2})\ket{s_{1},s_{2},s_{2}}, \label{q2}
\end{align}
and,
\begin{align}
\bra{\mathbf{k}_{1}',s_{1}',\mathbf{k}_{2}',s_{2}',\mathbf{k}_{3}',s_{3}'}\text{Tr}_{a}(\mathcal{U}^{(1)}\varrho_{0}\mathcal{U}^{(1)\dagger})&\ket{\mathbf{k}_{1},s_{1},\mathbf{k}_{2},s_{2},\mathbf{k}_{3},s_{3}}=2(ev_{F})^{2}\delta_{\mathbf{k}_{i},\mathbf{k}_{i}'}\delta_{\mathbf{k}_{j},\mathbf{k}_{j}'}\sum_{i,j=1,2,3;\nu} \mathcal{R}_{ij}(\mathbf{k}_{2}-\mathbf{k}_{2}') \notag\\
& \quad\quad\quad\quad\times \int_{0}^{t}\int_{0}^{t}dt_{1}dt_{2}\bra{s_{1}',s_{2}',s_{2}'}\sigma_{-\nu}^{(i)}(t_{1})\varrho_{e}\sigma_{-\nu}^{(j)\dagger}(t_{2})\ket{s_{1},s_{2},s_{2}}.
\end{align}
The result are shown in Eq (\ref{rr}).
\section{Derivation of the density matrix elements} \label{EM1}
The time evolution of the system's density matrix $\varrho(t)$ (\ref{rhot2}) is governed by a master equation derived from the influence functional formalism. The matrix elements in the sublattice basis $\ket{s_{1},s_{2},s_{3}}$ are computed by evaluating the double commutator structure that arises from the second-order perturbation in the system-field interaction. The crucial term takes the form
\begin{align}
\bra{s_{1}',s_{2}',s_{3}'}&\mathcal{R}_{ii}(\sigma_{-\nu}^{(i)}\sigma_{-\nu}^{(i)}\varrho_{e}-2\sigma_{-\nu}^{(i)}\varrho_{e}\sigma_{-\nu}^{(i)\dagger}+\varrho_{e}\sigma_{-\nu}^{(i)\dagger}\sigma_{-\nu}^{(i)\dagger})+\mathcal{R}_{ij}(i \neq j)(\sigma_{-\nu}^{(i)}\sigma_{-\nu}^{(j)}\varrho_{e}-2\sigma_{-\nu}^{(i)}\varrho_{e}\sigma_{-\nu}^{(j)\dagger}+\varrho_{e}\sigma_{-\nu}^{(i)\dagger}\sigma_{-\nu}^{(j)\dagger})\ket{s_{1},s_{2},s_{3}} \notag\\
=&-2\eta_{11}\mathcal{R}_{11}e^{i\Delta \phi_{11}}-2\eta_{22}\mathcal{R}_{22}e^{i\Delta \phi_{22}}-2\eta_{33}\mathcal{R}_{33}e^{i\Delta \phi_{33}}-2\mathcal{R}_{12}\Big[ \vartheta_{1}e^{i\Delta \phi_{21}}+
\vartheta_{2} e^{i\Delta \phi_{12}}-\frac{1}{2}\vartheta_{3}e^{i\phi_{12}}-\frac{1}{2}\vartheta_{4}e^{-i \phi_{12}^{'}}\Big]\notag\\
&-2\mathcal{R}_{13}\Big[ \zeta_{1}e^{i\Delta \phi_{31}}+ \zeta_{2}e^{i\Delta \phi_{13}}-\frac{1}{2}\zeta_{3}e^{i\phi_{13}}-\frac{1}{2}\zeta_{4}e^{-i\phi_{13}^{'}}\Big]-2\mathcal{R}_{23}\Big[ \lambda_{1}e^{i\Delta \phi_{32}}+\lambda_{2}e^{i\Delta \phi_{23}}-\frac{1}{2}\lambda_{3}e^{i\phi_{23}}-\frac{1}{2}\lambda_{4}e^{-i\phi_{12}^{'}}\Big].
\end{align}
The coefficients $\eta_{ii}$, $\vartheta_{i}$, $\zeta_{i}$, and $\lambda_{i}$ are expressed as follows
\begin{align}
&\eta_{11}=\delta_{s_{2},+}\delta_{s_{3},+}  \delta_{s_{2}',+}\delta_{s_{3}',+}(1+s_{1}s_{1}^{'}), \quad \eta_{22}=\delta_{s_{1},+}\delta_{s_{3},+}  \delta_{s_{1}',+}\delta_{s_{3}',+}(1+s_{2}s_{2}^{'}), \quad \eta_{33}=\delta_{s_{1},+}\delta_{s_{2},+}  \delta_{s_{1}',+}\delta_{s_{2}',+}(1+s_{3}s_{3}^{'}), \notag\\
&\vartheta_{1}=\delta_{s_{1},+}\delta_{s_{3},+} \delta_{s_{2}',+}\delta_{s_{3}',+}s_{1}^{'}s_{2}+\delta_{s_{2},+}\delta_{s_{3},+}  \delta_{s_{1}',+}\delta_{s_{3}',+}, \quad \vartheta_{2}=\delta_{s_{1},+}\delta_{s_{3},+}\delta_{s_{2}',+}\delta_{s_{3}',+}+\delta_{s_{2},+}\delta_{s_{3},+}  \delta_{s_{1}',+}\delta_{s_{3}',+}s_{1}s_{2}^{'}, \notag\\
&\vartheta_{3}=\delta_{s_{1},+}\delta_{s_{2},+} \delta_{s_{3},+}\delta_{s_{3}',+}+\delta_{s_{1}',+}\delta_{s_{2}',+}  \delta_{s_{3}',+}\delta_{s_{3},+}s_{1}s_{2}, \quad \vartheta_{4}=\delta_{s_{1},+}\delta_{s_{2},+}  \delta_{s_{3},+}\delta_{s_{3}',+}s_{1}'s_{2}'+\delta_{s_{1}',+}\delta_{s_{2}',+}  \delta_{s_{3}',+}\delta_{s_{3},+}, \notag \\
&\zeta_{1}=\delta_{s_{1},+}\delta_{s_{2},+}  \delta_{s_{2}',+}\delta_{s_{3}',+}s_{1}'s_{3}+\delta_{s_{2},+}\delta_{s_{3},+}  \delta_{s_{1}',+}\delta_{s_{2}',+}, \quad \zeta_{2}=\delta_{s_{1},+}\delta_{s_{2},+}  \delta_{s_{2}',+}\delta_{s_{3}',+}+\delta_{s_{2},+}\delta_{s_{3},+}  \delta_{s_{1}',+}\delta_{s_{2}',+}s_{1}s_{3}', \notag\\
&\zeta_{3}=\delta_{s_{1}',+}\delta_{s_{2}',+}  \delta_{s_{3}',+}\delta_{s_{2},+}s_{1}s_{3}+\delta_{s_{1},+}\delta_{s_{2},+}  \delta_{s_{3},+}\delta_{s_{2}',+}, \quad \zeta_{4}=\delta_{s_{1}',+}\delta_{s_{2}',+}  \delta_{s_{3}',+}\delta_{s_{2},+}+\delta_{s_{1},+}\delta_{s_{2},+}  \delta_{s_{3},+}\delta_{s_{2}',+}s_{1}'s_{3}', \notag\\
&\lambda_{1}=\delta_{s_{1},+}\delta_{s_{2},+}  \delta_{s_{1}',+}\delta_{s_{3}',+}s_{3}s_{2}'+\delta_{s_{1},+}\delta_{s_{3},+}  \delta_{s_{1}',+}\delta_{s_{2}',+}, \quad \lambda_{2}=\delta_{s_{1},+}\delta_{s_{2},+}  \delta_{s_{1}',+}\delta_{s_{3}',+}+\delta_{s_{1},+}\delta_{s_{3},+}  \delta_{s_{1}',+}\delta_{s_{2}',+}s_{2}s_{3}', \notag\\
&\lambda_{3}=\delta_{s_{1},+}\delta_{s_{2},+}  \delta_{s_{3},+}\delta_{s_{1}',+}+\delta_{s_{1}',+}\delta_{s_{2}',+}  \delta_{s_{3}',+}\delta_{s_{1},+}s_{2}s_{3}, \quad \lambda_{4}=\delta_{s_{1},+}\delta_{s_{2},+}  \delta_{s_{3},+}\delta_{s_{1}',+}s_{2}'s_{3}'+\delta_{s_{1}',+}\delta_{s_{2}',+}  \delta_{s_{3}',+}\delta_{s_{1},+},
\end{align}
where the phases are defined by $\phi_{ij}(\phi_{ij}') = \phi_{i} + \phi_{j} (\phi_{i}' + \phi_{j}')$ and $\Delta \phi_{ij} = \phi_{i} - \phi_{j}'$. The angles $\phi_{i}$ and $\phi_{i}'$ correspond to the initial and final wave vectors $\mathbf{k}_i$ and $\mathbf{k}_{i}'$ from Eq. (\ref{kk}). The resulting density matrix element is given in Eq. (\ref{rhot2}).

\end{document}